\documentclass[twocolumn,showpacs,preprintnumbers,amsmath,amssymb]{revtex4-1}

\usepackage{graphicx,latexsym}
\usepackage{amsmath,amssymb,amsfonts}
\usepackage{bm}
\usepackage{braket}
\usepackage{mathrsfs}

\newcommand{\up}{\uparrow}
\newcommand{\down}{\downarrow}
\newcommand{\m}{\mathcal}
\newcommand{\ep}{\varepsilon}
\begin{document}
\title{Nonequilibrium BCS-BEC crossover and unconventional FFLO superfluid in a strongly interacting driven-dissipative Fermi gas}
\author{Taira Kawamura\email{tairakawa@g.ecc.u-tokyo.ac.jp}$^1$ and Yoji Ohashi$^2$}
\affiliation{$^1$Department of Basic Science, The University of Tokyo, Meguro, Tokyo 153-8902, Japan}
\affiliation{$^2$Department of Physics, Keio University, 3-14-1 Hiyoshi, Kohoku-ku, Yokohama 223-8522, Japan}
\date{\today}
\begin{abstract}
We present a theoretical review of the recent progress in nonequilibrium BCS (Bardeen-Cooper-Schrieffer)-BEC (Bose-Einstein condensation) crossover physics. As a paradigmatic example, we consider a strongly interacting driven-dissipative two-component Fermi gas where the nonequilibrium steady state is tuned by adjusting the chemical potential difference between two reservoirs that are coupled with the system. As a powerful theoretical tool to deal with this system, we employ the Schwinger-Keldysh Green's function technique. We systematically evaluate the superfluid transition, as well as the single-particle properties, in the nonequilibrium BCS-BEC crossover region, by adjusting the chemical potential difference between the reservoirs and the strength of an $s$-wave pairing interaction associated with a Feshbach resonance. In the weak-coupling BCS side, the chemical potential difference is shown to imprint a two-step structure on the particle momentum distribution, leading to an anomalous enhancement of pseudogap, as well as the emergence of exotic Fulde-Ferrell-Larkin-Ovchinnikov-type superfluid instability. Since various nonequilibrium situations have recently been realized in ultracold Fermi gases, the theoretical understanding of nonequilibrium BCS-BEC crossover physics would become increasingly important in this research field.
\end{abstract}
\maketitle
\tableofcontents
\section{Introduction}

\subsection{Overview of conventional thermal equilibrium BCS-BEC crossover}

An \cite{Loftus2002} ultracold Fermi gas is an extremely dilute gas of charge-neutral Fermi atoms, which is cooled down to $\m{O}$(nK). The strength of a pairing interaction between Fermi atoms can experimentally be tuned by adjusting a Feshbach resonance \cite{Loftus2002, Dieckmann2002, Regal2003, Chin2010, PethickBook, Bloch2008}, where two Fermi atoms form a quasi-molecular boson, and it dissociates into two Fermi atoms again. Within the second-order perturbation theory with respect to the Fershbach resonance, one obtains the effective interaction between Fermi atoms, given by~\cite{PethickBook}
\begin{equation}
\label{eq.UFR}
U^{\rm FR}_{\rm eff} = -g^2 \frac{1}{2\nu}.
\end{equation}
Here, $g$ is a coupling constant of a Feshbach resonance, and $2\nu$ is the energy difference between the intermediate molecular state (closed channel) and the atomic states before and after the scattering event (open channel). The energy $2\nu$ is referred to as the threshold energy of a Feshbach resonance. Since the atomic hyperfine states in the closed channel are different from those in the open channel, their Zeeman energies also become different under an external magnetic field $B$. This naturally leads to the $B$-dependent threshold energy $2\nu$. Thus, one can tune the strength of the Feshbach-induced effective interaction $U_{\rm eff}^{\rm FR}$ in Eq.~\eqref{eq.UFR}, by adjusting the magnitude $B$ of an external magnetic field.

\begin{figure}[t]
\centering
\includegraphics[width=8cm]{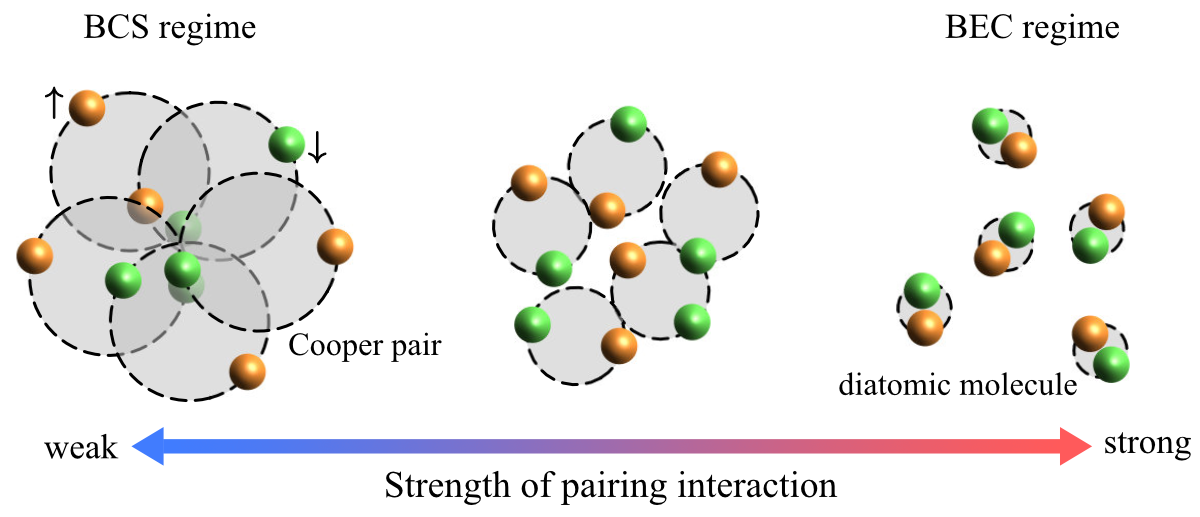}
\caption{Illustration of BCS-BEC crossover in a thermal equilibrium Fermi gas. Largely overlapping Cooper pairs in the weak-coupling BCS regime gradually shrink with increasing the strength of a pairing interaction, to eventually become the BEC of diatomic molecules.}
\label{fig.image} 
\end{figure}

\begin{figure}[t]
\centering
\includegraphics[width=8cm]{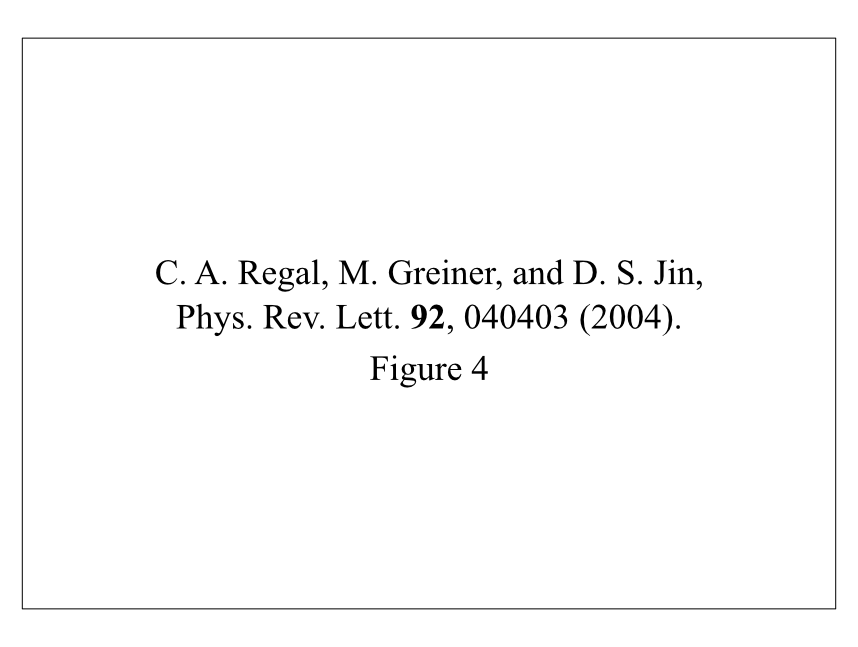}
\caption{Observed superfluid phase transition in a $^{40}{\rm K}$ Fermi gas~\cite{Regal2004}. The temperature $T$ is normalized by the Fermi temperature $T_{\rm F}=0.35{\rm \mu K}$. $\Delta B=B-B_{\rm res}$ is an applied external magnetic field, which is measured from the Feshbach resonance field $B_{\rm res}\simeq 202$ G. This axis physically represents the strength of an $s$-wave pairing interaction associated with a Feshbach resonance.  $N_0/N$ is the number of condensed Fermi atoms, being normalized by the total number $N$ of Fermi atoms. This experiment regards the region with $N_0>0$ as the superfluid phase, so that the superfluid phase transition temperature $T_{\rm c}$ exists around the sky-blue area in the $T$-$\Delta B$ plane. Adapted from Ref.~\cite{Regal2004}}
\label{fig.Exp} 
\end{figure}

An advantage of this tunable interaction is the realization of the crossover phenomenon from the Bardeen-Cooper-Schrieffer (BCS) state of Cooper pairs to the Bose-Einstein condensation (BEC) of diatomic molecules, as schematically shown in Fig.~\ref{fig.image}~\cite{Sademelo1993, Haussmann1993, Haussmann1994, Pistolesi1994, Ohashi2002, Regal2004, Giorgini2008, Ohashi2020}. Figure~\ref{fig.Exp} shows the first observation of the superfluid phase transition, as well as the BCS-BEC crossover phenomenon in a $^{40}{\rm K}$ Fermi gas. In this experiment, $^{40}{\rm K}$ atoms in two different hyperfine states (which are frequently described by pseudospin $\up$ and $\down$ in the literature) are trapped in a harmonic potential, and Cooper pairs are formed between them by a Feshbach-induced tunable $s$-wave pairing interaction. In Fig.~\ref{fig.Exp}, $\Delta B= B-B_{\rm res}$ is an external magnetic field, measured from the Feshbach resonance field $B_{\rm res}$. Physically, $\Delta B$ is directly related to the strength of a pairing interaction, that is, the decrease of $\Delta B$ corresponds to the increase of the interaction strength. In particular, the region $\Delta B > 0$ ($\Delta B < 0$) is the weak-coupling BCS (strong-coupling BEC) side. In this experiment, the superfluid state is identified as the parameter region where the number $N_0$ of condensed Cooper pairs takes a non-zero value. Thus, the superfluid phase transition temperature $T_{\rm c}$ exists around the sky-blue area in the $T$-$\Delta B$ plane in Fig.~\ref{fig.Exp}. We thus find that, starting from the weak-coupling regime ($\Delta B >0$), $T_{\rm c}$ gradually increases with increasing the strength of pairing interaction, to approach a constant value in the strong-coupling regime ($\Delta B <0$), which is known as a typical BCS-BEC crossover behavior of $T_{\rm c}$~\cite{Sademelo1993, Haussmann1993, Haussmann1994, Pistolesi1994, Ohashi2002, Regal2004, Giorgini2008, Ohashi2020}.

Soon after the realization of the above-mentioned $^{40}$K superfluid Fermi gas, the superfluid phase transition, as well as the BCS-BEC crossover phenomenon, has also been realized in a $^6$Li Fermi gas~\cite{Zwierlein2004, Kinast2004, Bartenstein2004}. At present, one can examine superfluid properties in the whole BCS-BEC crossover region by using these two kinds of Fermi gases in cold-atom physics. We also briefly note that the BCS-BEC crossover has recently been discussed in condensed matter physics, such as superconductors like FeSe and ZrNCl~\cite{Kasahara2014, Hashimoto2020, Nakagawa2021, Suzuki2022}, as well as astrophysics, such as neutron-star interior~\cite{Stein2012, Stein2016, Wyk2018, Ohashi2020}.

\subsection{Strongly interacting Fermi gas in a driven-dissipative steady state}

In cold-atom physics, since the achievement of the superfluid phase transition in $^{40}$K and $^6$Li Fermi gases, strong-coupling properties in the BCS-BEC crossover region have mainly been studied in the {\it thermal equilibrium} case. This is because the usual experimental situation of a trapped Fermi gas is well isolated from the environment and also in the (quasi)equilibrium state. However, a strong interest in nonequilibrium properties of this strongly interacting quantum system has recently emerged, being fueled by the rapid development of nonequilibrium quantum many-body physics \cite{Houck2012, Carusotto2013, Daley2014, Sanvitto2016}. For example, quench dynamics~\cite{Amico2018, Liu2021, Dyke2021, Dyke2024} and transport phenomena~\cite{Brown2019, Nichols2019} in a strongly interacting Fermi gas have been experimentally studied. Periodically-driven systems~\cite{Li2019, Shkedrov2022} and open systems with particle loss~\cite{Honda2023} have also been realized in an ultracold Fermi gas.

\begin{figure}[t]
\centering
\includegraphics[width=8cm]{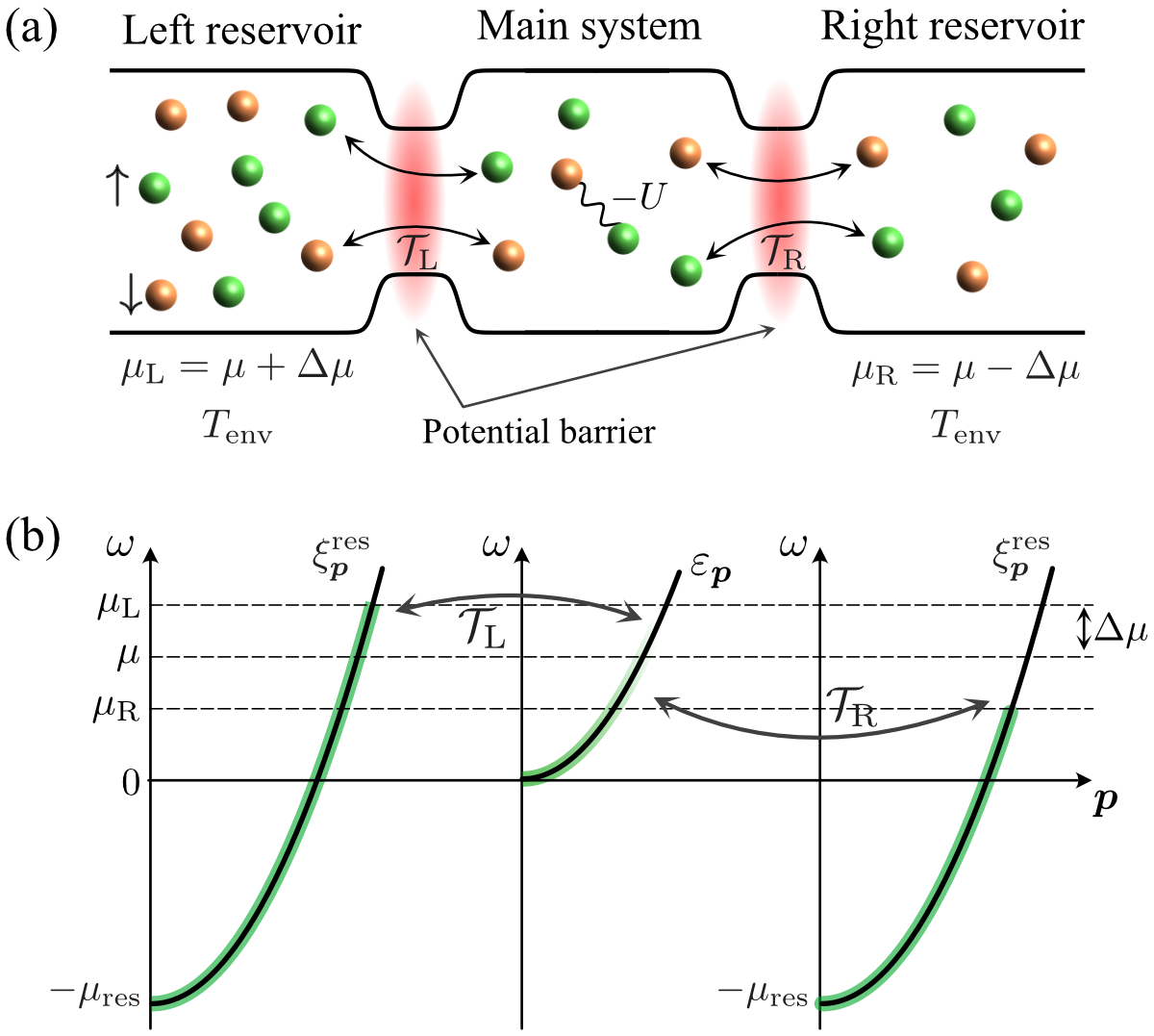}
\caption{(a) Model driven-dissipative Fermi gas with a tunable paring interaction $-U$~\cite{Kawamura2020_JLTP, Kawamura2020, Kawamura2022, Kawamura2023}. The main system is coupled with the left ($\alpha={\rm L}$) and the right ($\alpha={\rm R}$) reservoirs, which are assumed to be in the thermal equilibrium state at the temperature $T_{\rm env}$ and the chemical potential $\mu_{\alpha}$. $\m{T}_\alpha$ describes the tunneling amplitude between the main system and the $\alpha$ reservoir. (b) Schematic energy band structure of the model. We measure the energy from the bottom ($\ep_{\bm{p}=0}$) of the energy band in the main system. In the $\alpha$ reservoir at $T_{\rm env}=0$, the energy band $\xi^{\rm res}_{\bm{q}}=\bm{p}^2/(2m) -\mu_{\rm res}$ is filled up to $\mu_\alpha$. When the reservoirs have different Fermi levels ($\Delta\mu\neq 0$), the main system is driven out of equilibrium due to the pumping and decay of Fermi atoms by the two reservoirs.}
\label{fig.model}
\end{figure}

Motivated by such a recent trend in cold-atom physics, in this review, we consider a driven-dissipative Fermi gas, as schematically shown in Fig.~\ref{fig.model}(a)~\cite{Kawamura2020_JLTP, Kawamura2020, Kawamura2022, Kawamura2023}: A three-dimensional two-component Fermi gas (main system) with a tunable pairing interaction $-U$ is coupled with left (L) and right (R) reservoirs, consisting of non-interacting Fermi gases. Both reservoirs are assumed to be huge compared to the main system and are in the thermal equilibrium state at temperature $T_{\rm env}$. The main system becomes out of equilibrium when one imposes the chemical potential difference between the reservoirs as $\mu_{\rm L}=\mu+\Delta\mu$ and $\mu_{\rm R}=\mu-\Delta\mu$, where $\mu_{\alpha={\rm L}, {\rm R}}$ is the Fermi chemical potential in the $\alpha$-resevoir [see Fig.\ref{fig.model}(b)]. Thus, by tuning the pairing interaction strength $-U$, as well as the chemical potential difference $\Delta\mu =[\mu_{\rm L} -\mu_{\rm R}]/2$ between the reservoirs, we can study {\it nonequilibrium} BCS-BEC crossover physics in the main system.

The model nonequilibrium Fermi gas in Fig.~\ref{fig.model}(a) is inspired by the recent transport experiment on a $^6{\rm Li}$ Fermi gas in a two-terminal configuration~\cite{Brantut2012, Krinner2015, Husmann2015, Nagy2016, Krinner2017}: In this experiment, a two-component Fermi atomic cloud is shaped in a two-terminal setup, where two reservoirs are connected by a mesoscopic channel. By extending this two-terminal setup to the case with multiple junctions, we can prepare a system coupled with multiple reservoirs, just like the model in Fig.~\ref{fig.model}(a). We note that a narrow repulsive potential barrier produced by a Gaussian beam~\cite{Luick2020, Kwon2020, Pace2021} is also useful to separate a Fermi gas cloud into multiple subsystems. Indeed, a Josephson junction of a $^6{\rm Li}$ Fermi gas is implemented with this technique. By extending the technique, we could divide a Fermi gas into the main system and the two reservoirs shown in Fig.~\ref{fig.model}(a).

To realize the model in Fig.~\ref{fig.model}(a), we need to apply an external magnetic field only to the main system, to adjust the Feshbach-induced pairing interaction $-U$ there. This could be done by using the technique of the spatial manipulation of interaction strength~\cite{Bauer2009, Yamazaki2010,  Fu2013, Jagannathan2016, Arunkumar2018, Arunkumar2019}. In particular, in a two-component $^6{\rm Li}$ Fermi gas, the combination of the magnetic Feshbach resonance technique and the optical control enables us to adjust a scattering length with high spatial resolution~\cite{Jagannathan2016, Arunkumar2018, Arunkumar2019}. Thus, using this technique, one can independently tune the interatomic interaction in the reservoirs and the main system.

\subsection{Outline of this article}

In this review article, we discuss the nonequilibrium BCS-BEC crossover in the model shown in Fig.~\ref{fig.model}(a), as a paradigmatic example of nonequilibrium quantum many-body phenomenon in strongly interacting Fermi gases. We explain that the driving and the dissipation lead to exotic states that are never obtained in the thermal equilibrium case.

This article is organized as follows. In Sec.~\ref{sec.thermal.BCS.BEC}, we quickly review the BCS-BEC crossover theory for a thermal equilibrium Fermi gas. We also explain single-particle properties in the crossover region. In Sec.~\ref{sec.neq.MF}, we deal with the driven-dissipative Fermi gas in Fig.~\ref{fig.model}(a), to consider the nonequilibrium superfluid transition within the mean-field approximation. We show that a nonequilibrium particle energy distribution induces a Fulde-Ferrell-Larkin-Ovchinnikov (FFLO) like unconventional Fermi superfluid in spite of the absence of any spin imbalance. We then proceed to the nonequilibrium BCS-BEC crossover phenomenon in Sec.~\ref{sec.NETMA.neq.BCS.BEC}. Here we extend the thermal equilibrium BCS-BEC crossover theory to the driven-dissipative Fermi gas. We then elucidate nonequilibrium properties of the strongly interacting driven-dissipative Fermi gas. We find that the FFLO-like Fermi superfluid obtained in Sec.~\ref{sec.neq.MF} is actually unstable against pairing fluctuations. In Sec.~\ref{sec.stable.NFFLO}, we present a possible route to stabilize this unconventional Fermi superfluid state.

Throughout this article, we set $\hbar=k_{\rm B}=1$, and the system volume is taken to be unity, for simplicity.

\section{BCS-BEC crossover in a thermal equilibrium Fermi gas \label{sec.thermal.BCS.BEC}}

As a prelude to nonequilibrium BCS-BEC crossover physics in a driven-dissipative Fermi gas, this section reviews single-particle properties of a {\it thermal equilibrium} Fermi gas in the BCS-BEC crossover region.


\subsection{Strong-coupling theory for a thermal equilibrium Fermi gas \label{sec.eq.BCS.BEC}}

To theoretically describe the observed BCS-BEC crossover behavior of $T_{\rm c}$ in Fig.~\ref{fig.Exp}, we need to go beyond the mean-field approximation, to include effects of strong pairing fluctuations. This has extensively been attempted by many researchers by various methods, such as quantum Monte Carlo method~\cite{Carlson2003, Astrakharchik2004, Wlazowski2013}, renormalization group method~\cite{Gubbels2008, Boettcher2012, Boettcher2013, Tanizaki2014}, functional integral method~\cite{Klimin2015}, virial expansion~\cite{Liu2010, Hu2011}, as well as diagrammatic method~\cite{NSR1985, Perali2002, Hu2008, Ohashi2002, Haussmann1993, Haussmann1994}. Among them, here we explain the $T$-matrix approximation (TMA)~\cite{Tsuchiya2009}, which is a strong-coupling theory based on the diagrammatic technique. 

Although the Feshbach-induced pairing interaction $U^{\rm FR}_{\rm eff}$  in Eq.~\eqref{eq.UFR} is quite different from the phonon-mediated one in conventional metallic superconductivity, we can still capture the essence of the BCS-BEC crossover phenomenon, by employing the ordinary BCS Hamiltonian \cite{Ohashi2020}, given by
\begin{align}
H 
&=
H_0 + H_{\rm int}
\notag\\
&=
\sum_{\bm{p}} \xi_{\bm{p}} a^\dagger_{\bm{p}, \sigma} a_{\bm{p}, \sigma} 
\notag\\
&
-U \sum_{\bm{p}, \bm{p}', \bm{q}} a^\dagger_{\bm{p}+\bm{q}/2, \up} a^\dagger_{-\bm{p}+\bm{q}/2, \down} a_{-\bm{p}' +\bm{q}/2, \down} a_{\bm{p}' +\bm{q}/2, \up}.
\label{eq.H}
\end{align}
Here, $a^\dagger_{\bm{p}, \sigma}$ creates a Fermi atom with momentum $\bm{p}$ and (pseudo)spin $\sigma=\up, \down$. $\xi_{\bm{p}} =\bm{p}^2/(2m) -\mu$ is the kinetic energy measured from the chemical potential $\mu$, where $m$  is an atomic mass. $-U$ ($<0$) is a contact-type pairing interaction, which is assumed to be tunable by a Feshbach resonance. To remove the ultraviolet divergence coming from the contact-type interaction, we conveniently measure the interaction strength in terms of the $s$-wave scattering length $a_s$~\cite{RanderiaBook}. The scattering length $a_s$ is related to the bare interaction $U$ as
\begin{equation}
\frac{4\pi a_{\rm s}}{m} = \frac{-U}{1 -U\sum_{\bm{p}}^{p_{\rm c}} \frac{1}{2\ep_{\bm{p}}}},
\label{eq.as.U}
\end{equation}
where $p_{\rm c}$ is a momentum cutoff. The weak-coupling BCS side and the strong-coupling BEC side are then characterized by $(p_{\rm F}a_{\rm s})^{-1} \lesssim 0$ and $(p_{\rm F}a_{\rm s})^{-1} \gtrsim 0$, respectively. Here, $p_{\rm F}=(3\pi^2 N)^{1/3}$  is the Fermi momentum of a two-component free gas with $N$ fermions.

In TMA, we perturbatively include effects of pairing interaction $H_{\rm int}$ in Eq.~\eqref{eq.H}. In the thermal equilibrium state, this is usually done by using the imaginary-time Matsubara Green's function technique~\cite{AGD}. However, for later convenience, we employ the real-time Keldysh Green's function theory~\cite{Rammer2007, Zagoskin2014, Stefanucci2013} in this article. As we will see later, this formalism naturally allows the application of TMA also to the nonequilibrium case.

We introduce the following $2\times 2$ matrix single-particle Green's function:
\begin{equation}
\hat{G}_{{\rm TMA}, \sigma}(\bm{p}, \omega)=
\begin{pmatrix}
G^{\rm R}_{{\rm TMA}, \sigma}(\bm{p}, \omega) & 
G^{\rm K}_{{\rm TMA}, \sigma}(\bm{p}, \omega) \\[4pt]
0 & 
G^{\rm A}_{{\rm TMA}, \sigma}(\bm{p}, \omega) 
\end{pmatrix}.
\label{eq.G.TMA}
\end{equation}
Here, the superscripts ``R", ``A", and ``K" represent the retarded, advanced, and Keldysh components, respectively. These are respectively, defined by~\cite{Rammer2007, Stefanucci2013, Zagoskin2014}
\begin{align}
&
\scalebox{0.98}{$\displaystyle
G^{\rm R}_{{\rm TMA}, \sigma}(\bm{p}, \omega) =
-i \int_{-\infty}^\infty dt e^{-i\omega t} \Theta(t) \braket{[a_{\bm{p}, \sigma}(t), a^\dagger_{\bm{p}, \sigma}(0)]_+}$}
,\label{eq.GR.TMA} \\
&
\scalebox{0.98}{$\displaystyle
G^{\rm A}_{{\rm TMA}, \sigma}(\bm{p}, \omega) =
i \int_{-\infty}^\infty dt e^{-i\omega t} \Theta(-t) \braket{[a_{\bm{p}, \sigma}(t), a^\dagger_{\bm{p}, \sigma}(0)]_+}$}
\label{eq.GA.TMA}
,\\
&
\scalebox{0.98}{$\displaystyle
G^{\rm K}_{{\rm TMA}, \sigma}(\bm{p}, \omega) =
-i \int_{-\infty}^\infty dt e^{-i\omega t}  \braket{[a_{\bm{p}, \sigma}(t), a^\dagger_{\bm{p}, \sigma}(0)]_-}$}
,\label{eq.GK.TMA}
\end{align}
where $a_{\bm{p}, \sigma}(t)$ is the annihilation operator of a Fermi atom in the Heisenberg representation, $[A, B]_\pm = AB \pm BA$, $\Theta(t)$ is the step function, and the expectation value $\braket{\cdots}$ is taken over the BCS Hamiltonian $H$ in Eq.~\eqref{eq.H}. We find from Eqs.~\eqref{eq.GR.TMA} and \eqref{eq.GA.TMA} that the retarded Green's function is related to the advanced one as
\begin{equation}
G^{\rm R}_{{\rm TMA}, \sigma}(\bm{p}, \omega) =
\big[ G^{\rm A}_{{\rm TMA}, \sigma}(\bm{p}, \omega) \big]^*.
\label{eq.GR.GA.rel}
\end{equation}
In addition, in the thermal equilibrium state, the Keldysh Green's function is related to the retarded one via the fluctuation-dissipation relation (FDR)~\cite{Rammer2007, Zagoskin2014, Stefanucci2013} as
\begin{equation}
G^{\rm K}_{{\rm TMA}, \sigma}(\bm{p}, \omega) = 2i {\rm Im}\big[ G^{\rm R}_{{\rm TMA}, \sigma}(\bm{p}, \omega)\big] \big[1 -2f(\omega)\big].
\label{eq.FDR.TMA}
\end{equation}
Here, $f(\omega) = [e^{\omega/T}+1]^{-1}$ is the Fermi-Dirac distribution function. Equations~\eqref{eq.GR.GA.rel} and \eqref{eq.FDR.TMA} mean that in the thermal equilibrium case, once we compute the retarded component $G^{\rm R}_{{\rm TMA}, \sigma}(\bm{p}, \omega)$, the other ones in $\hat{G}_{{\rm TMA}, \sigma}(\bm{p}, \omega)$ are immediately obtained from $G^{\rm R}_{{\rm TMA}, \sigma}(\bm{p}, \omega)$. However, this is not the case when the system is out of equilibrium. In this case, since the FDR in Eq.~\eqref{eq.FDR.TMA} no longer holds, the retarded and the Keldysh Green's functions have to be evaluated independently, as we will explain in Sec.~\ref{sec.formalism.NETMA}.

\begin{figure}[t]
\centering
\includegraphics[width=8cm]{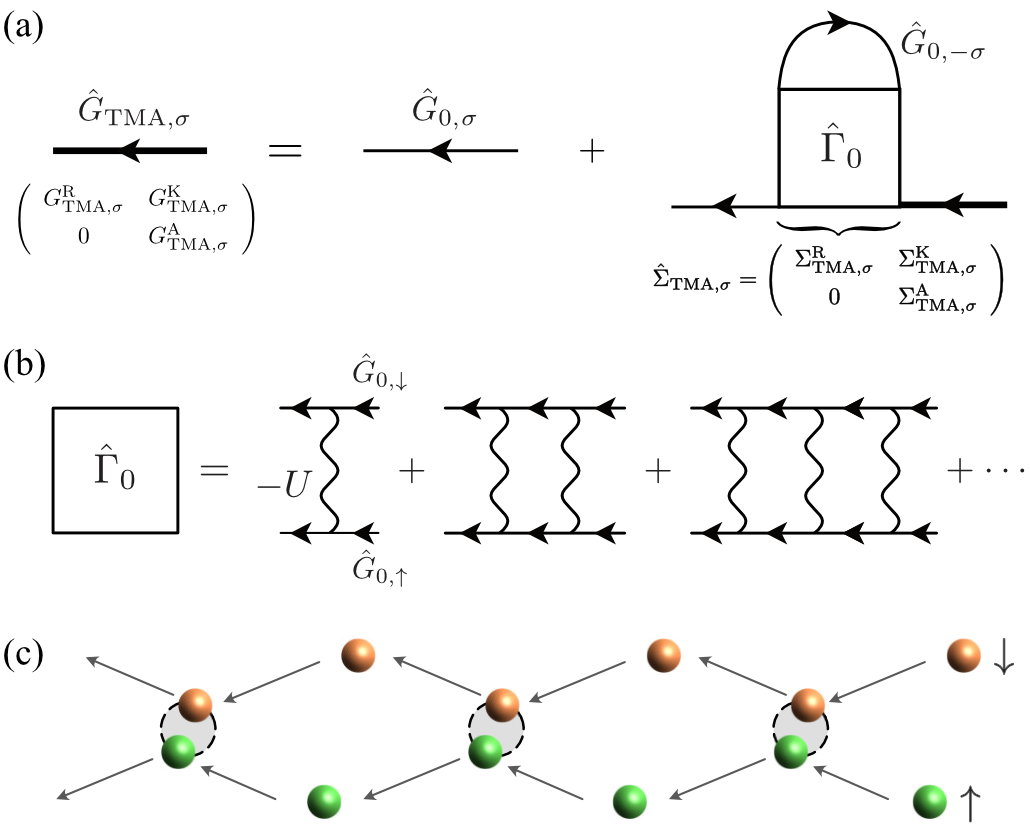}
\caption{(a) Dyson equation for $2\times 2$ matrix TMA Green's function $\hat{G}_{{\rm TMA}, \sigma}$ (thick solid line). $\hat{\Sigma}_{{\rm TMA}, \sigma}$ is the self-energy correction in TMA. (b) Particle-particle scattering matrix $\hat{\Gamma}_0$ in TMA. The wavy line represents the pairing interaction $-U$. (c) Illustration of pairing fluctuations described by $\hat{\Gamma}_0$.}
\label{fig.TMA.diagram} 
\end{figure}

The $2\times 2$ matrix Green's function $\hat{G}_{{\rm TMA}, \sigma}(\bm{p}, \omega)$ in Eq.~\eqref{eq.G.TMA} obey the Dyson equation~\cite{Rammer2007, Zagoskin2014, Stefanucci2013}, 
\begin{align}
\hat{G}_{{\rm TMA}, \sigma}(\bm{p}, \omega)
&=
\hat{G}_{0, \sigma}(\bm{p}, \omega) 
\notag\\
&\hspace{-1cm}+
\hat{G}_{0, \sigma}(\bm{p}, \omega)
\hat{\Sigma}_{{\rm TMA}, \sigma}(\bm{p}, \omega)
\hat{G}_{{\rm TMA}, \sigma}(\bm{p}, \omega),
\label{eq.Dyson.TMA}	
\end{align}
which is diagrammatically drawn as Fig.~\ref{fig.TMA.diagram}(a). Here,
\begin{align}
\hat{G}_{0, \sigma}(\bm{p}, \omega)
&=
\begin{pmatrix}
\hat{G}^{\rm R}_{0, \sigma}(\bm{p}, \omega) &
\hat{G}^{\rm K}_{0, \sigma}(\bm{p}, \omega) \\[4pt]
0 &
\hat{G}^{\rm A}_{0, \sigma}(\bm{p}, \omega) 
\end{pmatrix}
\notag\\[4pt]
&=
\begin{pmatrix}
\frac{1}{\omega +i\delta -\xi_{\bm{p}}} &
-2i \pi \delta(\omega -\xi_{\bm{p}}) \big[1 -2f(\omega)\big] \\
0 &
\frac{1}{\omega -i\delta -\xi_{\bm{p}}}
\end{pmatrix}
\label{eq.G0.Keldysh}	
\end{align}
is the bare Green's function in the case of a non-interacting Fermi gas, where $\delta$ is an infinitesimally small positive number. 

In the Dyson equation~\eqref{eq.Dyson.TMA}, the self-energy 
\begin{equation}
\hat{\Sigma}_{{\rm TMA}, \sigma}(\bm{p}, \omega) =
\begin{pmatrix}
\Sigma^{\rm R}_{{\rm TMA}, \sigma}(\bm{p}, \omega) & 
\Sigma^{\rm K}_{{\rm TMA}, \sigma}(\bm{p}, \omega) \\[4pt]
0 & \Sigma^{\rm A}_{{\rm TMA}, \sigma}(\bm{p}, \omega) 
\end{pmatrix}
\end{equation}
describes effects of the strong paring interaction $-U$. The self-energy $\hat{\Sigma}_{{\rm TMA}, \sigma}(\bm{p}, \omega)$ in TMA is diagrammatically drawn as Fig.~\ref{fig.TMA.diagram}(a), which gives~\cite{Kawamura2020_JLTP, Kawamura2020}
\begin{align}
&
\Sigma^{\rm R}_{{\rm TMA}, \sigma}(\bm{p}, \omega)=
\big[\Sigma^{\rm A}_{{\rm TMA}, \sigma}(\bm{p}, \omega) \big]^*
\notag\\[4pt]
&=
-\frac{i}{2} \sum_{\bm{q}} \int_{-\infty}^\infty \frac{d\nu}{2\pi} \Big[
\Gamma^{\rm R}_0(\bm{q}, \nu) G^{\rm K}_{0, -\sigma}(\bm{q} -\bm{p}, \nu -\omega) 
\notag\\
&\hspace{2.6cm}+
\Gamma^{\rm K}_0(\bm{q}, \nu) G^{\rm A}_{0, -\sigma}(\bm{q} -\bm{p}, \nu -\omega)
\Big],\label{eq.self.R.TMA}
\\[4pt]
&
\Sigma^{\rm K}_{{\rm TMA}, \sigma}(\bm{p}, \omega)= 
2i {\rm Im} \big[\Sigma^{\rm R}_{{\rm TMA}, \sigma}(\bm{p}, \omega) \big] \big[ 1 -2f(\omega) \big].
\end{align}
Here,  ``$-\sigma$" means the opposite component to $\sigma$. In Eq.~\eqref{eq.self.R.TMA}, $\Gamma^{\rm R (K)}_0(\bm{q}, \nu)$ is the retarded (Keldysh) component of the particle-particle scattering matrix,
\begin{equation}
\hat{\Gamma}_0(\bm{q}, \nu) =
\begin{pmatrix}
\Gamma^{\rm R}_0(\bm{q}, \nu) & 
\Gamma^{\rm K}_0(\bm{q}, \nu) \\[4pt]
0 & \Gamma^{\rm A}_0(\bm{q}, \nu)
\end{pmatrix},
\label{eq.TMA.Tmat}
\end{equation}
which is given by the series of the ladder-type diagrams shown in Fig.~\ref{fig.TMA.diagram}(b). These ladder diagrams physically describe pairing fluctuations, that is, the sequence of pair formation and dissociation of Fermi atoms, as schematically shown in Fig.~\ref{fig.TMA.diagram}(c). The summation of the ladder diagrams in Fig.\ref{fig.TMA.diagram}(b) gives
\begin{align}
&
\Gamma^{\rm R}_0(\bm{q}, \nu) = 
\big[\Gamma^{\rm A}_0(\bm{q}, \nu)\big]^*=
\frac{-U}{1+U\Pi^{\rm R}_0(\bm{q}, \nu)}
\label{eq.TMA.TR}
,\\[4pt]
&\Gamma^{\rm K}_0(\bm{q}, \nu)= 2i {\rm Im}\big[\Gamma^{\rm R}_0(\bm{q}, \nu)\big]\big[1 +2n_{\rm B}(\omega) \big],
\label{eq.TMA.TK}
\end{align}
where $n_{\rm B}(\omega) = [e^{\omega/T}-1]^{-1}$ is the Bose distribution function. In Eq.~\eqref{eq.TMA.TR}, 
\begin{align}
\Pi^{\rm R}_0(\bm{q}, \nu)&=
\frac{i}{2}\sum_{\bm{p}} \int_{-\infty}^\infty \frac{d\omega}{2\pi}
\notag\\
&\hspace{0.2cm}\times \big[
G^{\rm R}_{0, \up}(\bm{p}+\bm{q}/2, \omega +\nu)
G^{\rm K}_{0, \down}(-\bm{p}+\bm{q}/2, -\omega)
\notag\\[4pt]
&\hspace{0.3cm}+
G^{\rm K}_{0, \up}(\bm{p}+\bm{q}/2, \omega +\nu)
G^{\rm R}_{0, \down}(-\bm{p}+\bm{q}/2, -\omega)
\big]
\notag\\[4pt]
&= 
\sum_{\bm{p}} \frac{f(\xi_{\bm{p}+\bm{q}/2}) +f(\xi_{-\bm{p}+\bm{q}/2}) -1}{\nu +i\delta -\xi_{\bm{p}+\bm{q}/2} -\xi_{-\bm{p}+\bm{q}/2}}.
\label{eq.TMA.Pi.R}
\end{align}
is the lowest order pair correlation function~\cite{Kawamura2020_JLTP, Kawamura2020}. In obtaining the second line in Eq.~\eqref{eq.TMA.Pi.R}, we have usesd Eq.~\eqref{eq.G0.Keldysh}.

The Dyson equation~\eqref{eq.Dyson.TMA} gives the dressed retarded (advanced) Green's function $G^{\rm R(A)}_{{\rm TMA}, \sigma}(\bm{p}, \omega)$ as
\begin{equation}
G^{\rm R(A)}_{{\rm TMA}, \sigma}(\bm{p}, \omega) =
\frac{1}{\omega -\xi_{\bm{p}} -\Sigma^{\rm R(A)}_{{\rm TMA}, \sigma}(\bm{p}, \omega)}.
\label{eq.TMA.fullG}
\end{equation}
The Keldysh component $G^{\rm K}_{{\rm TMA}, \sigma}(\bm{p}, \omega)$ is then immediately obtained from the FDR in Eq.~\eqref{eq.FDR.TMA}.

In TMA, physical quantities are obtained from the dressed Green's function $\hat{G}_{{\rm TMA}, \sigma}(\bm{p}, \omega)$ in Eq.~\eqref{eq.G.TMA}. The total number $N$ of Fermi atoms is evaluated from the Keldysh component $G^{\rm K}_{{\rm TMA}, \sigma}(\bm{p}, \omega)$: Noting the definition of the Keldysh component in Eq.~\eqref{eq.GK.TMA}, we have
\begin{align}
N 
&= \sum_{\sigma= \up, \down} \sum_{\bm{p}} \braket{a^\dagger_{\bm{p}, \sigma} a_{\bm{p}, \sigma}}	
\notag\\
&= -\frac{1}{2}\sum_{\sigma= \up, \down} \sum_{\bm{p}} \braket{\big[a_{\bm{p}, \sigma}, a^\dagger_{\bm{p}, \sigma}\big]_{-}} +\frac{1}{2}
\notag\\
&= -\frac{i}{2}\sum_{\sigma= \up, \down} \sum_{\bm{p}} \int_{-\infty}^\infty \frac{d\omega}{2\pi} G^{\rm K}_{{\rm TMA}, \sigma}(\bm{p}, \omega) +\frac{1}{2}.
\label{eq.N.TMA}
\end{align}
The substitution of Eq.~\eqref{eq.FDR.TMA} into Eq.~\eqref{eq.N.TMA} gives
\begin{align}
N 
&=
\sum_{\sigma= \up, \down} \sum_{\bm{p}} \int_{-\infty}^\infty d\omega A_{{\rm TMA}, \sigma}(\bm{p}, \omega) f(\omega)
\notag\\
&=
\sum_{\sigma= \up, \down} \int_{-\infty}^\infty d\omega \rho_{{\rm TMA}, \sigma}(\omega) f(\omega).
\label{eq.N.TMA.2}
\end{align}
Here, 
\begin{align}
&
A_{{\rm TMA}, \sigma}(\bm{p}, \omega) = 
-\frac{1}{\pi}{\rm Im}\big[G^{\rm R}_{{\rm TMA}, \sigma}(\bm{p}, \omega)\big]
\notag\\[4pt]
&=
\frac{1}{\pi} \frac{-{\rm Im} \Sigma^{\rm R}_{{\rm TMA}, \sigma}(\bm{p}, \omega)}{\big[\omega -\xi_{\bm{p}} -{\rm Re}\Sigma^{\rm R}_{{\rm TMA}, \sigma}(\bm{p}, \omega)\big]^2 +\big[{\rm Im}\Sigma^{\rm R}_{{\rm TMA}, \sigma}(\bm{p}, \omega)\big]^2 },
\label{eq.SW.TMA}
\\
&
\rho_{{\rm TMA}, \sigma}(\omega) = 
\sum_{\bm{p}} A_{{\rm TMA}, \sigma}(\bm{p}, \omega).
\label{eq.DOS.TMA}
\end{align}
are the single-particle spectral function and the density of states, respectively~\cite{Rammer2007, Zagoskin2014, Stefanucci2013}.

One sees from the expression for the spectral function in Eq.~\eqref{eq.SW.TMA} how the pairing interaction affects single-particle excitations: The pairing interaction gives rise to the energy shift ${\rm Re}\Sigma^{\rm R}_{{\rm TMA}, \sigma}(\bm{p}, \omega)$, as well as the broadening of the linewidth ${\rm Im}\Sigma^{\rm R}_{{\rm TMA}, \sigma}(\bm{p}, \omega)$. We will explain these effects in more detail in Sec.~\ref{sec.SW.DOS.eq}.

In the TMA scheme, the strong-coupling effects associated with the pairing interaction are incorporated into the theory by solving the number equation~\eqref{eq.N.TMA.2} to determine the chemical potential $\mu$ for a given number $N$ of Fermi atoms and the temperature $T$ ($\geq T_{\rm c}$). The superfluid phase transition temperature $T_{\rm c}$ is determined by solving the number equation~\eqref{eq.N.TMA.2}, together with the Thouless criterion \cite{Thouless1960}. As shown by Kadanoff and Martin \cite{Kadanoff1961, KadanoffBook}, the system experiences superfluid instability, when the retarded particle-particle scattering matrix $\Gamma^{\rm R}_0(\bm{q}, \nu)$ in Eq.~\eqref{eq.TMA.TR} has a pole at $\bm{q}=\bm{q}_{\rm pair}$ and $\nu= \mu_{\rm pair}$, that is, 
\begin{equation}
\big[	\Gamma_0^{\rm R}(\bm{q}=\bm{q}_{\rm pair}, \nu=\mu_{\rm pair}) \big]^{-1}=0.
\label{eq.Thouless}
\end{equation}
We note that the momentum $\bm{q}_{\rm pair}$ and the energy $\mu_{\rm pair}$ physically describe the center-of-mass momentum and the energy of a Cooper pair, respectively. Since $\Gamma^{\rm R}_0(\bm{q}, \nu)$ in Eq.~\eqref{eq.TMA.TR} is a complex function, Eq.~\eqref{eq.Thouless} actually consists of two equations,
\begin{align}
& {\rm Re}\big[\Gamma^{\rm R}_0(\bm{q}=\bm{q}_{\rm pair}, \nu=\mu_{\rm pair})\big]^{-1}=0,	 \label{eq.KM.Re}
\\
& {\rm Im}\big[\Gamma^{\rm R}_0(\bm{q}=\bm{q}_{\rm pair}, \nu=\mu_{\rm pair})\big]^{-1}=0.
\end{align}
The latter equation is solved analytically: Substituting $\Pi^{\rm R}_0(\bm{q}, \nu)$ in Eq.~\eqref{eq.TMA.Pi.R} into $\Gamma^{\rm R}_0(\bm{q}, \nu)$ in Eq.~\eqref{eq.TMA.TR}, one has
\begin{align}
&
\delta(\mu_{\rm pair} -\xi_{\bm{p}+\bm{q}_{\rm pair}/2} -\xi_{\bm{p}+\bm{q}_{\rm pair}/2})
\notag\\[4pt]
&\times
\sum_{\bm{p}}\frac{f(\xi_{\bm{p}+\bm{q}_{\rm pair}/2}) +f(\xi_{-\bm{p}+\bm{q}_{\rm pair}/2}) -1}{\xi_{\bm{p}+\bm{q}_{\rm pair}/2} +\xi_{-\bm{p}+\bm{q}_{\rm pair}/2}} = 0. 
\label{eq.Thouless0.Im}
\end{align}
Since $f(x)+f(-x)=1$, Eq.~\eqref{eq.Thouless0.Im} is satisfied only when $\mu_{\rm pair}=0$. Substituting this into Eq.~\eqref{eq.KM.Re}, we have
\begin{equation}
\frac{1}{U} = \sum_{\bm{p}} \frac{1 -f(\xi_{\bm{p}+\bm{q}_{\rm pair}/2}) -f(\xi_{-\bm{p}+\bm{q}_{\rm pair}/2})}{\xi_{\bm{p}+\bm{q}_{\rm pair}/2} +\xi_{-\bm{p}+\bm{q}_{\rm pair}/2}},
\label{eq.KM.Re2}
\end{equation}
which is just the well-known Thouless criterion (or the gap equation)~\cite{Thouless1960}. In Eq.~\eqref{eq.KM.Re2}, $\bm{q}_{\rm pair}$ is chosen so as to obtain the highest $T_{\rm c}$. In the thermal equilibrium and spin-balanced ($N_\up=N_\down$) case, we obtain $\bm{q}_{\rm pair}=0$ because Cooper pairs must have zero center-of-mass momentum in this case. Setting $\bm{q}_{\rm pair}=0$ in Eq.~\eqref{eq.KM.Re2}, we have
\begin{equation}
\frac{1}{U} = \sum_{\bm{p}} \frac{1 -2f(\xi_{\bm{p}})}{2\xi_{\bm{p}}}.
\label{eq.KM.Re3}
\end{equation}

\begin{figure}[t]
\centering
\includegraphics[width=8cm]{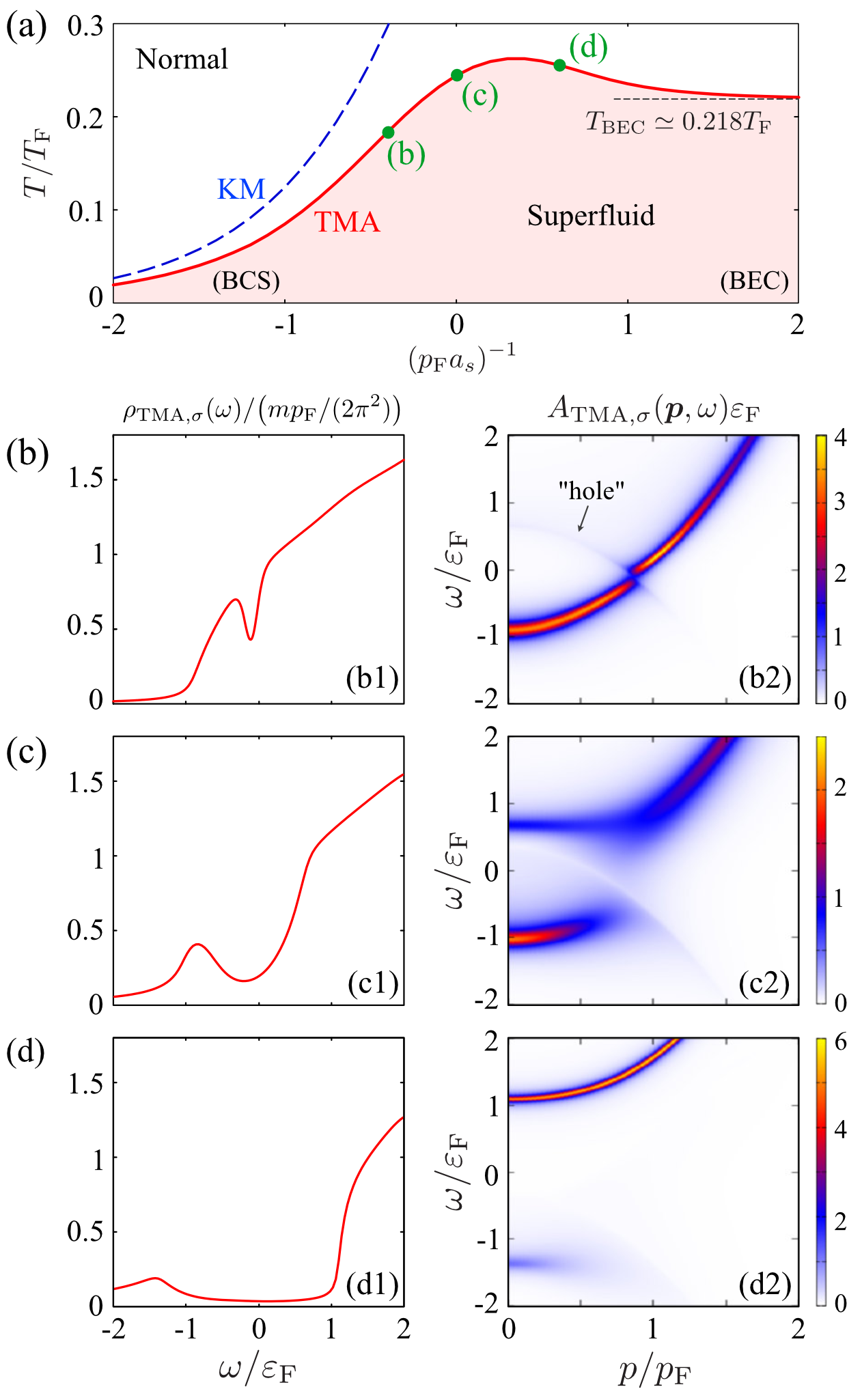}
\caption{(a) Calculated $T_{\rm c}$ in a thermal equilibrium Fermi gas in the BCS-BEC crossover region. ``TMA" (solid line) and ``KM" (dashed line), respectively, show the results of TMA and the mean-field KM theory. (b)-(d) Single-particle properties in the BCS-BEC crossover region. (b1)-(d1) Density of states $\rho_{{\rm TMA},\sigma}(\omega)$ in Eq.~\eqref{eq.DOS.TMA}. (b2)-(d2) Spectral function $A_{{\rm TMA}, \sigma}(\bm{p}, \omega)$ in Eq.~\eqref{eq.SW.TMA}. Each panel shows the result at (b)-(d) in panel (a). In panel (b2), ``hole" is the peak line along the hole dispersion $\omega = -\xi_{\bm{p}}$.}
\label{fig.TMA.DOS} 
\end{figure}

Figure~\ref{fig.TMA.DOS}(a) shows $T_{\rm c}$ obtained by solving the TMA coupled equations~\eqref{eq.N.TMA.2} and \eqref{eq.KM.Re3}. For comparison, we also show the results in the case when the Thouless criterion in Eq.~\eqref{eq.KM.Re3} is solved for the fixed value of the chemical potential $\mu=\ep_{\rm F}$. The latter is just the mean-field approximation, which is also referred to as the Kadanoff-Martin (KM) theory in the literature~\cite{Kadanoff1961, KadanoffBook}. Figure~\ref{fig.TMA.DOS}(a) indicates that the behavior of the calculated $T_{\rm c}$ in TMA agrees well with the experimental result shown in Fig.~\ref{fig.Exp}. That is, starting from the weak-coupling BCS regime, $T_c$ gradually increases with increasing the interaction strength, to approach a constant value in the BEC regime. Since all Fermi atoms form tightly bound molecules in the extreme BEC limit, this value just equals the BEC phase transition temperature
\begin{equation}
T_{\rm BEC} = \frac{2\pi}{\zeta(3/2)^{2/3}} \frac{N_{\rm M}^{2/3}}{M_{\rm M}} \simeq 0.218 T_{\rm F}
\label{eq.Tc.BEC}
\end{equation}
in an ideal gas of $N_{\rm M}=N/2$ bosons with the molecular mass $M_{\rm M}=2m$. In Eq.~\eqref{eq.Tc.BEC}, $\zeta(3/2) \simeq 2.612$ is the Riemann zeta function. 

On the other hand, the mean-field KM theory cannot properly describe the behavior of $T_{\rm c}$ in the strong-coupling BEC regime [see the dashed line in Fig.~\ref{fig.TMA.DOS}(a)]. This is simply because the KM theory ignores pairing fluctuations, as well as the formation of diatomic molecules in this regime.

\subsection{Single-particle properties in the thermal equilibrium BCS-BEC crossover regime \label{sec.SW.DOS.eq}}

In the BCS-BEC crossover region, a characteristic strong-coupling phenomenon appearing in the single-particle excitations is the pseudogap, where fluctuations in the Cooper channel play an essential role~\cite{Perali2002, Tsuchiya2009, Watanabe2010, Mueller2011, Magierski2011, Wlazlowski2013}. To explain this phenomenon, we first recall the non-interacting case, where the single-particle spectral function $A_{0, \sigma}(\bm{p}, \omega)$ and the density of states $\rho_{0, \sigma}(\omega)$ in the non-interacting Fermi gas are, respectively, obtained from $G^{\rm R}_{0, \sigma}(\bm{p}, \omega)$ in Eq.~\eqref{eq.G0.Keldysh} as
\begin{align}
& A_{0, \sigma}(\bm{p}, \omega) =
-\frac{1}{\pi}{\rm Im}\big[G^{\rm R}_{0, \sigma}(\bm{p}, \omega)\big] =\delta(\omega -\xi_{\bm{p}}) \label{eq.SW.free}
,\\
& \rho_{0, \sigma}(\omega) =
\sum_{\bm{p}}A_{0, \sigma}(\bm{p}, \omega) = \frac{(2m)^{3/2}}{2\pi^2} \sqrt{\omega+\mu}.
\label{eq.DOS.free}
\end{align}
As seen from these expressions, the spectral function $A_{0, \sigma}(\bm{p}, \omega)$ has the peak line along the particle dispersion $\omega=\xi_{\bm{p}}$. The single-particle density of states $\rho_{0, \sigma}(\omega)$ is a monotonically increasing function of $\omega$. These simple structures are remarkably modified in the BCS-BEC crossover region due to the presence of strong pairing fluctuations, as shown in Fig.~\ref{fig.TMA.DOS}(b)-(d). Although the superfluid order parameter vanishes at $T_{\rm c}$, Fig.~\ref{fig.TMA.DOS}(b1) shows that the single-particle density of states $\rho_{{\rm TMA}, \sigma}(\omega)$ still exhibits a dip structure around the Fermi level ($\omega=0$). This so-called pseudogap structure~\cite{Perali2002, Tsuchiya2009, Watanabe2010, Mueller2011, Magierski2011, Wlazlowski2013} becomes more remarkable, as one passes through the BCS-BEC crossover region [see Figs.~\ref{fig.TMA.DOS}(c1) and \ref{fig.TMA.DOS}(d1)].

To quickly understand the role of pairing fluctuations in the pseudogap phenomenon, we approximate the retarded self-energy $\Sigma^{\rm R}_{{\rm TMA}, \sigma}(\bm{p}, \omega)$ in Eq.~\eqref{eq.self.R.TMA} in the following manner: Noting that $\Gamma^{\rm R, K}_0(\bm{q}=0, \nu=0)$ diverges when the Thouless criterion in Eq.~\eqref{eq.Thouless} is satisfied at $T_{\rm c}$, one approximates the self-energy to
\begin{equation}
\Sigma^{\rm R}_{{\rm TMA}, \sigma}(\bm{p}, \omega) \simeq
-\Delta_{\rm PG}^2 G^{\rm A}_{0, -\sigma}(-\bm{p}, -\omega).
\label{eq.PGapp.self.TMA}
\end{equation}
Here,
\begin{equation}
\Delta^2_{\rm PG}= -\frac{i}{2}\sum_{\bm{q}} \int_{-\infty}^\infty \frac{d\nu}{2\pi}  \Gamma^{\rm K}_0(\bm{q}, \nu),
\label{eq.Delta.PG.0}
\end{equation}
describes effects of pairing fluctuations, which is also referred to as the pseudogap parameter in the literature~\cite{Tsuchiya2009}. Substituting Eq.~\eqref{eq.PGapp.self.TMA} into the Dyson equation~\eqref{eq.Dyson.TMA}, we have
\begin{equation}
G^{\rm R}_{{\rm TMA}, \sigma}(\bm{p}, \omega) \simeq
\frac{1}{\omega +i\delta -\xi_{\bm{p}} -\Delta_{\rm PG}^2 \dfrac{1}{\omega +i\delta +\xi_{\bm{p}}}}.
\label{eq.PGapp.GR.TMA}
\end{equation}
Equation~\eqref{eq.PGapp.GR.TMA} indicates that pairing fluctuations described by the pseudogap parameter $\Delta_{\rm PG}$ brings about a coupling phenomenon between the particle band $(\omega =\xi_{\bm{p}})$ and the hole band $(\omega = -\xi_{\bm{p}})$. Evaluating the single-particle spectral function in Eq.~\eqref{eq.SW.TMA} in this so-called pseudogap approximation, one obtains
\begin{align}
&
A_{{\rm TMA}, \sigma}(\bm{p}, \omega) 
\notag\\
&\simeq
\frac{1}{2}\left[1 +\frac{\xi_{\bm{p}}}{\sqrt{\xi^2_{\bm{p}} +\Delta_{\rm PG}^2}}\right] \delta\Big(\omega -\sqrt{\xi_{\bm{p}}^2 +\Delta_{\rm PG}^2}\Big) 
\notag\\
&\hspace{0.5cm}+
\frac{1}{2}\left[1 -\frac{\xi_{\bm{p}}}{\sqrt{\xi^2_{\bm{p}} +\Delta_{\rm PG}^2}}\right] \delta\Big(\omega +\sqrt{\xi_{\bm{p}}^2 +\Delta_{\rm PG}^2}\Big).
\label{eq.PGapp.SW}	
\end{align}
Equation~\eqref{eq.PGapp.SW} just has the same form as the single-particle spectral function in the mean-field BCS theory where the superfluid order parameter is replaced by the pseudogap parameter $\Delta_{\rm PG}$. Thus, Eq.~\eqref{eq.PGapp.SW} has the excitation gap $\Delta_{\rm PG}$ around $\omega=0$. From the viewpoint of the above-mentioned particle-hole coupling by pairing fluctuations, the level repulsion between the particle- and hole-band around $\omega=0$ opens this energy gap [see also Fig.~\ref{fig.TMA.DOS}(b1)]. While the pseudogap approximation gives a clear single-particle excitation gap as in the BCS superfluid state, quasiparticle lifetime associated with particle-particle scatterings actually rounds this gap structure to some extent. Because of this, this phenomenon appears as a dip in the single-particle density of states, as seen in Fig.~\ref{fig.TMA.DOS}(b1).

The pseudogap develops with increasing the interaction strength, reflecting the enhancement of pairing fluctuations, as shown in Fig.~\ref{fig.TMA.DOS}(c1) and (c2). In the strong-coupling BEC regime, the density of states $\rho_{{\rm TMA}, \sigma}(\omega)$, as well as the spectral function $A_{{\rm TMA}, \sigma}(\bm{p}, \omega)$ has a clear energy gap, rather than the pseudogap. These gapped excitation spectra reflect the formation of tightly-bounded diatomic molecules~\cite{Tsuchiya2009}. The energy gap corresponds to the binding energy $2|\mu|= 1/(m a_{\rm s})^2$ of a two-body bound molecule in the strong-coupling BEC limit.

In the current stage of cold Fermi gas physics, it is still difficult to directly observe the single-particle spectral function, as well as the density of states. Regarding this, however, we note that the photoemission spectrum (PES) $L_{{\rm TMA}, \sigma}(\bm{p}, \omega)$ is observable~\cite{Stewart2008, Sagi2015, Torma2016}, which is related to the spectral function $A_{{\rm TMA}, \sigma}(\bm{p}, \omega)$ as~\cite{Torma2016}
\begin{equation}
L_{{\rm TMA}, \sigma}(\bm{p}, \omega) \propto p^2 f(\omega) A_{{\rm TMA}, \sigma}(\bm{p}, \omega),
\label{eq.PES.TMA}
\end{equation}
where $f(\omega)$ is the Fermi-Dirac distribution function. In a sense, PES may be viewed as the occupied spectral function. Thus, the structural changes of the spectral function in the BCS-BEC crossover region can be observed through the photoemission-type experiment~\cite{Sagi2015, Ota2017}. 

\section{Nonequilibrium superfluid transition in the driven-dissipative Fermi gas: Mean-field approach \label{sec.neq.MF}}

We now proceed to the nonequilibrium case.  In Sec.~\ref{sec.model}, we present the model Hamiltonian for the driven-dissipative Fermi gas in Fig.~\ref{fig.model}(a). In Sec.~\ref{sec.noneq.nonint}, we explain how the couplings with the reservoirs affect the single-particle properties of the main system. To apply the mean-field KM theory to the driven-dissipative Fermi gas, we extend the Thouless criterion to the nonequilibrium case in Sec.~\ref{sec.noneq.Thouless}. Using the nonequilibrium Thouless criterion, we study the nonequilibrium superfluid phase transition within the mean-field approximation in Sec.~\ref{sec.NFFLO.NMF}.

\subsection{Model Hamiltonian for the driven-dissipative Fermi gas \label{sec.model}}
The model driven-dissipative Fermi gas in Fig.~\ref{fig.model}(a) is described by the Hamiltonian~\cite{Kawamura2020_JLTP, Kawamura2020, Kawamura2022}
\begin{equation}
H = H_{\rm sys} + H_{\rm env} + H_{\rm mix}.
\end{equation}
Here,
\begin{align}
H_{\rm sys} 
&= 
\sum_{\sigma=\up, \down}\sum_{\bm{p}} \ep_{\bm{p}} a^\dagger_{\bm{p}, \sigma} a_{\bm{p}, \sigma} 
\notag\\
&\hspace{0.5cm}
-U \sum_{\bm{p}, \bm{p}', \bm{q}} a^\dagger_{\bm{p}+\bm{q}/2, \up} a^\dagger_{-\bm{p}+\bm{q}/2, \down} a_{-\bm{p}'+\bm{q}/2, \down} a_{\bm{p}'+\bm{q}/2, \up}
\label{eq.Hsys.main}
\end{align}
describes the main system (nonequilibrium interacting Fermi gas) in Fig.~\ref{fig.model}(a), where $a^\dagger_{\bm{p}, \sigma}$ is the creation operator of a Fermi atom with pseudospin $\sigma=\up, \down$ in the main system. $\ep_{\bm{p}}=\bm{p}^2/(2m)$ is the kinetic energy of a Fermi atom with an atomic mass $m$. $-U$ is a tunable pairing interaction associated with a Feshbach resonance. As in the thermal equilibrium case discussed in Sec.~\ref{sec.eq.BCS.BEC}, we measure the interaction strength in terms of the $s$-wave scattering length $a_s$, which is related to $-U$ via Eq.~\eqref{eq.as.U}.
 
The left ($\alpha={\rm L}$) and right ($\alpha={\rm R}$) reservoirs in Fig.~\ref{fig.model}(a) are described by
\begin{equation}
H_{\rm env} = \sum_{\alpha={\rm L}, {\rm R}}
\sum_{\sigma=\up, \down} \sum_{\bm{p}} \xi^{\rm res}_{\bm{p}} c^{\alpha\dagger}_{\bm{p}, \sigma} c_{\bm{p}, \sigma}^\alpha.
\end{equation}
Here, $c^{\alpha}_{\bm{q}, \sigma}$ is an anhiration operator of a Fermi atom in the $\alpha$ reservoir, and $\xi^{\rm res}_{\bm{p}}= \ep_{\bm{p}} -\mu_{\rm res}$ is the single-particle energy in the reservoirs [see Fig.~\ref{fig.model}(b)]. We assume that the reservoirs are huge compared to the main system and always in the thermal equilibrium state at the common (environment) temperature $T_{\rm env}$. Under this assumption, the Fermi atoms in the reservoirs obey the ordinary Fermi-Dirac distribution function,
\begin{equation}
f_{\rm env}(\omega) = \frac{1}{e^{\omega/T_{\rm env}}+1}.
\label{eq.Fermi}
\end{equation}

The tunnelings of the Fermi atoms between the main system and reservoirs are described by the tunneling Hamiltonian $H_{\rm mix}$, given by
\begin{align}
&
H_{\rm mix}= 
\sum_{\alpha={\rm L}, {\rm R}} \sum_{j=1}^{N_{\rm t}} \sum_{\sigma=\up, \down}\sum_{\bm{p}, \bm{q}} 
\notag\\
&\hspace{0.6cm}\times
\Big[e^{i\mu_\alpha t} \m{T}_{\alpha, \bm{q}, \bm{p}} c^{\alpha\dagger}_{\bm{q}, \sigma} a_{\bm{p}, \sigma} e^{-i \bm{q}\cdot \bm{R}_j^\alpha} e^{-i \bm{p}\cdot \bm{r}_j^\alpha} +{\rm H.c.}\Big],
\label{eq.Hmix}
\end{align}
where $\m{T}_{\alpha, \bm{q}, \bm{p}}$ is a tunneling matrix element between the main system and the $\alpha$ ($=$L, R) reservoir. For simplicity, we ignore the momentum and $\alpha$ dependence of the tunneling matrix element, to set $\m{T}_{{\rm L}, \bm{q}, \bm{p}}=\m{T}_{{\rm R}, \bm{q}, \bm{p}} \equiv \m{T}$. In the tunneling Hamiltonian in Eq.~\eqref{eq.Hmix}, the atom tunneling is assumed to occur between randomly distributing spatial positions $\bm{R}_i^\alpha$ in the $\alpha$ reservoir and $\bm{r}_i^\alpha$ in the main system ($i=1,\cdots, N_{\rm t} \gg 1$). The introduction of the random tunneling points is just a simple theoretical trick to study the bulk properties of the main system \cite{Kawamura2020_JLTP, Kawamura2020, Kawamura2022, Kawamura2023}. After taking the spatial average over the randomly distributing tunneling positions, the translational invariance of the main system is recovered.

The factor $e^{i\mu_\alpha t}$ in Eq.~\eqref{eq.Hmix} describes the situation where the energy band $\xi_{\bm{q}}$ in the $\alpha$ reservoir is occupied up to their respective Fermi levels $\mu_\alpha =\mu \pm \Delta \mu$ when $T_{\rm env}=0$, as schematically shown in Fig.~\ref{fig.model}(b). Due to this factor, Fermi atoms are injected into and extracted from the main system when $\mu_{\rm L} \neq \mu_{\rm R}$.

We note that the temperature in the main system is not well defined in the nonequilibrium case when $\Delta\mu \neq 0$. In this case, the superfluid instability in the main system is controlled by the temperature $T_{\rm env}$ in the thermal equilibrium reservoirs. To emphasize this, in what follows, we write the superfluid transition temperature in the driven-dissipative Fermi gas as $T_{\rm env}^{\rm c}$.

\subsection{Nonequilibrium properties of the driven-dissipative non-interacting Fermi gas \label{sec.noneq.nonint}}
\begin{figure}[t]
\centering
\includegraphics[width=8cm]{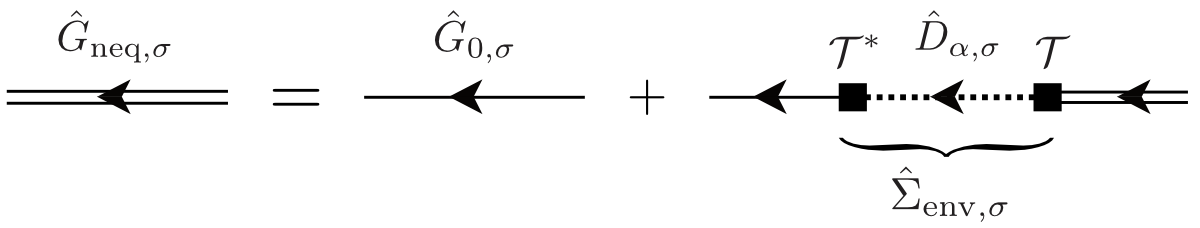}
\caption{Dyson equation for $2\times 2$ matrix nonequilibrium Green's function $\hat{G}_{{\rm neq}, \sigma}$ (double solid line). The self-energy $\hat{\Sigma}_{{\rm env}, \sigma}$ describes the system-reservoir coupling effects in the second-order Born approximation with respect to the tunneling amplitude $\m{T}$ (solid square). $\hat{D}_{\alpha, \sigma}$ (dotted line) denotes the non-interacting Green's function in the $\alpha$ ($=$L, R) reservoir.}
\label{fig.Genv}
\end{figure}

When the main system is in the nonequilibrium steady state due to the couplings with the reservoirs, the Fermi atoms in the main system obey a nonequilibrium energy distribution $f_{\rm neq}(\omega)$ that has a different structure from the ordinary Fermi-Dirac distribution function. To obtain the distribution $f_{\rm neq}(\omega)$ in the {\it absence} of pairing interaction ($U=0$), we conveniently introduce the $2\times 2$ matrix nonequilibrium Green's function in the main system, given by
\begin{equation}
\hat{G}_{{\rm neq}, \sigma}(\bm{p}, \omega) =
\begin{pmatrix}
G^{\rm R}_{{\rm neq}, \sigma}(\bm{p}, \omega) & 
G^{\rm K}_{{\rm neq}, \sigma}(\bm{p}, \omega) \\[4pt]
0 & G^{\rm A}_{{\rm neq}, \sigma}(\bm{p}, \omega)
\end{pmatrix},
\label{eq.Gneq}
\end{equation}
which obeys the Dyson equation,
\begin{align}
\hat{G}_{{\rm neq}, \sigma}(\bm{p}, \omega)
&=
\hat{G}_{0, \sigma}(\bm{p}, \omega) 
\notag\\
&+
\hat{G}_{0, \sigma}(\bm{p}, \omega)
\hat{\Sigma}_{{\rm env}, \sigma}(\bm{p}, \omega)
\hat{G}_{{\rm neq}, \sigma}(\bm{p}, \omega).
\label{eq.Dyson1}
\end{align}
The Dyson equation~\eqref{eq.Dyson1} is diagrammatically drawn as Fig.~\ref{fig.Genv}. In Eq.~\eqref{eq.Dyson1}, $\hat{G}_{0, \sigma}(\bm{p}, \omega)$ is the bare Green's function in the absence of the system-reservoir couplings, given in Eq.~\eqref{eq.G0.Keldysh}. We emphasize that the distribution function $f(\omega)$ in $\hat{G}_{0, \sigma}(\bm{p}, \omega)$ has nothing to do with the distribution function $f_{\rm env}(\omega)$ in the reservoirs in Eq.~\eqref{eq.Fermi}. We can regard $f(\omega)$ in $\hat{G}_{0, \sigma}(\bm{p}, \omega)$ as the energy distribution in the {\it isolated} Fermi gas, before the main system is connected to the reservoirs. As we will see below, the nonequilibrium distribution $f_{\rm neq}(\omega)$ in the main system does not depend on $f(\omega)$, when the system relaxes into the nonequilibrium steady state due to the couplings with the reservoirs.

In Eq.~\eqref{eq.Dyson1}, the self-energy $\hat{\Sigma}_{{\rm env}, \sigma}(\bm{p}, \omega)$ describes effects of system-reservoir couplings. Within the second-order Born approximation with respect to the tunneling matrix $\m{T}$, the self-energy is diagrammatically given in Fig.~\ref{fig.Genv}. Evaluating this diagrammatic equation, we obtain~\cite{Kawamura2020, Kawamura2022}
\begin{align}
&
\hat{\Sigma}_{{\rm env}, \sigma}(\bm{p}, \omega)=
\begin{pmatrix}
\Sigma^{\rm R}_{{\rm env}, \sigma}(\bm{p}, \omega) &  
\Sigma^{\rm K}_{{\rm env}, \sigma}(\bm{p}, \omega)\\[4pt]
0&
\Sigma^{\rm A}_{{\rm env}, \sigma}(\bm{p}, \omega) 
\end{pmatrix}
\notag\\
&=
N_{\rm t} |\m{T}|^2 \sum_{\alpha={\rm L}, {\rm R}} \sum_{\bm{q}}
\hat{D}_{\alpha, \sigma}(\bm{q}, \omega -\mu_\alpha)
\notag\\
&=
\begin{pmatrix}
-2i\gamma & -2i\gamma \left[\tanh\left(\frac{\omega-\mu_{\rm L}}{2T_{\rm env}}\right) +\tanh\left(\frac{\omega-\mu_{\rm R}}{2T_{\rm env}}\right) \right] \\[6pt]
0 &2i\gamma 
\end{pmatrix}.
\label{eq.self.env}
\end{align}
For the derivation of Eq.~\eqref{eq.self.env}, we refer to Ref.~\cite{Kawamura2020}. Here,
\begin{equation}
\scalebox{0.96}{$\displaystyle
\hat{D}_{\alpha, \sigma}(\bm{q}, \omega)=
\begin{pmatrix}
\frac{1}{\omega +i\delta -\xi^{\rm res}_{\bm{q}}} &
-2i \pi \delta(\omega -\xi^{\rm res}_{\bm{q}}) \big[1 -2f_{\rm env}(\omega)\big] \\[6pt]
0 &
\frac{1}{\omega -i\delta -\xi^{\rm res}_{\bm{q}}}
\end{pmatrix}$}
\end{equation}
is the non-interacting $2\times 2$ matrix Green's function in the $\alpha$ ($=$ L, R) reservoir, and
\begin{equation}
\gamma= \pi N_{\rm t} \rho |\m{T}|^2,
\label{eq.gam.def}
\end{equation}
describes the quasiparticle damping originating from the system-reservoir coupling. In Eq.~\eqref{eq.gam.def}, $\rho=\rho_\alpha$ is the single-particle density of states in the $\alpha$ reservoir, where we have ignored the $\alpha$ dependence of this quantity, for simplicity. We have also ignored the $\omega$ dependence of $\rho$, which is sometimes referred to as the wide-band limit approximation in the literature \cite{Stefanucci2013}. This approximation is justified in the case when the reservoirs are so huge that the energy dependence of their density of states around $\omega=\mu$ can be ignored, within the variation of $\Delta \mu$. 

The substitution of Eq.~\eqref{eq.self.env} into the Dyson equation~\eqref{eq.Dyson1} gives
\begin{align}
&
\hat{G}_{{\rm neq}, \sigma}(\bm{p}, \omega)
\notag\\
&\hspace{0.2cm}=
\begin{pmatrix}
\frac{1}{\omega -\ep_{\bm{p}} +2i\gamma} &
\frac{-4i\gamma [1 -f_{\rm env}(\omega -\mu_{\rm L}) -f_{\rm env}(\omega -\mu_{\rm R})]}{[\omega -\ep_{\bm{p}}]^2 +4\gamma^2}
\\[6pt]
0 &
\frac{1}{\omega -\ep_{\bm{p}} -2i\gamma} 
\end{pmatrix}.
\label{eq.Gneq2}
\end{align}
Here, the Fermi-Dirac function $f_{\rm env}(\omega)$ in the reservoirs is given in Eq.~\eqref{eq.Fermi}. As mentioned previously, while the nonequilibrium steady-state Green's function $\hat{G}_{{\rm neq}, \sigma}(\bm{p}, \omega)$ depends on $f_{\rm env}(\omega)$, it is not affected by the distribution function $f(\omega)$ in the initial state of the main system. This is because the initial memory of the isolated Fermi gas is wiped out due to the coupling with the reservoirs.

The energy distribution function $f_{\rm neq}(\omega)$ in the main system is obtained from the Keldysh component of $\hat{G}_{{\rm neq}, \sigma}(\bm{p}, \omega)$ in Eq.~\eqref{eq.Gneq2}. Although the FDR, like Eq.~\eqref{eq.FDR.TMA}, does not hold between the retarded and the Keldysh components in the nonequilibrium state ($\Delta\mu \neq 0$), they still possess a similar relation: The Keldysh component in Eq.~\eqref{eq.Gneq2} can be written as
\begin{align}
G^{\rm K}_{{\rm neq}, \sigma}(\bm{p}, \omega) = 2i {\rm Im}\big[ G^{\rm R}_{{\rm neq}, \sigma}(\bm{p}, \omega)\big] \big[1 -2f_{\rm neq}(\omega)\big],
\end{align}
where
\begin{equation}
f_{\rm neq}(\omega) = \frac{1}{2}\big[f(\omega -\mu_{\rm L}) +f(\omega -\mu_{\rm R}) \big].
\label{eq.fneq.w}
\end{equation}
By analogy with the FDR in the thermal equilibrium case, we can interpret $f_{\rm neq}(\omega)$ in Eq.~\eqref{eq.fneq.w} as the nonequilibrium energy distribution function in the driven-dissipative Fermi gas. The distribution function $f_{\rm neq}(\omega)$ has a clear two-step structure (at $\omega=\mu_{\rm L}$ and $\omega=\mu_{\rm R}$) when $T_{\rm env} \ll \Delta \mu$, originating from the different Fermi levels between the left and right reservoirs. We note that such a nonequilibrium energy distribution with the two-step structure has experimentally been observed in mesoscopic wires~\cite{Pothier1996, Pothier1997, Anthore2003}, as well as carbon nanotubes~\cite{Chen2009}, under a bias voltage $V$ (which corresponds to the chemical potential bias $\Delta \mu$ in the present model driven-dissipative Fermi gas).

The couplings with the reservoirs affect, not only the energy distribution,  but also the spectral function in the main system. The single-particle spectral function $A_{{\rm neq}, \sigma}(\bm{p}, \omega)$ is obtained from the retarded component of $\hat{G}_{{\rm neq}, \sigma}(\bm{p}, \omega)$ as
\begin{equation}
\scalebox{0.98}{$\displaystyle
A_{{\rm neq}, \sigma}(\bm{p}, \omega) =  -\frac{1}{\pi} {\rm Im}\big[G^{\rm R}_{{\rm neq}, \sigma}(\bm{p}, \omega) \big] = \frac{1}{\pi}\frac{2\gamma}{[\omega -\ep_{\bm{p}}]^2 +4\gamma^2}.$}
\label{eq.SW.neq}
\end{equation}
A comparison of $A_{{\rm neq}, \sigma}(\bm{p}, \omega)$ in Eq.~\eqref{eq.SW.neq} and $A_{0, \sigma}(\bm{p}, \omega)$ in Eq.~\eqref{eq.SW.free} shows that the system-reservoir couplings give rise to the linewidth $2\gamma$ in the excitation spectrum. The linewidth physically means the finite lifetime $\tau$ $(\sim 1/\gamma)$ of the excitation mode $\ep_{\bm{p}}$ in the driven-dissipative Fermi gas, due to the atom tunneling between the main system and the reservoirs.

\begin{figure}[t]
\centering
\includegraphics[width=6.5cm]{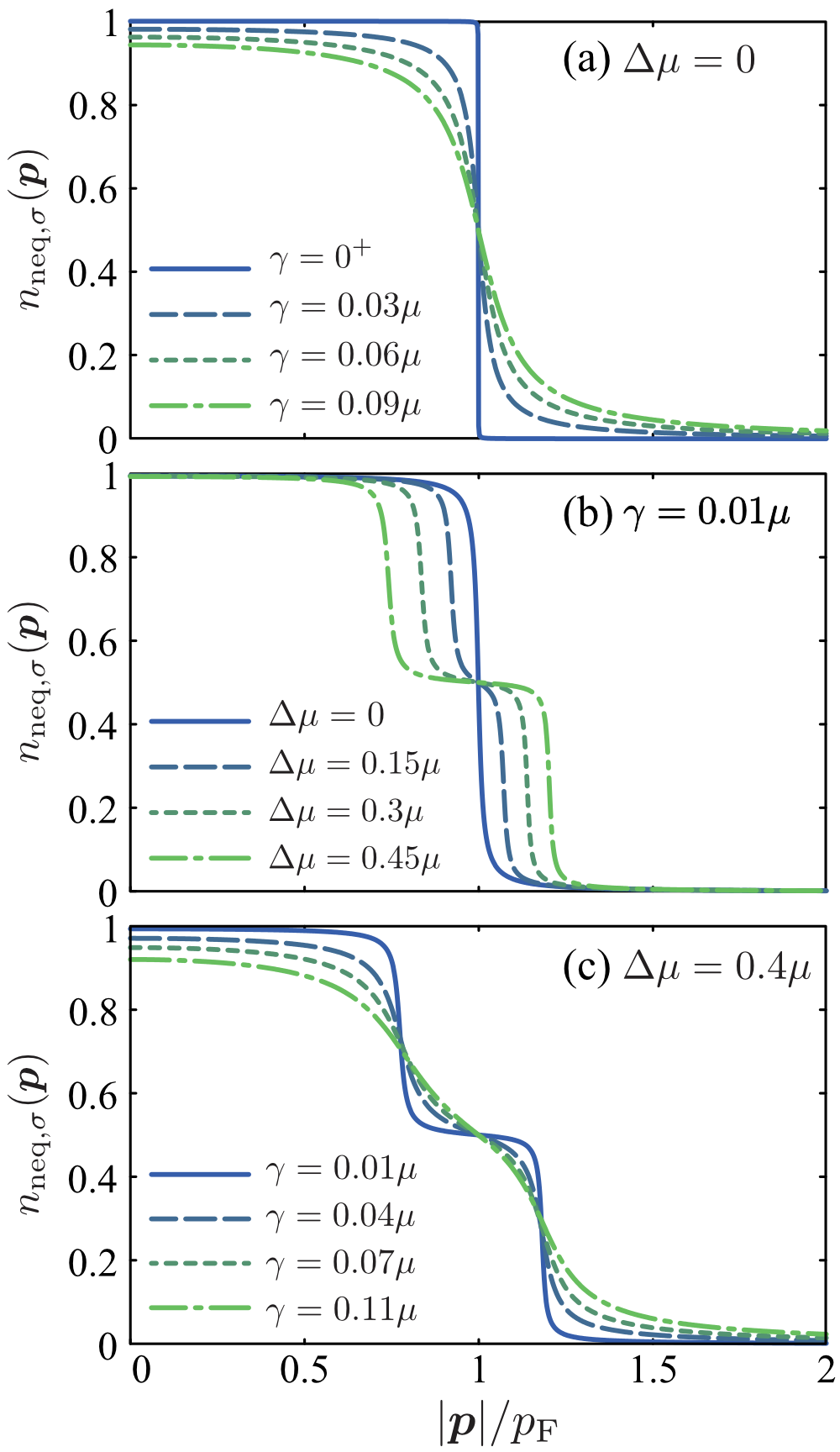}
\caption{Calculated nonequilibrium momentum distribution $n_{{\rm neq}, \sigma}(\bm{p})$ as a function of $|\bm{p}|$ in the driven-dissipative Fermi gas. Panels (a) and (c) show the effects of the system-reservoir coupling strength $\gamma$ in the presence and the absence of the chemical potential bias $\Delta\mu$, respectively. Panel (b) shows effects of $\Delta\mu$, when $\gamma=0.01\mu$. We set $T_{\rm env}=0$.}
\label{fig.neq}
\end{figure}

The momentum distribution $n_{{\rm neq}, \sigma}(\bm{p})=\braket{a^\dagger_{\bm{p}, \sigma} a_{\bm{p}, \sigma}}$ is obtained from the Keldysh component $G^{\rm K}_{{\rm neq}, \sigma}(\bm{p}, \omega)$ as
\begin{align}
n_{{\rm neq}, \sigma}(\bm{p}) 
&= -\frac{i}{2} \int_{-\infty}^\infty \frac{d\omega}{2\pi} G^{\rm K}_{{\rm neq}, \sigma}(\bm{p}, \omega) +\frac{1}{2}
\notag\\
&= \int_{-\infty}^\infty d\omega A_{{\rm neq}, \sigma}(\bm{p}, \omega) f_{\rm neq}(\omega).
\label{eq.neq.momentum}
\end{align}
Equation~\eqref{eq.neq.momentum} indicates that $n_{{\rm neq}, \sigma}(\bm{p})$ involves information about the single-particle spectral function $A_{{\rm neq}, \sigma}(\bm{p}, \omega)$ in Eq.~\eqref{eq.SW.neq}, as well as the energy distribution function $f_{{\rm neq}, \sigma}(\omega)$ in Eq.~\eqref{eq.fneq.w}.

Figure.~\ref{fig.neq} shows the calculated $n_{{\rm neq}, \sigma}(\bm{p})$. In the zero bias case $(\Delta\mu=0)$ shown in Fig.~\ref{fig.neq}(a), the effects of system-reservoir couplings are dominated by the quasi-particle damping described by $\gamma$ in Eq.~\eqref{eq.gam.def}. Since the quasi-particle peak in the single-particle spectral function $A_{{\rm neq},\sigma}(\bm{p}, \omega)$ is broadened by this damping, the smearing of the sharp Fermi edge at $|\bm{p}|\simeq p_{\rm F}$ becomes more remarkably with increasing the value of $\gamma$, as shown in Fig.~\ref{fig.neq}(a).

Once non-zero bias $(\Delta\mu\neq 0)$ is imposed, one finds in Fig.~\ref{fig.neq}(b) that the momentum distribution $n_{{\rm neq},\sigma}(\bm{p})$ exhibits the two-step structure expected in Eq.~\eqref{eq.fneq.w}. However, as shown in Fig.~\ref{fig.neq}(c), the two-step structure imprinted on the momentum distribution $n_{{\rm neq},\sigma}(\bm{p})$ becomes obscure as $\gamma$ increases due to the broadening of the quasi-particle peak in $A_{{\rm neq}, \sigma}(\bm{p}, \omega)$. Although we do not show it explicitly, the two-step structure is also rounded by thermal excitations as the temperature $T_{\rm env}$ increases. Because of these, the conditions for the two-step structure to appear in the momentum distribution $n_{{\rm neq}, \sigma}(\bm{p})$ are $\Delta \mu \gg \gamma$ and $\Delta \mu \gg T_{\rm env}$.

\subsection{Thouless criterion for the driven-dissipative Fermi gas \label{sec.noneq.Thouless}}

\begin{figure}[t]
\centering
\includegraphics[width=8cm]{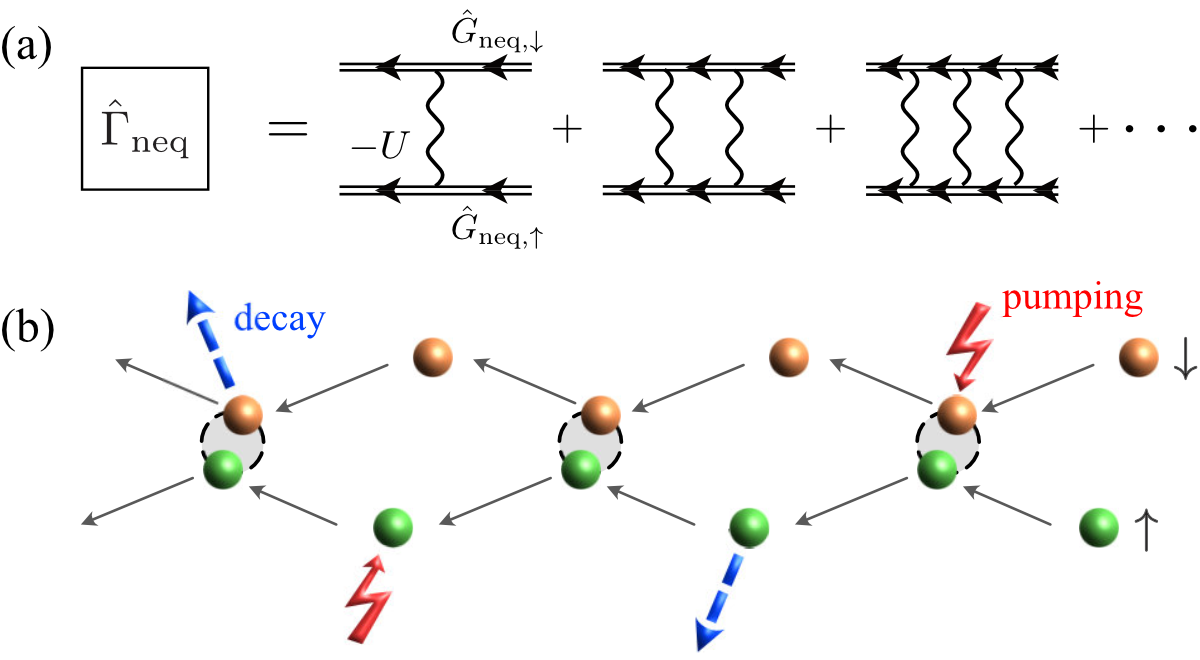}
\caption{(a) Particle-particle scattering matrix $\hat{\Gamma}_{\rm neq}$ in the driven-dissipative Fermi gas. The nonequilibrium Green's function $\hat{G}_{{\rm neq}, \sigma}$ (double solid line) is diagrammatically given in Fig.~\ref{fig.Genv}. $\hat{\Gamma}_{\rm neq}$ physically describes paring fluctuations in the presence of pumping and decay of Fermi atoms, as schematically shown in panel (b).}
\label{fig.Tmat.neq}
\end{figure}

To extend the Thouless criterion in Eq.~\eqref{eq.Thouless} to the nonequilibrium steady state, we evaluate the particle-particle scattering matrix $\hat{\Gamma}_{\rm neq}(\bm{q}, \nu)$ in the driven-dissipative Fermi gas. As in the thermal equilibrium case, $\hat{\Gamma}_{\rm neq}(\bm{q}, \nu)$ is also given by the series of the ladder diagrams shown in Fig.~\ref{fig.Tmat.neq}(a). While $\hat{\Gamma}_0(\bm{q}, \nu)$ involves the bare Green's function $\hat{G}_{0, \sigma}(\bm{p}, \omega)$ in Eq.~\eqref{eq.G0.Keldysh}, $\hat{\Gamma}_{\rm neq}(\bm{q}, \nu)$ involves the nonequilibrium Green's function $\hat{G}_{{\rm neq}, \sigma}(\bm{p}, \omega)$ in Eq.~\eqref{eq.Gneq2}, to take into account the system-reservoir coupling effects. In this sense, $\hat{\Gamma}_{\rm neq}(\bm{q}, \nu)$ physically describes pairing fluctuations in the presence of pumping and decay of Fermi atoms, as schematically shown in Fig.~\ref{fig.Tmat.neq}(b).

Summing up the ladder diagrams in Fig.~\ref{fig.Tmat.neq}(a), we have~\cite{Kawamura2020_JLTP, Kawamura2020}
\begin{align}
&
\hat{\Gamma}_{\rm neq}(\bm{q}, \nu)=
\begin{pmatrix}
\Gamma_{\rm neq}^{\rm R} &\Gamma_{\rm neq}^{\rm K}\\[4pt]
0 & \Gamma_{\rm neq}^{\rm A}
\end{pmatrix}(\bm{q}, \nu)
\notag\\
&=
\frac{-U}{1-U\hat{\Pi}_{\rm neq}(\bm{q}, \nu)}
\notag\\
&=
\begin{pmatrix}
\frac{-U}{1 +U\Pi^{\rm R}_{\rm neq}(\bm{q}, \nu)} &  
\frac{U^2 \Pi^{\rm K}_{\rm neq}(\bm{q}, \nu)}{[1 +U\Pi^{\rm R}_{\rm neq}(\bm{q}, \nu)][1 +U\Pi^{\rm A}_{\rm neq}(\bm{q}, \nu)]} \\[4pt]
0 & \frac{-U}{1 +U\Pi^{\rm A}_{\rm neq}(\bm{q}, \nu)}
\end{pmatrix}.
\label{eq.Tmat.neq}
\end{align}
Here,
\begin{equation}
\hat{\Pi}_{\rm neq}(\bm{q}, \nu) =
\begin{pmatrix}
\Pi^{\rm R}_{\rm neq}(\bm{q}, \nu) & \Pi^{\rm K}_{\rm neq}(\bm{q}, \nu) \\[4pt]
0 & \Pi^{\rm A}_{\rm neq}(\bm{q}, \nu)
\end{pmatrix}
\end{equation}
is the lowest-order pair correlation function in the driven-dissipative Fermi gas. The retarded (advanced) component $\Pi^{\rm R(A)}_{\rm neq}(\bm{q}, \nu) $ is obtained  by simply replacing $G^{\rm R, A, K}_{0, \sigma}(\bm{p}, \omega)$ with $G^{\rm R, A, K}_{{\rm neq}, \sigma}(\bm{p}, \omega)$ in Eq.~\eqref{eq.TMA.Pi.R}, which reads
\begin{align}
&
\Pi^{\rm R}_{\rm neq}(\bm{q}, \nu)=
\big[\Pi^{\rm A}_{\rm neq}(\bm{q}, \nu) \big]^*
=
\frac{i}{2}\sum_{\bm{p}} \int_{-\infty}^\infty \frac{d\omega}{2\pi}
\notag\\
&\hspace{0.3cm}\times\big[
G^{\rm R}_{{\rm neq}, \up}(\bm{p}+\bm{q}/2, \omega +\nu)
G^{\rm K}_{{\rm neq}, \down}(-\bm{p}+\bm{q}/2, -\omega)
\notag\\[4pt]
&\hspace{0.5cm}+
G^{\rm K}_{{\rm neq}, \up}(\bm{p}+\bm{q}/2, \omega +\nu)
G^{\rm R}_{{\rm neq}, \down}(-\bm{p}+\bm{q}/2, -\omega)
\big] \label{eq.Pi.R.neq}.
\end{align}
In the thermal equilibrium state, the Keldysh component $\Pi^{\rm K}_{\rm neq}(\bm{q}, \nu)$ is immediately obtained from the retarded component via the FDR. However, this is not the case for the driven-dissipative Fermi gas. In this nonequilibrium case, the Keldysh component $\Pi^{\rm K}_{\rm neq}(\bm{q}, \nu)$ is independently evaluated from the following expression \cite{Kawamura2020_JLTP, Kawamura2020}:
\begin{align}
&
\Pi^{\rm K}_{\rm neq}(\bm{q}, \nu)=
\frac{i}{2}\sum_{\bm{p}} \int_{-\infty}^\infty \frac{d\omega}{2\pi}
\notag\\
&\hspace{0.35cm}\times
\big[
G^{\rm R}_{{\rm neq}, \up}(\bm{p}+\bm{q}/2, \omega +\nu)
G^{\rm R}_{{\rm neq}, \down}(-\bm{p}+\bm{q}/2, -\omega)
\notag\\[4pt]
&\hspace{0.5cm}+
G^{\rm A}_{{\rm neq}, \up}(\bm{p}+\bm{q}/2, \omega +\nu)
G^{\rm A}_{{\rm neq}, \down}(-\bm{p}+\bm{q}/2, -\omega)
\notag\\[4pt]
&\hspace{0.5cm}+
G^{\rm K}_{{\rm neq}, \up}(\bm{p}+\bm{q}/2, \omega +\nu)
G^{\rm K}_{{\rm neq}, \down}(-\bm{p}+\bm{q}/2, -\omega)
\big]. \label{eq.Pi.K.neq}
\end{align}

As in the thermal equilibrium case, the superfluid phase transition temperature $T^{\rm c}_{\rm env}$ is determined from the retarded particle-particle scattering matrix $\Gamma^{\rm R}_{\rm neq}(\bm{q}, \nu)$. The nonequilibrium version of the pole equation~\eqref{eq.Thouless} is given by
\begin{equation}
\big[\Gamma^{\rm R}_{\rm neq}(\bm{q}=\bm{q}_{\rm pair}, \nu=\mu_{\rm pair})\big]^{-1}=0,
\label{eq.KM1}
\end{equation}
which actually consists of the following two equations:
\begin{align}
& {\rm Re}\big[\Gamma^{\rm R}_{\rm neq}(\bm{q}=\bm{q}_{\rm pair}, \nu=\mu_{\rm pair})\big]^{-1}=0,
\label{eq.NKM.Re}
\\
& {\rm Im}\big[\Gamma^{\rm R}_{\rm neq}(\bm{q}=\bm{q}_{\rm pair}, \nu=\mu_{\rm pair})\big]^{-1}=0.
\label{eq.NKM.Im}
\end{align}
In the symmetric coupling case ($\m{T}_{\rm L}=\m{T}_{\rm R}\equiv T$) we are considering, the latter equation can be solved analytically: When $\Pi^{\rm R}_{\rm neq}(\bm{q}, \nu)$ in Eq.~\eqref{eq.Pi.R.neq} is substituted into the retarded component $\Gamma^{\rm R}_{\rm neq}(\bm{q}, \nu)$ in Eq.~\eqref{eq.Tmat.neq}, Eq.~\eqref{eq.NKM.Im} is reduced to
\begin{widetext}
\begin{equation}
0 = \sum_{\eta=\pm}\sum_{\bm{p}}\int_{-\infty}^\infty\frac{d\omega}{2\pi}\frac{\eta\left[\tanh\left(\frac{\omega +\eta[\mu_{\rm L}-\mu_{\rm pair}/2]}{2T_{\rm env}} \right) +\tanh\left(\frac{\omega +\eta[\mu_{\rm R}-\mu_{\rm pair}/2]}{2T_{\rm env}} \right)\right]}{\big[(\omega +\ep_{\bm{p}+\bm{q}_{\rm pair}/2} -\mu_{\rm pair}/2)^2 +4\gamma^2\big]\big[(\omega -\ep_{-\bm{p}+\bm{q}_{\rm pair}/2} +\mu_{\rm pair}/2)^2 +4\gamma^2\big]},
\end{equation}
which is satisfied only when $\mu_{\rm pair}=2\mu$.  We note that in the thermal equilibrium case, the imaginary part of the pole equation~\eqref{eq.Thouless} is satisfied when $\mu_{\rm pair}=0$ rather than $\mu_{\rm pair}=2\mu$. The difference comes from the different origins of energy in the thermal equilibrium and the present nonequilibrium case. Since we now measure the energy from the bottom of the energy band ($\ep_{\bm{p}=0}$) in the main system [see Fig.~\ref{fig.model}(b)], a Cooper pair should have nonzero energy $\mu_{\rm pair}=2\mu$. [The factor $2$ comes from the fact that a Cooper pair is formed by {\it two} Fermi atoms.] 

Substituting this solution $\mu_{\rm pair}=2\mu$ into Eq.~\eqref{eq.NKM.Re}, we have 
\begin{equation}
\frac{1}{U} = \gamma \sum_{\bm{p}} \int_{-\infty}^\infty \frac{d\omega}{2\pi} \frac{\big[2\omega +\ep_{\bm{p}+\bm{q}_{\rm pair}/2} -\ep_{-\bm{p}+\bm{q}_{\rm pair}/2}\big] \left[\tanh\left(\frac{\omega +\Delta \mu}{2T_{\rm env}}\right) +\tanh\left(\frac{\omega -\Delta \mu}{2T_{\rm env}}\right)\right]}{\big[(\omega +\ep_{\bm{p}+\bm{q}_{\rm pair}/2} -\mu)^2 +4\gamma^2\big]\big[(\omega -\ep_{-\bm{p}+\bm{q}_{\rm pair}/2} +\mu)^2 +4\gamma^2\big]},
\label{eq.neq.Thouless}
\end{equation}
\end{widetext}
which is the nonequilibrium version of Eq.~\eqref{eq.KM.Re2}, and is referred to as the nonequilibrium Thouless criterion in what follows. In Eq.~\eqref{eq.neq.Thouless}, $\bm{q}_{\rm pair}$ is determined so as to obtain the highest $T_{\rm env}^{\rm c}$. As noted in Sec.~\ref{sec.eq.BCS.BEC}, $\bm{q}_{\rm pair}$ physically describes the center-of-mass momentum of a Cooper pair. Thus, when $\bm{q}_{\rm pair}=0$, the main system transitions to the uniform nonequilibrium BCS-type superfluid state, where the Cooper pairs have zero center-of-mass momentum. On the other hand, when $\bm{q}_{\rm pair}\neq 0$, an inhomogeneous superfluid state is realized, being characterized by a spatially oscillating superfluid order parameter (symbolically written
as $\Delta(\bm{r})= \Delta e^{i\bm{q}_{\rm pair}\cdot \bm{r}}$). This is analogous to the Fulde-Ferrell-Larkin-Ovchinnikov (FFLO) state in a superconductor under an external magnetic field~\cite{Fulde1964, Larkin1964, Takada1969}.

\subsection{Nonequilibrium FFLO superfluid transition in the mean-field approximation \label{sec.NFFLO.NMF}}

\begin{figure}[t]
\centering
\includegraphics[width=8.5cm]{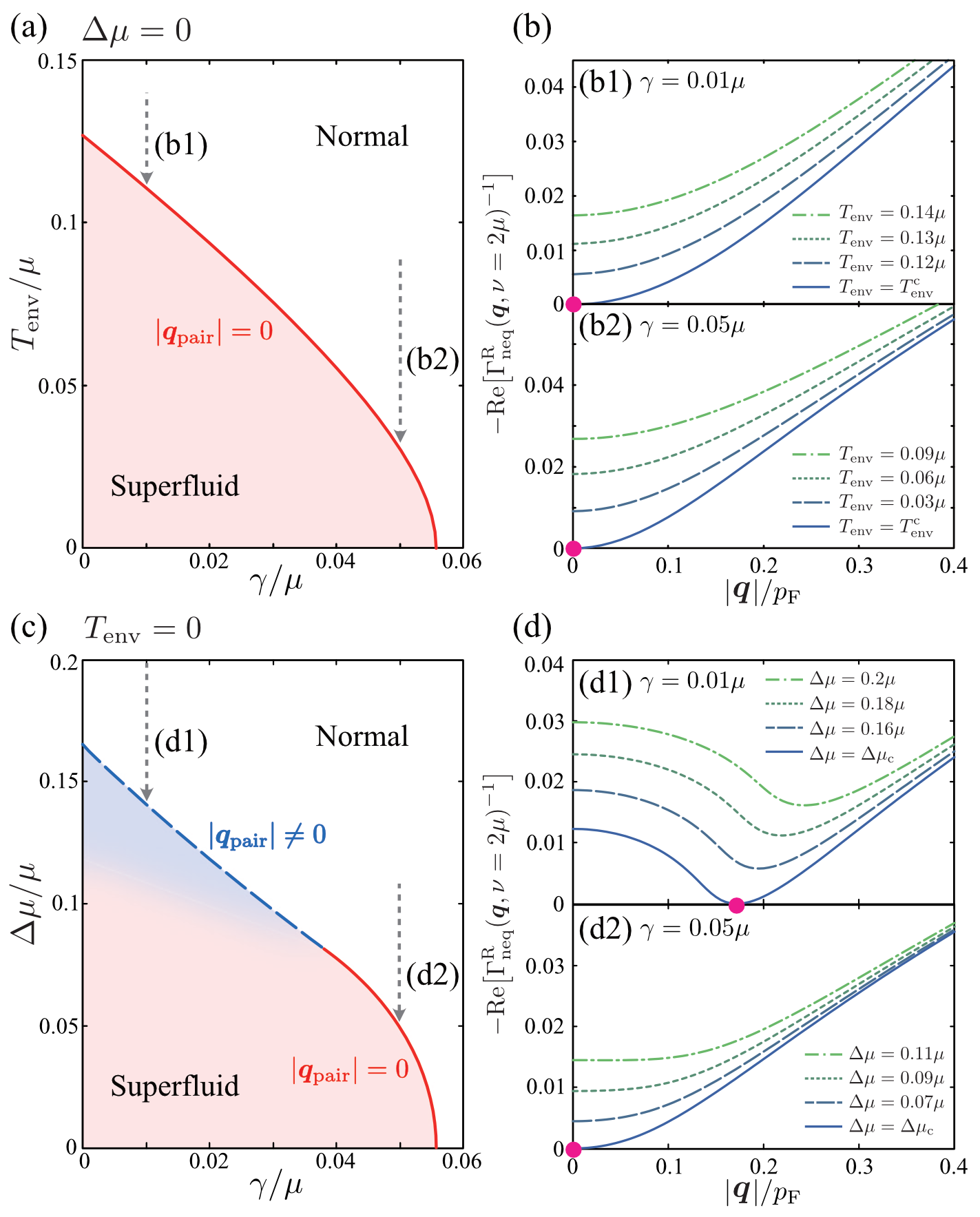}
\caption{(a) and (c): Superfluid transition line obtained by solving the nonequilibrium Thouless criterion in Eq.~\eqref{eq.neq.Thouless} in the weak-coupling BCS regime when $(p_{\rm F}a_s)^{-1}=-1$. The system experiences the BCS-type (FFLO-type) superfluid instability on the solid (dashed) line. (b) and (d): Calculated inverse particle-particle scattering matrix $\Gamma^{\rm R}_{\rm neq}(\bm{q}, \nu=2\mu)$ as a function of $|\bm{q}|$. We note that ${\rm Im}\Gamma^{\rm R}_{\rm neq}(\bm{q}, \nu=2\mu)=0$. Each panel shows the results along the paths depicted in panels (a) and (c). The pole of $\Gamma^{\rm R}_{\rm neq}(\bm{q}, \nu=2\mu)$ (solid circle) is obtained at $|\bm{q}|\neq 0$ ($|\bm{q}|= 0$) on the nonequilibrium FFLO (BCS) transition line.}
\label{fig.NThouless1}
\end{figure}

Figure.~\ref{fig.NThouless1}(a) shows the mean-field results of the superfluid phase transition temperature as a function of the system-reservoir coupling strength $\gamma$ in the zero bias case ($\Delta\mu=0$). This figure is obtained by solving Eq.~\eqref{eq.neq.Thouless} with a fixed value of $\mu$. (This is an extension of the mean-field KM theory explained in Sec.~\ref{sec.eq.BCS.BEC} to the driven-dissipative Fermi gas.) As in the well-known thermal equilibrium case, the superfluid phase is suppressed by thermal fluctuations as the temperature $T_{\rm env}$ increases. The superfluid phase is also suppressed as the system-reservoir coupling strength $\gamma$ increases. This is simply because, as shown in Fig.~\ref{fig.neq}(a), the system-reservoir coupling rounds the Fermi edge in the momentum distribution $n_{{\rm neq}, \sigma}(\bm{p})$ so that $\gamma$ plays a similar role to the temperature.

In the zero bias case $(\Delta\mu=0)$ shown in Fig.~\ref{fig.NThouless1}(a), the pole of $\Gamma^{\rm R}_{\rm neq}(\bm{q}, 2\mu)$ always appears at $\bm{q}=0$, when $T_{\rm env}=T_{\rm env}^{\rm c}$ [see Fig.~\ref{fig.NThouless1}(b)]. As explained previously, this means that the occurrence of the BCS-type uniform superfluid transition at this temperature, being accompanied by BEC of Cooper pairs with zero center-of-mass momentum.

The situation changes when the main system is in the nonequilibrium steady state by the non-zero chemical potential bias $\Delta \mu \neq 0$. Figure~\ref{fig.NThouless1}(c) is the phase diagram of the driven-dissipative Fermi gas in terms of the system-reservoir coupling strength $\gamma$ and the chemical potential bias $\Delta\mu$, when $T_{\rm env}=0$. When $\Delta\mu$ is very large, the system is in the normal state due to the strong pumping and decay of Fermi atoms. As $\Delta \mu$ decreases and the nonequilibrium effects are weakened, the system experiences superfluid instability. As shown in Fig.~\ref{fig.NThouless1}(d1), while the pole of $\Gamma^{\rm R}_{\rm neq}(\bm{q}, 2\mu)$ appears at $\bm{q}=0$ when $\gamma / \mu \gtrsim 0.04$, it appears at $|\bm{q}|=|\bm{q}_{\rm pair}|$ ($> 0$) when $\gamma / \mu \lesssim 0.04$. While the former case ($\bm{q}=0$) means the BCS-type superfluid transition at $T_{\rm env}^{\rm c}$, the latter indicates the FFLO-type superfluid transition associated with Cooper pairs with non-zero center-of-mass momentum $\bm{q}_{\rm pair}\neq 0$.

\begin{figure}[t]
\centering
\includegraphics[width=7.5cm]{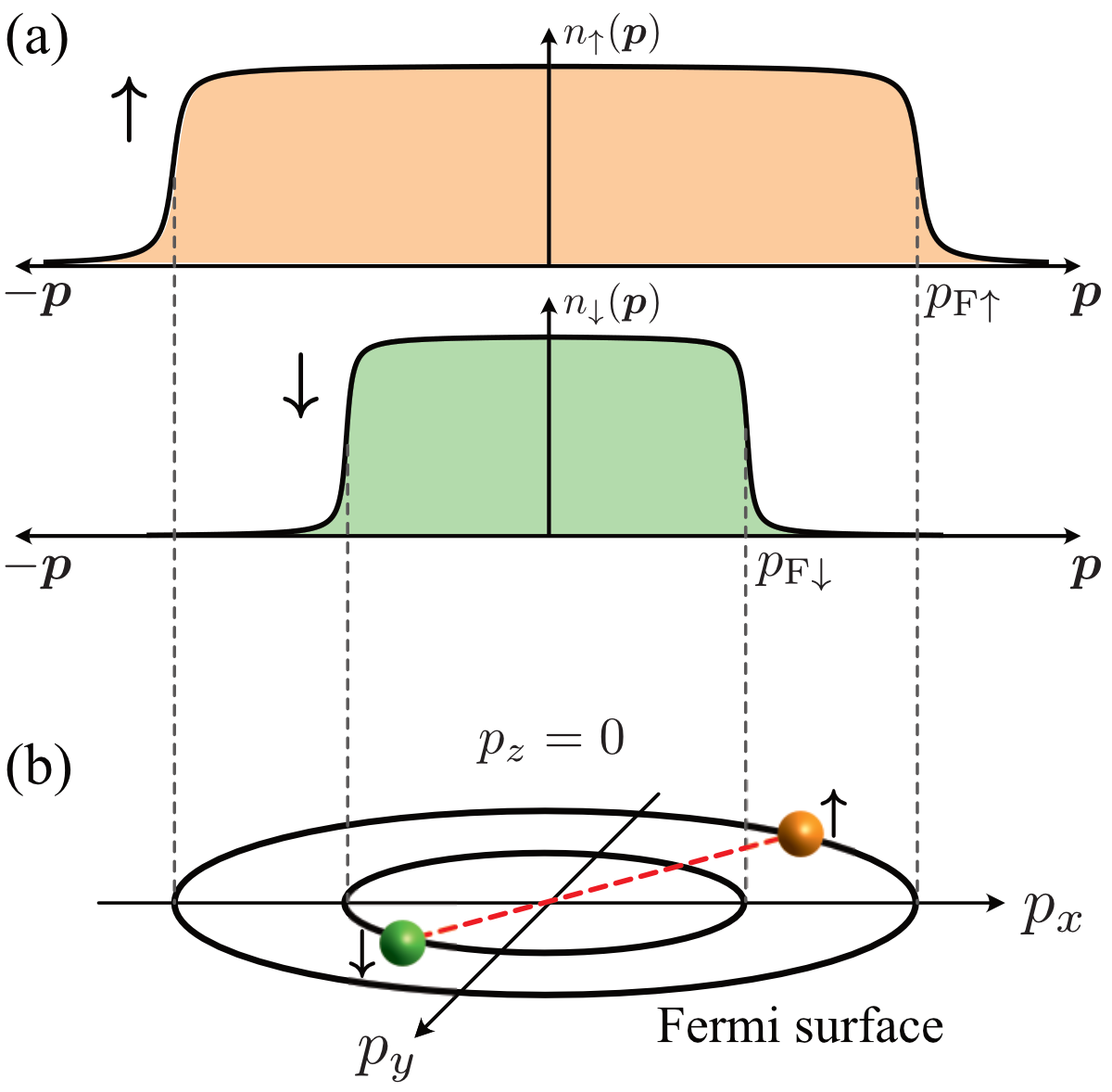}
\caption{Cooper-pair formation in the FFLO state in a thermal equilibrium Fermi gas with spin imbalance ($N_\up \neq N_\down$). (a) Momentum distribution $n_{\sigma=\up, \down}(\bm{p})$. Since the system is in the thermal equilibrium state, $n_{\sigma}(\bm{p})$ is simply given by the Fermi-Dirac distribution function. However, the spin imbalance leads to $p_{{\rm F}\up}\neq p_{{\rm F}\down}$ (where $p_{{\rm F}\sigma}$ is the Fermi momentum in the pseudospin $\sigma=\up, \down$ component). (b) The misalignment between $\sigma=\up$ and $\down$ Fermi surfaces naturally induces a Cooper pair with non-zero center-of-mass momentum, $|\bm{q}_{\rm pair}| \simeq p_{{\rm F} \up} -p_{{\rm F} \down}\neq 0$.}
\label{fig.Cooper1}
\end{figure}

The FFLO state was first proposed in metallic superconductivity under an external magnetic field \cite{Fulde1964, Larkin1964, Takada1969}, and was later discussed in spin-imbalanced ($N_\up \neq N_\down$) ultracold Fermi gas \cite{Hu2006, Parish2007, Liao2010, Chevy2010, Kinnunen2018}, as well as color superconductivity in quantum chromodynamics \cite{Casalbuoni2004}. In these systems, the FFLO state is induced by the misalignment of the Fermi surfaces. For example, as schematically shown in Fig~\ref{fig.Cooper1}(a), the Fermi momenta $p_{{\rm F}\sigma}$ of the pseudospin $\sigma =\up$ and $\sigma = \down$ Fermi atoms are different from each other in a Fermi gas with the spin imbalance $N_\up \neq N_\down$. Since Cooper pairs are formed by Fermi atoms near each Fermi surface, Cooper pairs acquire nonzero center-of-mass momentum $|\bm{q}_{\rm pair}| \simeq p_{{\rm F}\up} -p_{{\rm F}\down}$ due to the Fermi surface mismatch, as schematically shown in Fig.~\ref{fig.Cooper1}(b).

However, we emphasize that the present driven-dissipative Fermi gas is a spin-{\it balanced} ($N_\up=N_\down$) system, so that there is no Fermi surface misalignment in the main system. In this sense, the mechanism of the FFLO-type superfluid in Fig.~\ref{fig.NThouless1}(c) is different from the conventional one discussed in metallic superconductivity under an external magnetic field.

\begin{figure}[t]
\centering
\includegraphics[width=7.5cm]{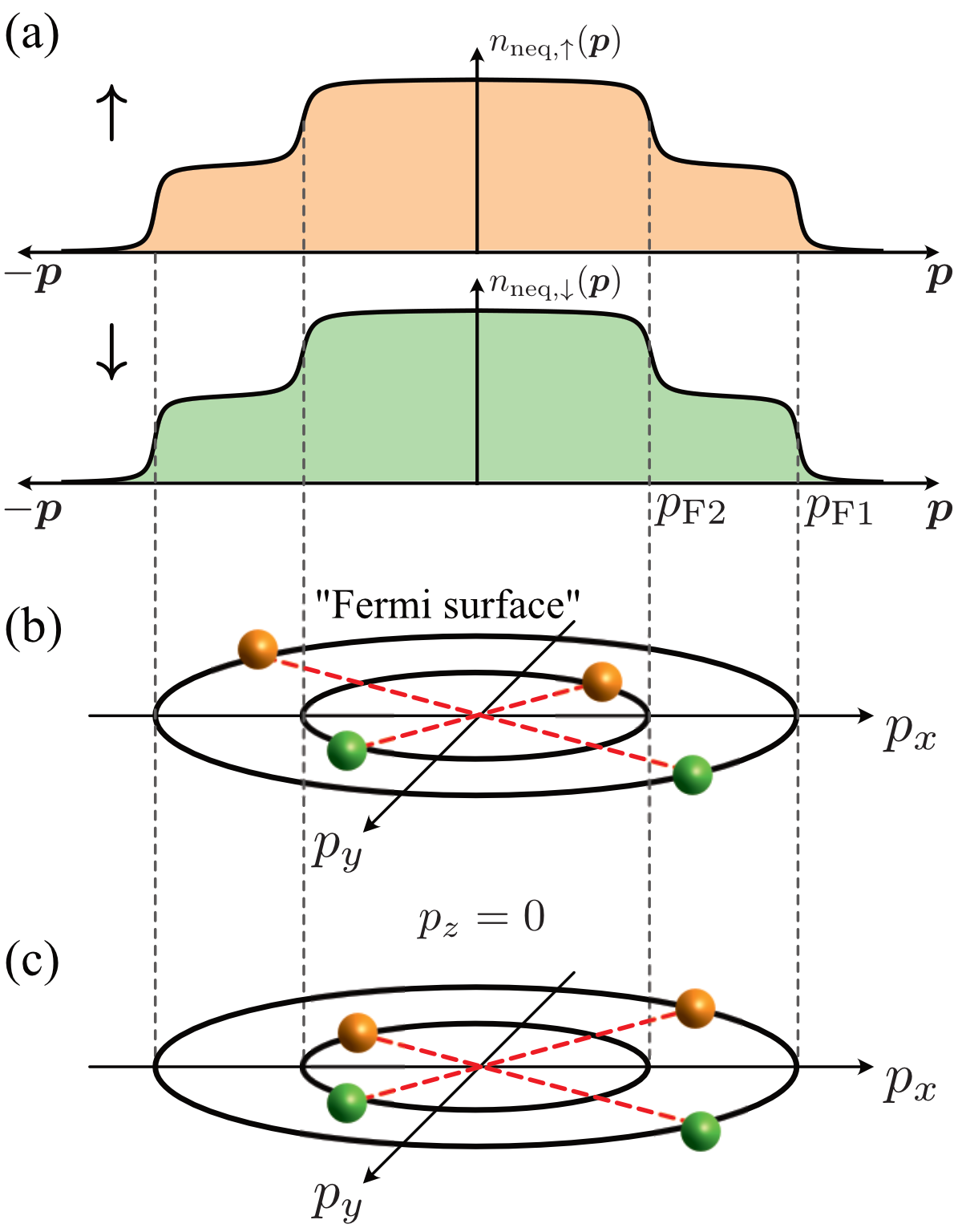}
\caption{Same figure as Fig.~\ref{fig.Cooper1} for the driven-dissipative Fermi gas. (a) The nonequilibrium momentum distribution $n_{{\rm neq}, \sigma}(\bm{p})$ in Eq.~\eqref{eq.neq.momentum}. Since the main system is the spin-{\it balanced} driven-dissipative Fermi gas, both pseudospin components ($\sigma=\up, \down$) obey the same momentum distribution, $n_{{\rm neq}, \up}(\bm{p})=n_{{\rm neq}, \down}(\bm{p})$. In each component, two Fermi edges imprinted on $n_{{\rm neq}, \sigma}(\bm{p})$ at $p=p_{{\rm F}1}$ and $p=p_{{\rm F2}}$ work like two ``Fermi surfaces" with different sizes. (b) and (c) indicate possible Cooper pairings between Fermi atoms around the ``Fermi surfaces". While Cooper pairs have zero center-of-mass momentum in (b), they have non-zero center-of-mass momentum in (c), which is similar to the FFLO Cooper pair shown in Fig.~\ref{fig.Cooper1}(b).}
\label{fig.Cooper2}
\end{figure}

The origin of this {\it nonequilibrium FFLO superfluid} is the nonequilibrium momentum distribution $n_{{\rm neq}, \sigma}(\bm{p})$ in Eq.~\eqref{eq.neq.momentum}~\cite{Kawamura2022, Kawamura2024}: When the momentum distribution $n_{{\rm neq}, \sigma}(\bm{p})$ has the clear two-step structure [see Fig.~\ref{fig.neq}(b)], the occupation sharply changes at $p_{{\rm F}1}=\sqrt{2m[\mu +\Delta\mu]}$ and at $p_{{\rm F}2}=\sqrt{2m[\mu -\Delta\mu]}$, as schematically shown in Fig.~\ref{fig.Cooper2}(a). Then if each edge structure at $p=p_{{\rm F}1}$ and $p=p_{{\rm F}2}$ work like the ordinary Fermi edge at the Fermi surface, each pseudospin component ($\sigma=\up, \down$) has two ``Fermi surfaces" with different sizes. Then, as schematically shown in Figs.~\ref{fig.Cooper2}(b) and (c), several types of Cooper pairings between Fermi atoms around the ``Fermi surfaces" would be possible. Among them, the Cooper pairs depicted in Fig.~\ref{fig.Cooper2}(c) have nonzero center-of-mass momentum $|\bm{q}_{\rm pair}| \simeq p_{{\rm F}1}-p_{{\rm F}2}$ $(\neq 0)$, which result in the nonequilibrium FFLO superfluid.

\section{Nonequilibrium BCS-BEC crossover in the driven-dissipative Fermi gas \label{sec.NETMA.neq.BCS.BEC}}

In the previous section, we discuss the nonequilibrium superfluid phase transition within the mean-field approximation. In this section, we go beyond the mean-field approximation to study the nonequilibrium BCS-BEC crossover.

\subsection{Strong-coupling theory for the driven-dissipative Fermi gas \label{sec.formalism.NETMA}}

\begin{figure}[t]
\centering
\includegraphics[width=8.5cm]{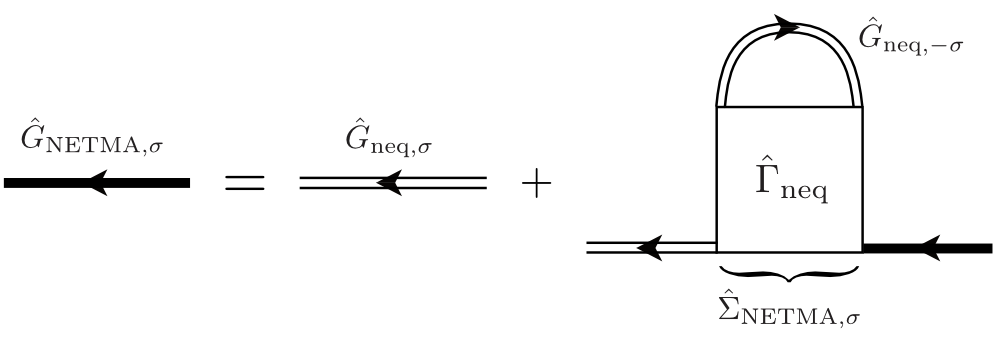}
\caption{Dyson equation for NETMA Green's function $\hat{G}_{{\rm NETMA}, \sigma}$ (thick solid line). The nonequilibrium Green's function $\hat{G}_{{\rm neq}, \sigma}$ (double solid line) is given in Fig.~\ref{fig.Genv}. The self-energy $\hat{\Sigma}_{{\rm NETMA}, \sigma}$ describes the nonequilibrium pairing fluctuation effects. The particle-particle scattering matrix $\hat{\Gamma}_{\rm neq}$ is given in Fig.~\ref{fig.Tmat.neq}(a).}
\label{fig.NETMA.Dyson}
\end{figure}

We extend the thermal-equilibrium TMA explained in Sec.~\ref{sec.eq.BCS.BEC} to the case when the system is out of equilibrium. In the nonequilibrium TMA (NETMA) scheme, while $T_{\rm env}^{\rm c}$-equation~\eqref{eq.neq.Thouless} obtained from the Thouless criterion is still valid, we need to include strong-coupling corrections to the number equation. For this purpose, we introduce the nonequilibrium Green's function in the main system, given by
\begin{equation}
\scalebox{0.97}{$\displaystyle
\hat{G}_{{\rm NETMA}, \sigma}(\bm{p}, \omega)=
\begin{pmatrix}
G^{\rm R}_{{\rm NETMA}, \sigma}(\bm{p}, \omega) & 
G^{\rm K}_{{\rm NETMA}, \sigma}(\bm{p}, \omega) \\[4pt]
0 &
G^{\rm A}_{{\rm NETMA}, \sigma}(\bm{p}, \omega)
\end{pmatrix}.$}
\label{eq.GNETMA.def}
\end{equation}
This $2\times 2$ matrix Green's function obeys the Dyson equation,
\begin{align}
&
\hat{G}_{{\rm NETMA}, \sigma}(\bm{p}, \omega)=
\hat{G}_{{\rm neq}, \sigma}(\bm{p}, \omega) 
\notag\\[4pt]
&\hspace{0.2cm}+
\hat{G}_{{\rm neq}, \sigma}(\bm{p}, \omega)
\hat{\Sigma}_{{\rm NETMA}, \sigma}(\bm{p}, \omega)
\hat{G}_{{\rm NETMA}, \sigma}(\bm{p}, \omega),
\label{eq.Dyson.NETMA}
\end{align} 
which is diagrammatically drawn as Fig.~\ref{fig.NETMA.Dyson}. In Eq.~\eqref{eq.Dyson.NETMA}, $\hat{G}_{{\rm neq}, \sigma}(\bm{p}, \omega)$ is given in Eq.~\eqref{eq.Gneq2}. (Note that it already involves the effects of system-reservoir couplings.) The self-energy $\hat{\Sigma}_{{\rm NETMA}, \sigma}(\bm{p}, \omega)$ describes the effects of pairing fluctuations within the TMA level, which is diagrammatically given by Fig.~\ref{fig.NETMA.Dyson}. Evaluating the last term in Fig.~\ref{fig.NETMA.Dyson}, we obtain~\cite{Kawamura2020_JLTP, Kawamura2020}
\begin{align}
&
\Sigma^{\rm R}_{{\rm NETMA}, \sigma}(\bm{p}, \omega)=
\big[\Sigma^{\rm A}_{{\rm NETMA}, \sigma}(\bm{p}, \omega)\big]^*
\notag\\
&= 
-\frac{i}{2} \sum_{\bm{q}} \int_{-\infty}^\infty \frac{d\nu}{2\pi}\Big[
\Gamma^{\rm R}_{\rm neq}(\bm{q}, \nu) G^{\rm K}_{{\rm neq}, -\sigma}(\bm{q} -\bm{p}, \nu- \omega) 
\notag\\
&\hspace{2cm}+
\Gamma^{\rm K}_{\rm neq}(\bm{q}, \nu) G^{\rm A}_{{\rm neq}, -\sigma}(\bm{q} -\bm{p}, \nu- \omega) 
\Big] \label{eq.sigR.NETMA}
,\\[6pt]
&
\Sigma^{\rm K}_{{\rm NETMA}, \sigma}(\bm{p}, \omega)
\notag\\
&= 
-\frac{i}{2} \sum_{\bm{q}} \int_{-\infty}^\infty \frac{d\nu}{2\pi}\Big[
\Gamma^{\rm A}_{\rm neq}(\bm{q}, \nu) G^{\rm R}_{{\rm neq}, -\sigma}(\bm{q} -\bm{p}, \nu- \omega) 
\notag\\
&\hspace{2cm}+
\Gamma^{\rm R}_{\rm neq}(\bm{q}, \nu) G^{\rm A}_{{\rm neq}, -\sigma}(\bm{q} -\bm{p}, \nu- \omega) 
\notag\\
&\hspace{2cm}+
\Gamma^{\rm K}_{\rm neq}(\bm{q}, \nu) G^{\rm K}_{{\rm neq}, -\sigma}(\bm{q} -\bm{p}, \nu- \omega)
\Big], \label{eq.sigK.NETMA}
\end{align}
where the particle-particle scattering matrix $\Gamma^{\rm R, A, K}_{\rm neq}(\bm{q}, \nu)$ is given in Eq.~\eqref{eq.Tmat.neq}. Substituting Eqs.~\eqref{eq.sigR.NETMA} and \eqref{eq.sigK.NETMA} into the Dyson equation~\eqref{eq.Dyson.NETMA}, we obtain each component in Eq.~\eqref{eq.GNETMA.def} as
\begin{align}
G^{\rm R}_{{\rm NETMA}, \sigma}(\bm{p}, \omega) 
&=
\big[G^{\rm A}_{{\rm NETMA}, \sigma}(\bm{p}, \omega) \big]^*
\notag\\
&=
\frac{1}{\omega -\ep_{\bm{p}}+2i\gamma -\Sigma^{\rm R}_{{\rm NETMA}, \sigma}(\bm{p}, \omega)}
,\\
G^{\rm K}_{{\rm NETMA}, \sigma}(\bm{p}, \omega) 
&=
\frac{\Sigma^{\rm K}_{{\rm NETMA}, \sigma}(\bm{p}, \omega) +\Sigma^{\rm K}_{{\rm env}, \sigma}(\bm{p}, \omega)}{|\omega -\ep_{\bm{p}} +2i\gamma -\Sigma^{\rm R}_{{\rm NETMA}, \sigma}(\bm{p}, \omega)|^2}.	
\end{align}

The number $N$ of Fermi atoms in the main system is computed from the Keldysh component $G^{\rm K}_{{\rm NETMA}, \sigma}(\bm{p}, \omega) $ as
\begin{widetext}
\begin{align}
N
&= 
-\frac{i}{2} \sum_{\sigma=\up, \down} \sum_{\bm{p}} \int_{-\infty}^\infty \frac{d\omega}{2\pi} G^{\rm K}_{{\rm NETMA}, \sigma}(\bm{p}, \omega) +\frac{1}{2}
\notag\\
&=
-2i \sum_{\bm{p}} \int_{-\infty}^\infty \frac{d\omega}{2\pi} \frac{2i\gamma \big[f(\omega -\mu_{\rm L}) +f(\omega -\mu_{\rm R})\big] +\Sigma^<_{{\rm NETMA}, \sigma}(\bm{p}, \omega)}{\big[\omega -\ep_{\bm{p}} -{\rm Re}\Sigma^{\rm R}_{{\rm NETMA}, \sigma}(\bm{p}, \omega) \big]^2 +\big[2\gamma -{\rm Im} \Sigma^{\rm R}_{{\rm NETMA}, \sigma}(\bm{p}, \omega)\big]^2}.
\label{eq.N.NETMA}
\end{align}
Here, we have introduced the lesser self-energy~\cite{Stefanucci2013, Rammer2007, Zagoskin2014},
\begin{equation}
\Sigma^<_{{\rm NETMA}, \sigma}(\bm{p}, \omega)= \frac{1}{2}
\big[-\Sigma^{\rm R}_{{\rm NETMA}, \sigma}(\bm{p}, \omega) +\Sigma^{\rm A}_{{\rm NETMA}, \sigma}(\bm{p}, \omega) +\Sigma^{\rm K}_{{\rm NETMA}, \sigma}(\bm{p}, \omega)\big].
\end{equation}
The number equation~\eqref{eq.N.NETMA} now includes both the strong-coupling effects associated with the tunable pairing interaction, as well as the system-reservoir coupling effects. As in the thermal equilibrium BCS-BEC crossover theory explained in Sec.~\ref{sec.eq.BCS.BEC}, we solve the number equation~\eqref{eq.N.NETMA}, together with the nonequilibrium Thouless criterion in Eq.~\eqref{eq.neq.Thouless}, to determine $\mu$ and $T_{\rm env}^{\rm c}$ for a given parameter set ($N$, $(p_{\rm F} a_s)^{-1}$, $\Delta \mu$). When $T_{\rm env}>T_{\rm env}^{\rm c}$, we only solve the number equation~\eqref{eq.N.NETMA}, to determine $\mu$.
\end{widetext}

Once $\mu$ is determined, the single-particle spectral function $A_{{\rm NETMA}, \sigma}(\bm{p}, \omega)$, the density of states $\rho_{{\rm NETMA}, \sigma}(\omega)$, as well as the photoemission spectrum $L_{{\rm NETMA}, \sigma}(\omega)$ are obtained from the NETMA Green's function as, respectively,~\cite{Kawamura2020}
\begin{align}
& A_{{\rm NETMA}, \sigma}(\bm{p}, \omega) =
-\frac{1}{\pi} {\rm Im}\big[G^{\rm R}_{{\rm NETMA}, \sigma}(\bm{p}, \omega) \big]
,\\[4pt]
& \rho_{{\rm NETMA}, \sigma}(\omega) = \sum_{\bm{p}} A_{{\rm NETMA}, \sigma}(\bm{p}, \omega)
,\\
& L_{{\rm NETMA}, \sigma}(\omega) = -i p^2 \big[-G^{\rm R}_{{\rm NETMA}, \sigma} +G^{\rm A}_{{\rm NETMA}, \sigma} 
\notag\\
&\hspace{3.4cm}
+G^{\rm K}_{{\rm NETMA}, \sigma}\big](\bm{p}, \omega).
\label{eq.PES.NETMA}
\end{align}

\subsection{Nonequilibrium BCS-BEC crossover}

\begin{figure}[t]
\centering
\includegraphics[width=8cm]{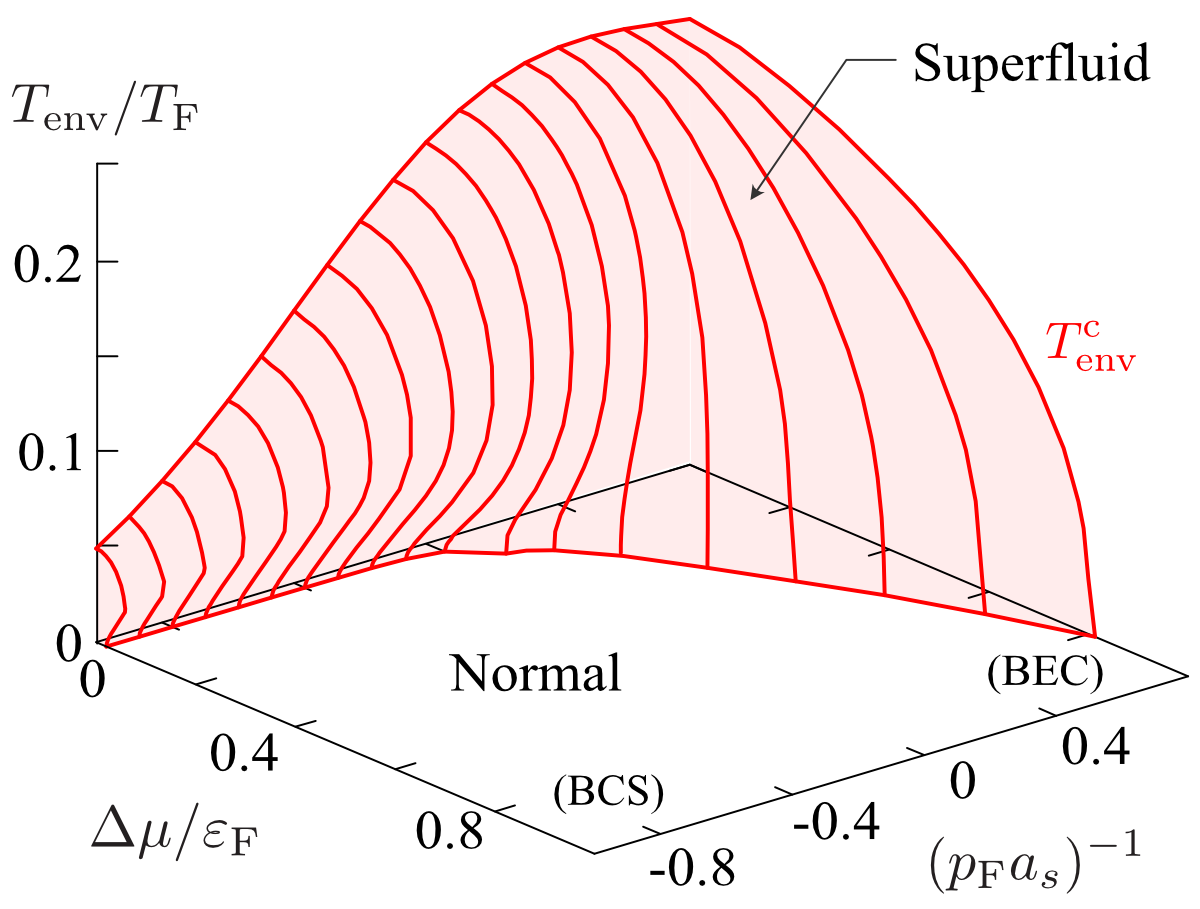}
\caption{Calculated $T_{\rm env}^{\rm c}$ in the driven-dissipative Fermi gas in the nonequilibrium BCS-BEC crossover region. We set $\gamma/\ep_{\rm F}=0.01$. (This value is also used in Figs.~\ref{fig.NBCS}, \ref{fig.NBCS.sus}, and \ref{fig.SW.PGapp}-\ref{fig.NBEC}.)}
\label{fig.NETMA.Tc}
\end{figure}

Figure~\ref{fig.NETMA.Tc} shows the superfluid phase transition temperature $T_{\rm env}^{\rm c}$ as a function of the chemical potential bias $\Delta\mu$, and the pairing interaction strength $(p_{\rm F}a_s)^{-1}$, in the nonequilibrium BCS-BEC crossover region obtained by NETMA. In the zero bias case $(\Delta\mu=0)$, the overall behavior of $T_{\rm env}^{\rm c}$ is the same as $T_{\rm c}$ in a thermal equilibrium Fermi gas shown in Fig.~\ref{fig.TMA.DOS}(a): Starting from the weak-coupling BCS regime, $T_{\rm env}^{\rm c}$ gradually increases with increasing the interaction strength, to eventually approach a constant value in the strong-coupling BEC regime.

\begin{figure*}[t]
\centering
\includegraphics[width=15cm]{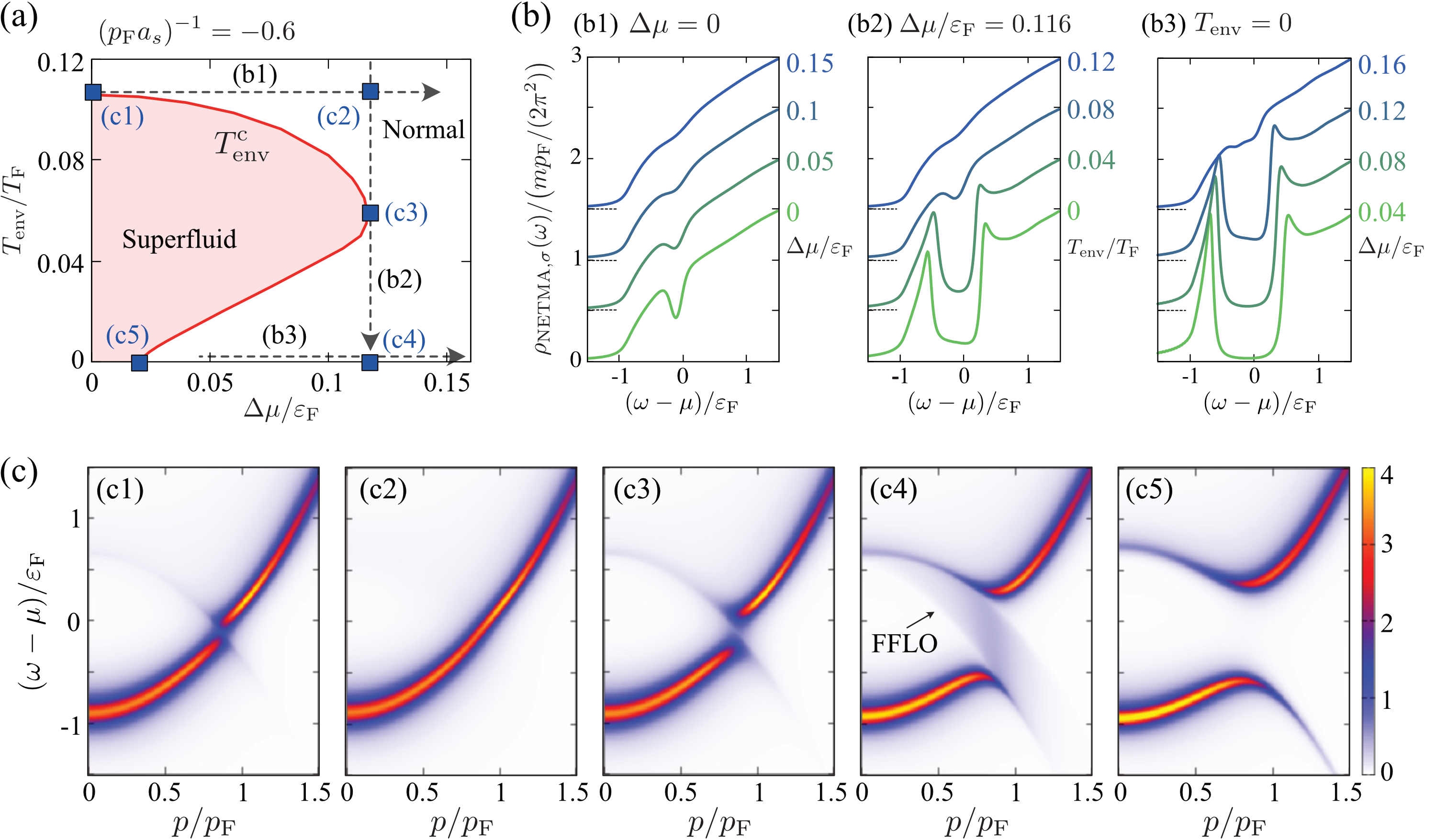}
\caption{(a) Calculated $T_{\rm env}^{\rm c}$ as a function of $\Delta\mu$, in the driven-dissipative Fermi gas in the BCS region at $(p_{\rm F}a_s)^{-1}=-0.6$. $T_{\rm env}^{\rm c}$ exhibits re-entrant behavior in this region. (b) Single-particle density of states $\rho_{{\rm NETMA}, \sigma}(\omega)$ along the paths (b1)-(b3) at panel (a). For visibility, we offset each result by 0.5. (c) Single-particle spectral function $A_{{\rm NETMA}, \sigma}(\bm{p}, \omega)$ at (c1)-(c5) in panel (a). In (c4), the broad peak structure  ``FFLO" originates from FFLO-type pairing fluctuations. The intensity of the spectral function is normalized by $\ep_{\rm F}^{-1}$.}
\label{fig.NBCS}
\end{figure*}

The main system is in the driven-dissipative steady state when the non-zero bias $\Delta\mu\neq 0$ is imposed. In this nonequilibrium case, $T_{\rm env}^{\rm c}$ is suppressed, as shown in Fig.~\ref{fig.NETMA.Tc}. In the strong-coupling BEC regime, $(p_{\rm F} a_s)^{-1}\gtrsim 0$, $T_{\rm env}^{\rm c}$ is monotonically suppressed with increasing $\Delta\mu$. On the other hand, in the weak-coupling BCS side, $(p_{\rm F} a_s)^{-1}\lesssim 0$, $T_{\rm env}^{\rm c}$ is found to exhibits re-entrant behavior. That is, with decreasing the temperature $T_{\rm env}$, although the main system experiences the superfluid phase transition at $T_{\rm env}=T_{\rm env}^{\rm c}$, the main system again becomes the normal state at low temperatures [see also Fig.~\ref{fig.NBCS}(a)].

Figure~\ref{fig.NETMA.Tc} indicates that, in NETMA, the nonequilibrium FFLO state, which is obtained in the nonequilibrium mean-field theory as explained in Sec.~\ref{sec.NFFLO.NMF}, does not appear in the whole BCS-BEC crossover region. That is, at $T_{\rm env}=T_{\rm env}^{\rm c}$, the nonequilibrium Thouless criterion in Eq.~\eqref{eq.neq.Thouless} is always satisfied at $\bm{q}_{\rm pair}=0$ in NETMA. In general, the strong-coupling theory and the mean-field KM theory are expected to give qualitatively the same results in the weak-coupling BCS regime because pairing fluctuations are sufficiently weak there. Indeed, in the thermal equilibrium case, $T_{\rm c}$ obtained by TMA approaches the mean-field result in the weak-coupling limit, as shown in Fig.~\ref{fig.TMA.DOS}(a). Nevertheless, in the present case, the nonequilibrium FFLO state predicted in the mean-field theory is not obtained in NETMA even in the weak-coupling BCS regime. 

\subsubsection{Weak-coupling BCS regime \label{sec.NETMA.weak}}

We discuss the mechanism of the re-entrant behavior of $T_{\rm env}^{\rm c}$ as well as the absence of the nonequilibrium FFLO state, in the weak-coupling BCS regime. Figure~\ref{fig.NBCS}(a) shows $T_{\rm env}^{\rm c}$ as a function of $\Delta\mu$. In Figs.~\ref{fig.NBCS}(b) and (c), we show the single-particle density of states $\rho_{{\rm NETMA}, \sigma}(\omega)$ along the paths (b1)-(b3) and the spectral function $A_{{\rm NETMA}, \sigma}(\bm{p}, \omega)$ at (c1)-(c5) in Fig.~\ref{fig.NBCS}(a).

When $\Delta\mu=0$, the pseudogap appears in the density of states, as shown in Fig.~\ref{fig.NBCS}(b1). We also find from Fig.~\ref{fig.NBCS}(c1) that the spectral function has the paek along the particle dispersion ($\omega=\ep_{\bm{p}}$), as well as the hole dispersion ($\omega = -\ep_{-\bm{p}} +2\mu$), which is the typical pseudogap structure explained in Sec.~\ref{sec.SW.DOS.eq}. Since the pseudogap phenomenon is induced by pairing fluctuations, it should be enhanced (suppressed) as we approach (move away from) $T_{\rm env}^{\rm c}$. Indeed, we see in Fig.~\ref{fig.NBCS}(b1) that the pseudogap gradually disappears, as one moves away from $T_{\rm env}^{\rm c}$. At (c2) in Fig.~\ref{fig.NBCS}(a), the peak structure along the hole dispersion disappears in the spectral function, as expected.

However, a very different behavior of the pseudogap is obtained when one moves along the path (b2) in Fig.~\ref{fig.NBCS}(a). In this case, although we first approach $T_{\rm env}^{\rm c}$ and then move away from $T_{\rm env}^{\rm c}$, the pseudogap monotonically develops up to the point (c4) in Fig.~\ref{fig.NBCS}(a), as seen in Fig.~\ref{fig.NBCS}(b2). This tendency is also seen in the single-particle spectral function: Indeed, Figs.~\ref{fig.NBCS}(c2)-(c4) show that the level repulsion between the particle and the hole dispersions monotonically becomes remarkable, even when one moves from (c3) to (c4) in Fig.~\ref{fig.NBCS}(a).

\begin{figure}[t]
\centering
\includegraphics[width=7cm]{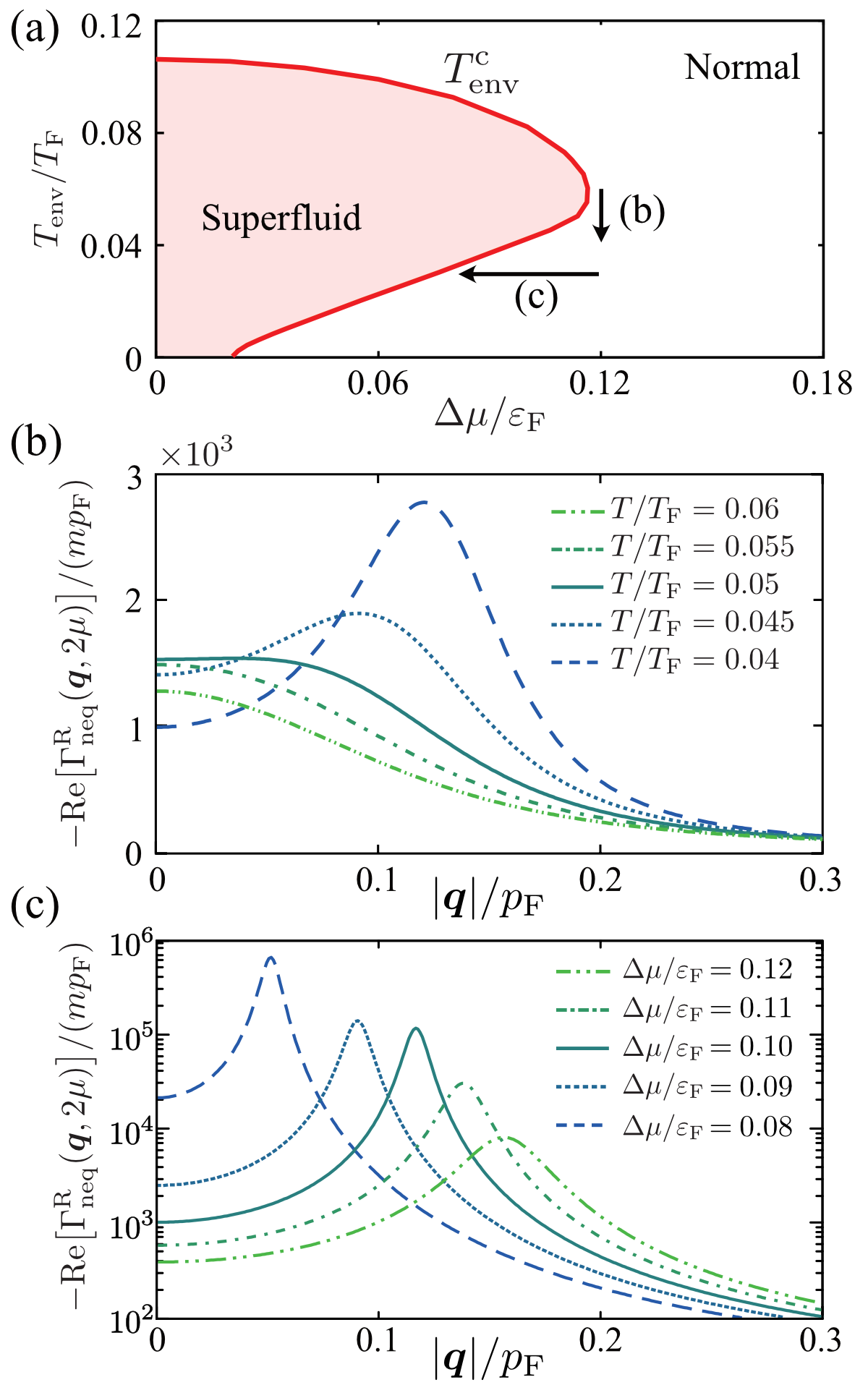}
\caption{(a) Phase diagram of the driven-dissipative Fermi gas at $(p_{\rm F}a_s)^{-1}=-0.6$. (b) and (c) show the real part of the particle-particle scattering matrix $-{\rm Re}\big[\Gamma^{\rm R}_{\rm neq}(\bm{q}, \nu=2\mu)\big]$ as a function of $|\bm{q}|$. Each panel shows the result along the path (b) and path (c) in panel (a).}
\label{fig.NBCS.sus}
\end{figure}

To understand the reason why the pseudogap phenomenon is remarkably seen at (c4) in Fig.~\ref{fig.NBCS}(a), we plot in Fig.~\ref{fig.NBCS.sus}(b) the real part of the particle-particle scattering matrix $-{\rm Re}\big[\Gamma^{\rm R}_{\rm neq}(\bm{q}, \nu=2\mu)\big]$ as a function of $|\bm{q}|$, which informs us of the strength of pairing fluctuations. (Note that $\bm{q}$ in this quantity physically has the meaning of the center-of-mass momentum of preformed Cooper pairs.) We find from this figure that, when one moves along the path (b) in Fig.~\ref{fig.NBCS.sus}(a), although pairing fluctuations in the BCS-type Cooper channel at $\bm{q}=0$ decreases, FFLO-type pairing fluctuations being characterized by $\bm{q}\neq 0$ becomes strong.

Since we are considering the spin-balanced case, the ordinary mechanism of the FFLO state does not work here. Instead, the anomalous enhancement of $-{\rm Re}\big[\Gamma^{\rm R}_{\rm neq}(\bm{q}, \nu=2\mu)\big]$ at $\bm{q}\neq 0$ seen in Fig.~\ref{fig.NBCS.sus}(b) is attributed to the two-step structure imprinted on the nonequilibrium momentum distribution $n_{{\rm neq},\sigma}({\bm{p}})$, as explained in Sec.~\ref{sec.NFFLO.NMF}. Since the thermal excitations are suppressed and the momentum distribution $n_{{\rm neq},\sigma}({\bm{p}})$ has the clear two-step structure at low temperatures, pairing fluctuations at $\bm{q}\neq 0$ are enhanced as $T_{\rm env}$ decreases. To conclude, the pronounced pseudogap seen around (c4) in Fig.~\ref{fig.NBCS}(a) is induced by strong FFLO-type pairing fluctuations that are enhanced by the two-step structure of the nonequilibrium momentum distribution $n_{{\rm neq},\sigma}({\bm{p}})$.

\begin{figure}[t]
\centering
\includegraphics[width=6.2cm]{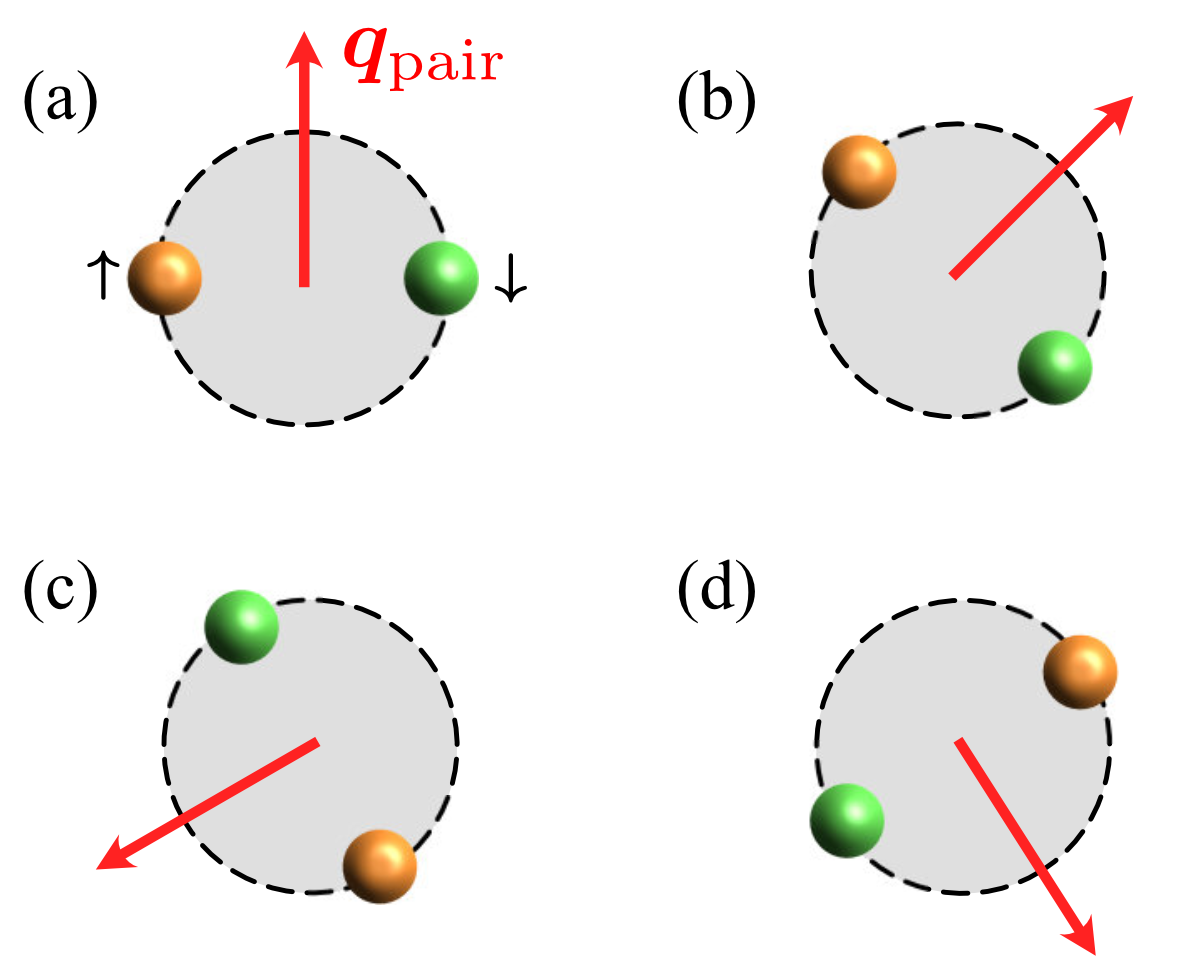}
\caption{Schematic illustration of a nonequilibrium FFLO Cooper pairs. All superfluid states associated with a Cooper pair depicted in (a)-(d) are degenerate due to the continuous rotational symmetry of the gas system.}
\label{fig.instability}
\end{figure}

Regarding the vanishing FFLO phase transition in the NETMA result, we recall that it has been shown in the thermal equilibrium state that the FFLO state is always unstable against rotational fluctuations in the spatially isotropic system, even in three dimensions~\cite{Kawamura2022_2, Shimahara1998, Ohashi2002_2, Radzihovsky2009, Radzihovsky2011, Jakubczyk2017}: When the system has a continuous rotational symmetry in space, the FFLO state is infinitely degenerate with respect to the direction of the FFLO-$\bm{q}_{\rm pair}$ vector, as schematically shown in Fig.~\ref{fig.instability}. This infinite degeneracy remarkably enhances FFLO pairing fluctuations, which completely destroy the FFLO long-range order. We note that a similar instability phenomenon is also known in the ordinary BCS superfluid in one and two dimensions, which is sometimes referred to as the Hohenberg-Mermin-Wagner theorem in the literature~\cite{Hohenberg1967, Mermin1966}. However, while the vanishing superfluid long-range order in one and two dimensions is also due to anomalously enhanced pairing fluctuations, in the FFLO case, the instability occurs even in three dimensions. Since the same instability mechanism also works in the present nonequilibrium case~\cite{Kawamura2023}, the FFLO-type superfluid phase transition seen in the mean-field phase diagram in Fig.~\ref{fig.NThouless1}(c) completely vanishes in the NETMA phase diagram.

We briefly note that, since the spatial isotropy of the system is crucial for the vanishing FFLO long-range order, the FFLO-type superfluid phase transition would be possible when this symmetry is removed from the present gas system, which we will examine in Sec.~\ref{sec.stable.NFFLO}.

As explained in Sec.~\ref{sec.NFFLO.NMF}, the magnitude of the center-of-mass momentum $|\bm{q}_{\rm pair}|$ of a nonequilibrium FFLO Cooper pair depends on the difference between the ``Fermi momenta" of the two effective ``Fermi surfaces".  That is, $|\bm{q}_{\rm pair}|\sim p_{\rm F1}-p_{\rm F2}=\sqrt{2m[\mu+\Delta\mu]} -\sqrt{2m[\mu-\Delta\mu]}$ (see Fig.~\ref{fig.Cooper2}). Indeed, we see in Fig.~\ref{fig.NBCS.sus}(c) that the peak position shifts toward $\bm{q}=0$ with decreasing $\Delta\mu$ along the path (c) in Fig.~\ref{fig.NBCS.sus}(a). When we arrive at $T_{\rm env}^{\rm c}$, $\Gamma^{\rm R}_{\rm neq}(\bm{q}, 2\mu)$ diverges at $\bm{q}=0$ (that is, $T_{\rm env}^{\rm c}$-equation~\eqref{eq.neq.Thouless} is satisfied at $\bm{q}=0$), and the main system transitions to the uniform nonequilibrium BCS superfluid state. 

The strong nonequilibrium FFLO pairing fluctuations also affect the single-particle excitation spectrum. As shown in Fig.~\ref{fig.NBCS}(c4), the spectral function at (c4) in Fig.~\ref{fig.NBCS}(a) exhibits a broad downward spectral peak, which is denoted as ``FFLO". To understand the mechanism of this anomalous spectral structure, we apply the pseudogap approximation to the NETMA self-energy $\Sigma^{\rm R}_{{\rm NETMA}, \sigma}(\bm{p}, \omega)$: As seen from Fig.~\ref{fig.NBCS.sus}(b), $\Gamma^{\rm R}_{\rm neq}(\bm{q}, \nu)$ has a large intensity at $(\bm{q}, \nu) = (\bm{q}_{\rm pair}, 2\mu)$ around (c4) in Fig.~\ref{fig.NBCS}(a). Using this, one approximates the self-energy in Eq.~\eqref{eq.sigR.NETMA} to \cite{Kawamura2020}
\begin{align}
&
\Sigma^{\rm R}_{{\rm NETMA}, \sigma}(\bm{p}, \omega)
\notag\\[4pt]
&\hspace{0.2cm}\simeq
-\Delta^2_{\rm PG} \braket{G^{\rm A}_{{\rm neq},-\sigma}(\bm{q}_{\rm pair} -\bm{p}, -\omega +2\mu)}_{\bm{q}_{\rm pair}}
\notag\\
&\hspace{0.2cm}=
-\frac{\Delta^2_{\rm PG}}{4|\bm{q}_{\rm pair}||\bm{p}|} \log\left(\frac{\omega +[|\bm{p}| -|\bm{q}_{\rm pair}|]^2 -\mu +2i\gamma}{\omega +[|\bm{p}| +|\bm{q}_{\rm pair}|]^2 -\mu +2i\gamma}\right).
\label{eq.self.PG}
\end{align}
Here, $\braket{\cdots}_{\bm{q}_{\rm pair}}$ represents the average over the direction of the $\bm{q}_{\rm pair}$, and 
\begin{equation}
\Delta^2_{\rm PG} = \frac{i}{2} \sum_{\bm{q}} \int_{-\infty}^\infty \frac{d\nu}{2\pi} \Gamma^{\rm K}_{\rm neq}(\bm{q}, \nu)
\end{equation}
is the pseudogap parameter.

\begin{figure}[t]
\centering
\includegraphics[width=7.5cm]{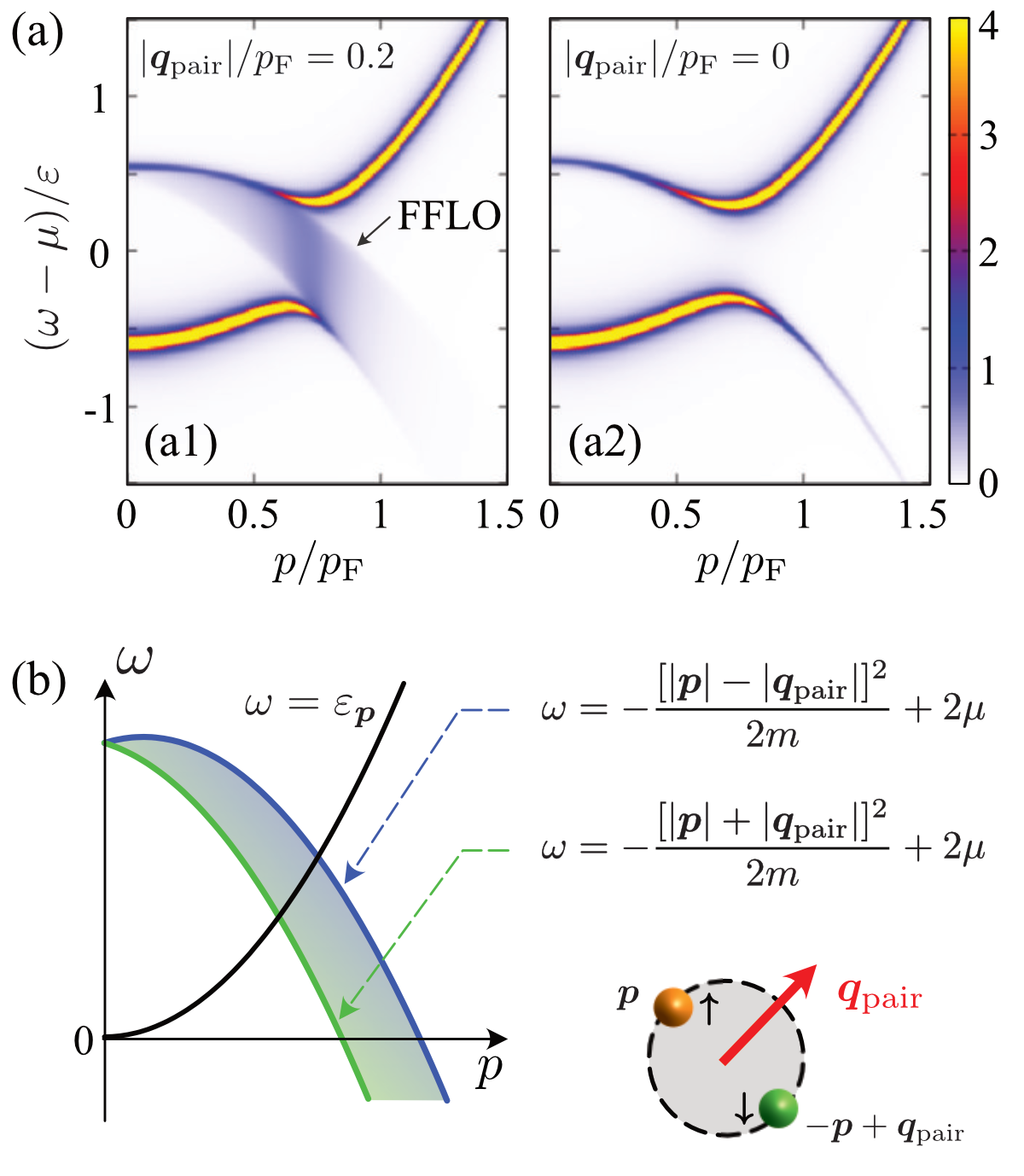}
\caption{(a) Single-particle spectral function in the pseudogap approximation. We set (a1) $\bm{q}_{\rm pair}\neq 0$ and (a2) $\bm{q}_{\rm pair}=0$. A broad downward spectral peak ``FFLO" in panel (a1) is consistent with Fig.~\ref{fig.NBCS}(c4). (b) Mechanism of the broad spectral peak. When a Cooper pair has nonzero center-of-mass momentum $\bm{q}_{\rm pair}$, the excitation modes along the particle dispersion $\omega = \ep_{\bm{p}}$ is coupled to the excitation modes along the hole dispersion $\omega = -\ep_{-\bm{p}+\bm{q}_{\rm pair}} +2\mu$ with {\it various directions} of $\bm{q}_{\rm pair}$.}
\label{fig.SW.PGapp}
\end{figure}

Figure~\ref{fig.SW.PGapp} shows the single-particle spectral function $A_{{\rm NETMA}, \sigma}(\bm{p}, \omega)$ computed with the approximated self-energy in Eq.~\eqref{eq.self.PG}. While we set $\bm{q}_{\rm pair}/p_{\rm F}=0$ in Fig.~\ref{fig.SW.PGapp}(a1), we take $\bm{q}_{\rm pair}/p_{\rm F}=0.2$ in Fig.~\ref{fig.SW.PGapp}(a2). Comparing these two cases, we find that the ``FFLO" structure seen in Fig.~\ref{fig.SW.PGapp}(a) is induced by nonequilibrium FFLO pairing fluctuations, because the broad peak structure appears only when $\bm{q}_{\rm pair}\neq 0$. The difference between the two cases ($\bm{q}_{\rm pair}=0$ and $\bm{q}_{\rm pair}\neq 0$) arises from the degrees of freedom with respect to the direction of $\bm{q}_{\rm pair}$ due to the spatial isotropy of the gas system: In the case of BCS-type pairing, where a Cooper pair is formed between Fermi atoms at opposite momenta $\bm{p}$ and $-\bm{p}$, the excitation modes along the particle dispersion $\omega = \ep_{\bm{p}}$ is coupled only to the modes along the hole dispersion $\omega = -\ep_{-\bm{p}}+2\mu$. As explained in Sec.~\ref{sec.SW.DOS.eq}, the level repulsion between these excitation modes results in the pseudogap. On the other hand, in the case of FFLO-type paring, a Cooper pair is formed between Fermi atoms at momenta $\bm{p}$ and $-\bm{p}+\bm{q}_{\rm pair}$. Then, the excitation modes along the particle dispersion $\omega = \ep_{\bm{p}}$ is coupled to the excitation modes along the hole dispersions $\omega = -\ep_{-\bm{p}+\bm{q}_{\rm pair}} +2\mu$. As schematically shown in Fig.~\ref{fig.SW.PGapp}(b), the latter is a continuous excitation mode between $\omega = -\big[|\bm{p}|+|\bm{q}_{\rm pair}|\big]^2/(2m) +2\mu$ and $\omega = -\big[|\bm{p}|-|\bm{q}_{\rm pair}|\big]^2/(2m) +2\mu$, because there is no restriction on the direction of $\bm{q}_{\rm pair}$ due to the spatial isotropy of the gas system. The coupling of the continuous spectrum with the excitation modes along the particle dispersion $\omega=\ep_{\bm{p}}$ results in the characteristic spectral structure ``FFLO" in Fig.~\ref{fig.NBCS}(c4).

\begin{figure}[t]
\centering
\includegraphics[width=8cm]{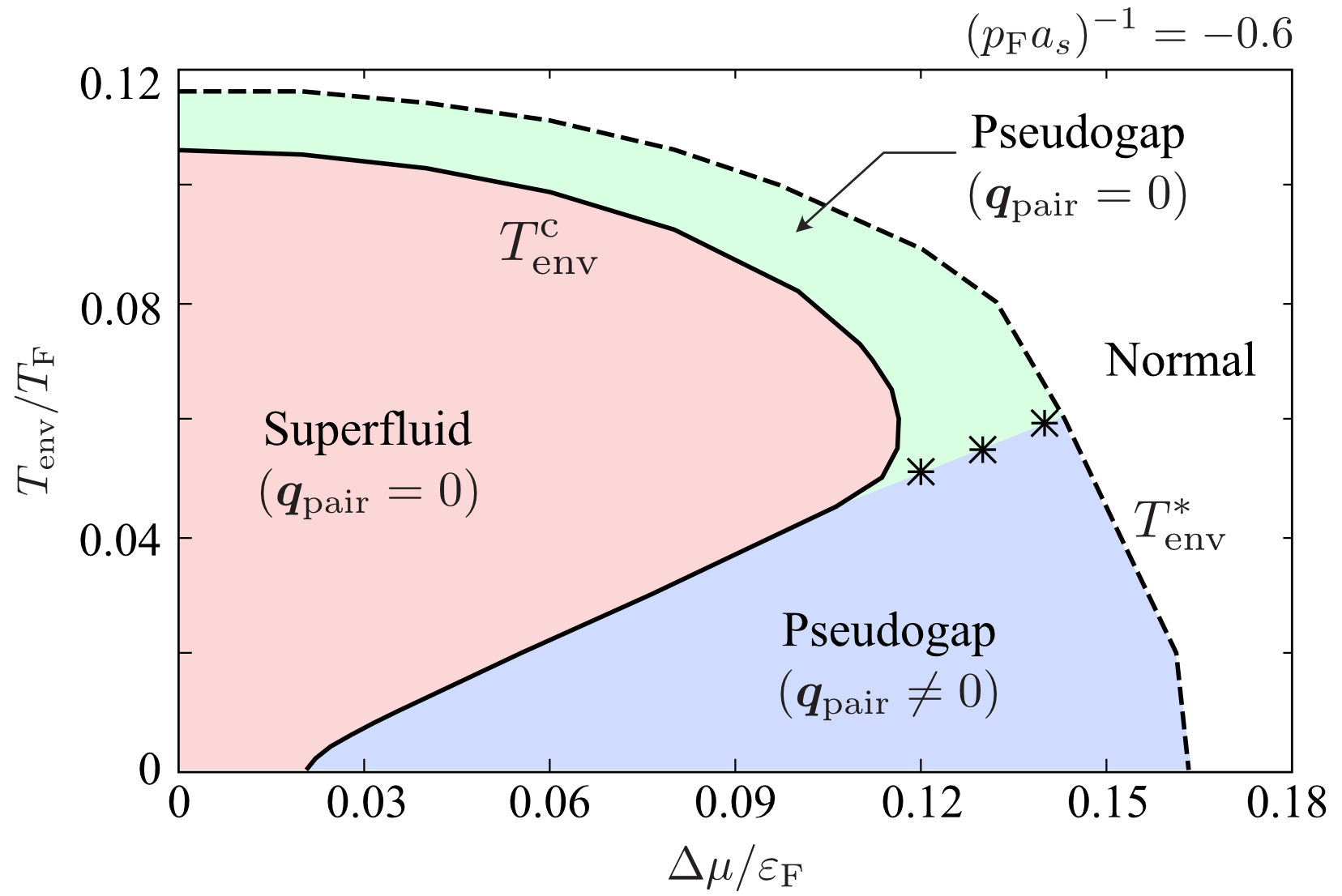}
\caption{Phase diagram of the driven-dissipative Fermi gas in the BCS regime, in terms of the temperature $T_{\rm env}$ and the chemical potential bias $\Delta\mu$. $T_{\rm env}^{\rm c}$ (solid line) is the superfluid phase transition temperature and $T^*_{\rm env}$ (dashed line) is the pseudogap temperature. The pseudogap phase between $T_{\rm env}^{\rm c}$ and $T_{\rm env}^*$ is divided into two regions: One is the pseudogap phase induced by BCS-type pairing fluctuations (green-shaded area). In this phase, $-{\rm Re}\big[\Gamma^{\rm R}_{\rm neq}(\bm{q}, 2\mu)\big]$ has the peak at $\bm{q}=0$. The other pseudogap phase is induced by nonequilibrium FFLO-type pairing fluctuations (blue-shaded area). In this region, the peak appears at $\bm{q}=\bm{q}_{\rm pair}$ ($\neq 0$).}
\label{fig.PD.weak}
\end{figure}

To summarize the discussion, the phase diagram of the driven-dissipative Fermi gas in the weak-coupling BCS regime is obtained as Fig.~\ref{fig.PD.weak}. In this phase diagram, $T^*_{\rm env}$ is the so-called pseudogap temperature~\cite{Tsuchiya2009}, where the pseudogap appears in the density of states $\rho_{{\rm NETMA}, \sigma}(\omega)$. The pseudogap phase between $T^{\rm c}_{\rm env}$ and $T^*_{\rm env}$ can be further divided into two regions:
\begin{enumerate}
\item BCS-type pseudogap phase ($\bm{q}_{\rm pair}=0$): The pseudogap is induced by BCS-type pairing fluctuations. In this phase, $-{\rm Re}\big[\Gamma^{\rm R}_{\rm neq}(\bm{q}, 2\mu)\big]$ has the peak at $\bm{q}=0$.
\item FFLO-type pseudogap phase ($\bm{q}_{\rm pair}\neq 0$): The pseudogap is induced by nonequilibrium FFLO-type pairing fluctuations. In this phase, the peak appears at $\bm{q}=\bm{q}_{\rm pair}$ ($\neq 0$) and the spectral function $A_{{\rm NETMA}, \sigma}(\bm{p}, \omega)$ exhibits the broad downward spectral peak.
\end{enumerate}
Since the nonequilibrium FFLO superfluid state is unstable against its own paring fluctuations, only the BCS-type uniform superfluid state is realized inside the region surrounded by $T_{\rm env}^{\rm c}$ in Fig.~\ref{fig.PD.weak}.

\subsubsection{Strong-coupling BEC regime \label{sec.NETMA.BEC}}

\begin{figure*}[t]
\centering
\includegraphics[width=15cm]{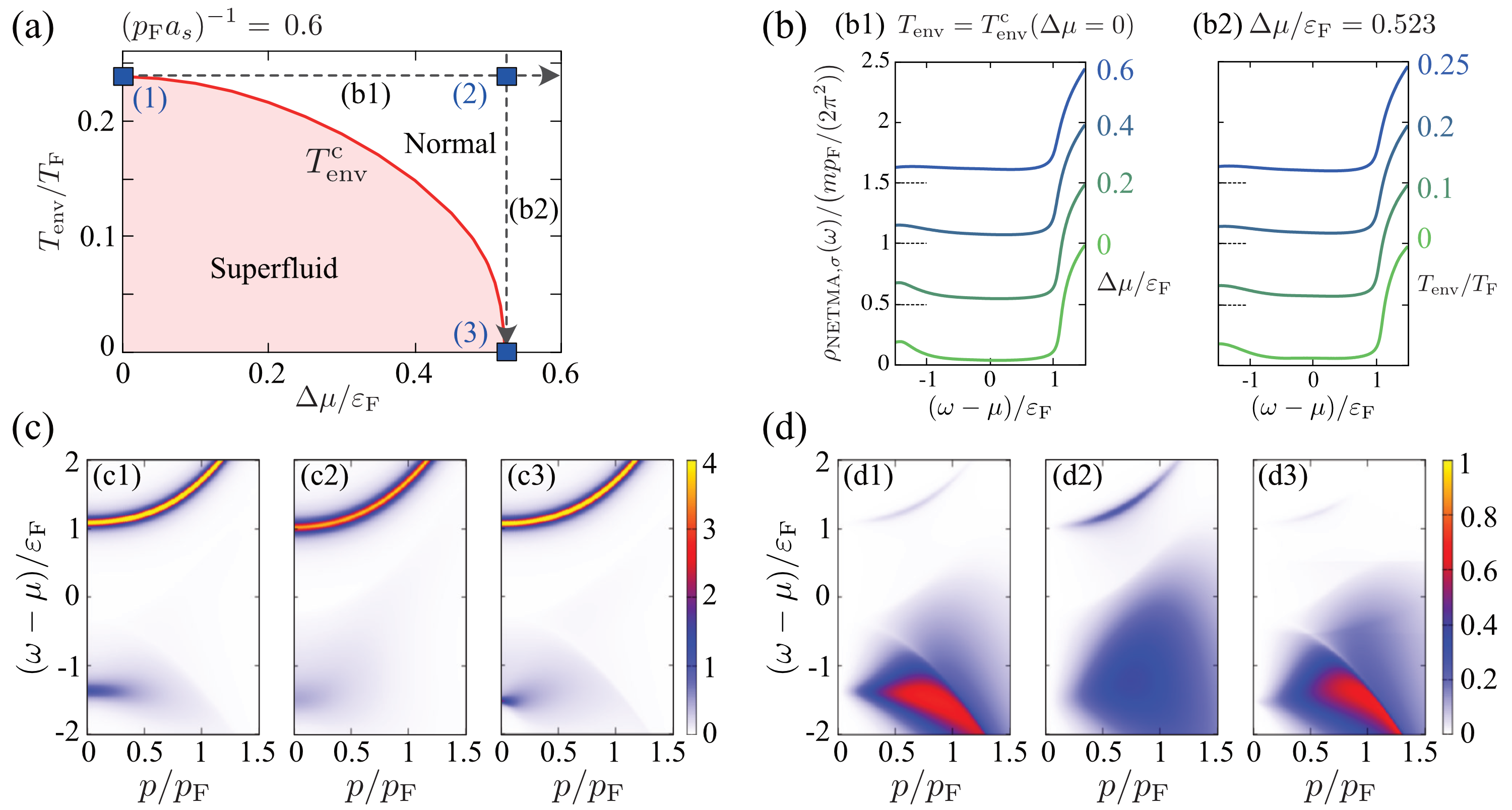}
\caption{(a) Calculated $T_{\rm env}^{\rm c}$ in the BEC regime, when $(p_{\rm F}a_s)^{-1}=0.6$. (b) Density of states $\rho_{{\rm NETMA}, \sigma}(\omega)$ along the paths (b1) and (b2) in panel (a). (c) Single-particle spectral function $A_{{\rm NETMA}, \sigma}(\bm{p}, \omega)$. (d) Photoemission spectrum $L_{{\rm NETMA}, \sigma}(\bm{p}, \omega)$. Panels (c1)-(c3) and (d1)-(d3) show the results at positions (1)-(3) in panel (a).}
\label{fig.NBEC}
\end{figure*}

We next consider the strong-coupling BEC regime. In this regime, as shown in Fig.~\ref{fig.NBEC}(a), $T_{\rm env}^{\rm c}$ does not exhibit re-entrant behavior, but monotonically decreases as the chemical potential bias $\Delta\mu$ increases. This is simply because the nonequilibrium momentum distribution $n_{{\rm neq}, \sigma}(\bm{p})$ does not exhibit the two-step structure in this regime, so that, in contrast to the weak-coupling BCS case shown in Fig.~\ref{fig.NBCS.sus}(b), the anomalous enhancement of FFLO-type pairing fluctuations does not occur.

Figure~\ref{fig.NBEC}(b) shows how the density of states $\rho_{{\rm NETMA}, \sigma}(\omega)$ varies when one moves along the paths (b1) and (b2) in Fig.~\ref{fig.NBEC}(a). Figure~\ref{fig.NBEC}(c) shows the single-particle spectral function $A_{{\rm NETMA}, \sigma}(\bm{p}, \omega)$ at positions (1)-(3) in Fig.~\ref{fig.NBEC}(a). These figures indicate that the single-particle excitation spectra do not change so significantly as in the weak-coupling BCS regime at least within the variations of $\Delta\mu$ and $T_{\rm env}$ along the paths (b1) and (b2) in Fig.~\ref{fig.NBEC}(a).

On the other hand, the nonequilibrium effects appear in the photoemission spectrum $L_{{\rm NETMA}, \sigma}(\bm{p}, \omega)$ in Eq.~\eqref{eq.PES.NETMA}. Comparing Figs.~\ref{fig.NBEC}(d1) with (d2), the occupancy of the excitation modes along the particle dispersion (upper branch) is higher in (d2). These excitation modes correspond to single-particle excitations of Fermi atoms due to the dissociation of diatomic molecules~\cite{Tsuchiya2009}. Thus, the higher intensity of the upper branch in Fig.~\ref{fig.NBEC}(d2) means that the diatomic molecules are more destroyed by strong pumping and decay of Fermi atoms in the large $\Delta\mu$ regime. 

We briefly note that in Fig.~\ref{fig.NBEC}(d3), the occupancy of the upper branch is lower than in (d2). This is because the thermal dissociation of diatomic molecules is suppressed in the low $T_{\rm env}$ region.

\section{Stable nonequilibrium FFLO superfluid in the driven-dissipative lattice Fermi gas \label{sec.stable.NFFLO}}

\subsection{Driven-dissipative lattice Fermi gas}
As explained in Sec.~\ref{sec.NETMA.weak}, the strong nonequilibrium FFLO pairing fluctuations resulting from the continuous rotational symmetry of the gas system completely destroy the FFLO-type long-range order in the driven-dissipative Fermi gas. However, this also indicates that, if the infinite degeneracy shown in Fig.~\ref{fig.instability} is lifted and FFLO-type pairing fluctuations can be suppressed, by removing the continuous rotational symmetry of the system, the nonequilibrium FFLO superfluid state may be stabilized.

\begin{figure}[t]
\centering
\includegraphics[width=8.5cm]{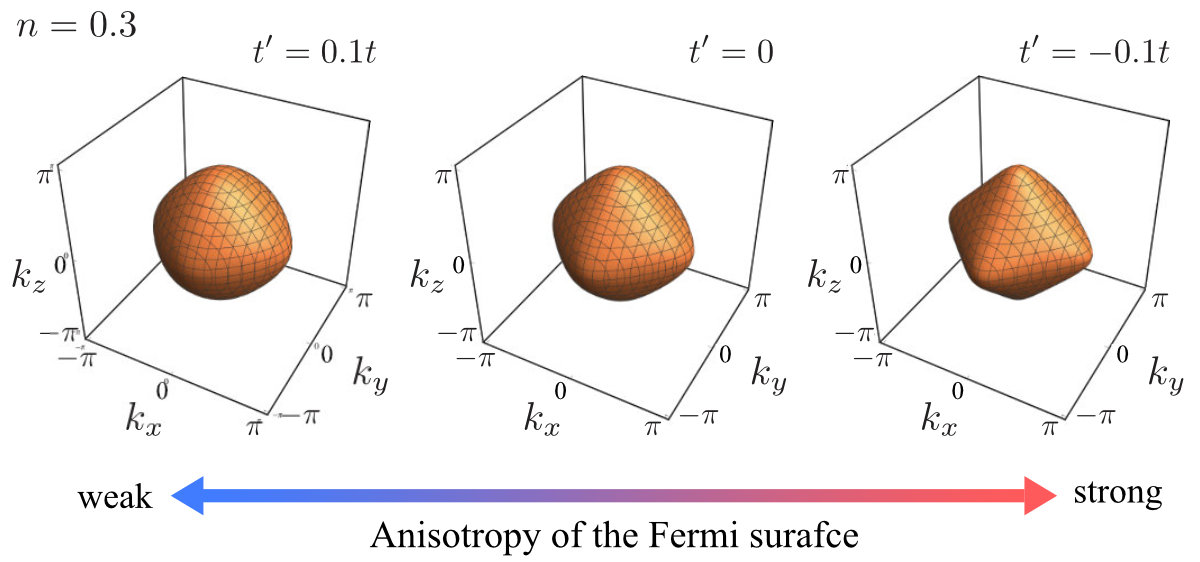}
\caption{Fermi surface of a free Fermi gas with the band dispersion $\ep_{\bm{k}}$ in Eq.~\eqref{eq.ep.lattice} and effects of the NNN hopping $t'$. We take $n=0.3$.}
\label{fig.FS}
\end{figure}

To explore this possibility, we consider the case when the main system is loaded on a three-dimensional cubic optical lattice~\cite{Greiner2002, Kohl2005, Jordens2008}. To model this situation, we describe the main system by the three-dimensional attractive Hubbard model~\cite{Tamaki2008}, given by
\begin{align}
H_{\rm sys} 
&= 
\sum_{\sigma=\up, \down}\sum_{\bm{k}} \ep_{\bm{k}} a^\dagger_{\bm{k}, \sigma} a_{\bm{k}, \sigma} 
\notag\\
&\hspace{0.2cm}
-U \sum_{\bm{k}, \bm{k}', \bm{q}} a^\dagger_{\bm{k}+\bm{q}/2, \up} a^\dagger_{-\bm{k}+\bm{q}/2, \down} a_{-\bm{k}'+\bm{q}/2, \down} a_{\bm{k}'+\bm{q}/2, \up}.
\label{eq.Hsys.lattice}
\end{align}
Here, $-U$ is the on-site $s$-wave pairing interaction, and the kinetic energy $\ep_{\bm{k}}$ of the lattice fermion has the form
\begin{align}
\ep_{\bm{k}}&= 
-2t \sum_{j=x, y, z}\big[ \cos(k_j)-1\big] 
\notag\\[4pt]
&\hspace{0.9cm}
-4t'\big[\cos(k_x)\cos(k_y) +\cos(k_y) \cos(k_z) 
\notag\\[4pt]
&\hspace{1.5cm}
+\cos(k_z) \cos(k_x)-3\big],
\label{eq.ep.lattice}
\end{align}
with $t$ and $t'$ being, respectively, the nearest-neighbor and the next-nearest-neighbor (NNN) hopping amplitude. In Eq.~\eqref{eq.ep.lattice}, the lattice constant $a$ is taken to be unity, for simplicity. In the following discussions, we use the NNN hopping $t'$ to tune the anisotropy of the Fermi surface: As shown in Fig.~\ref{fig.FS}, the Fermi surface shape gradually deviates from the spherical one as $t'$ increases. Thus, by adjusting the value of $t'$, we can control the spatial anisotropy of the main system.

\subsection{Hartree self-energy in the driven-dissipative lattice Fermi gas}

In the absence of the optical lattice, the Hartree self-energy ($\propto U$) vanishes because $U\to 0$ in the limit of $p_{\rm c} \to \infty$, where $p_{\rm c}$ is the momentum cutoff in Eq.~\eqref{eq.as.U}~\cite{Ohashi2020}. On the other hand, in the case of the lattice Fermi gas, $U \neq 0$ for a given $s$-wave scattering length $a_{\rm s}$, due to the finite bandwidth of the energy band $\ep_{\bm{k}}$ in Eq.~\eqref{eq.ep.lattice}. Thus, the Hartree term gives a nonzero contribution in the lattice system, and we need to incorporate it into the theory.

\begin{figure}[t]
\centering
\includegraphics[width=8cm]{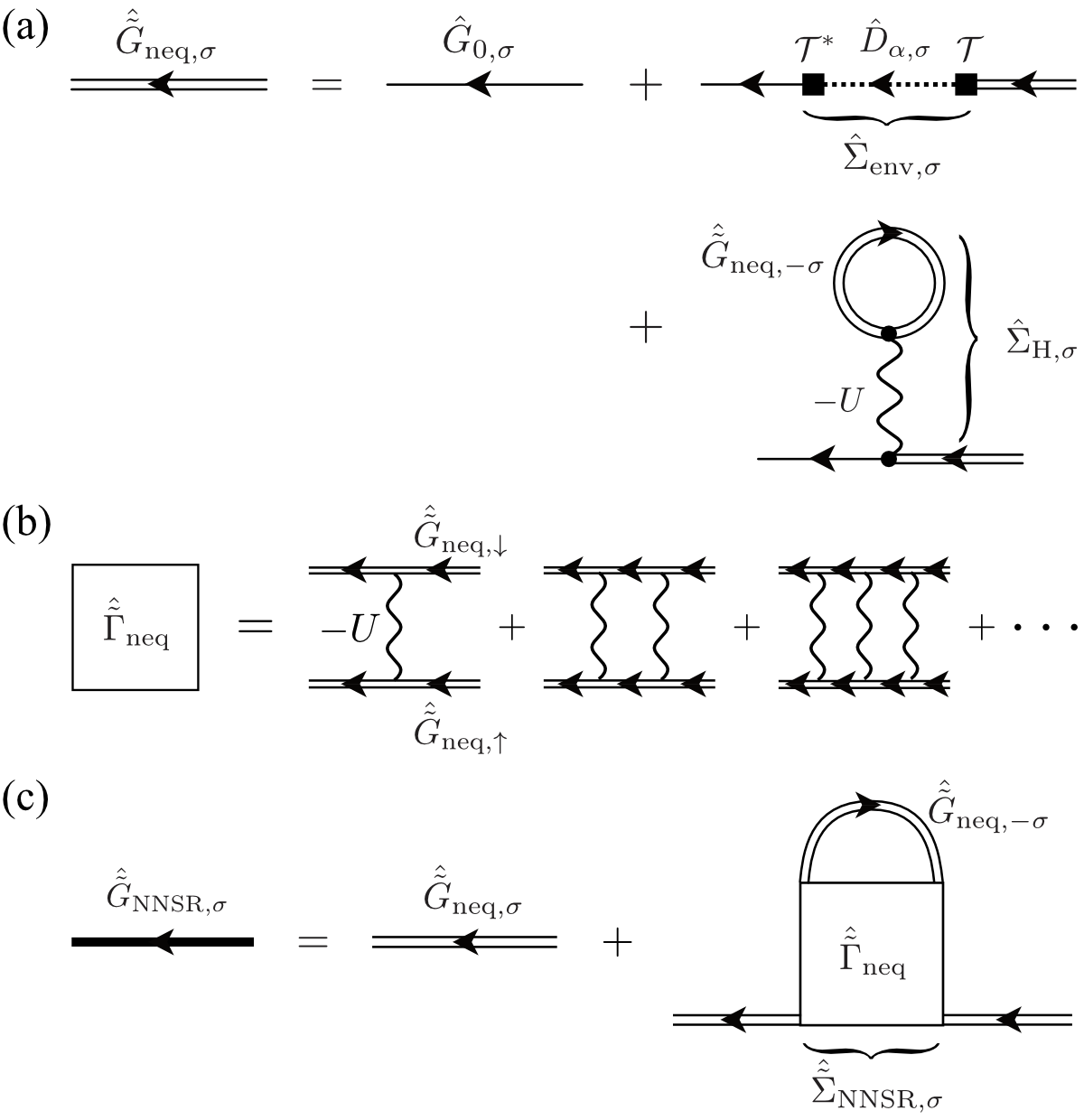}
\caption{(a) Dyson equation for the nonequilibrium Green's function $\hat{\tilde{G}}_{{\rm neq}, \sigma}$ in the presence of the optical lattice. (b) Particle-particle scattering matrix $\hat{\tilde{\Gamma}}_{\rm neq}$ in the driven-dissipative lattice Fermi gas. (c) Diagrammatic representation of the NENSR Green's function $\hat{\tilde{G}}_{{\rm NENSR}, \sigma}$ in Eq.~\eqref{eq.Dyson.NENSR}.}
\label{fig.NENSR.diagram}
\end{figure}

The Hartree self-energy $\hat{\Sigma}_{{\rm H}, \sigma}(\bm{k}, \omega)$ is diagrammatically given in  Fig.~\ref{fig.NENSR.diagram}(a), which gives~\cite{Kawamura2023}
\begin{align}
\hat{\Sigma}_{{\rm H}, \sigma}(\bm{k}, \omega) &= 
\begin{pmatrix}
\Sigma_{{\rm H}, \sigma}^{\rm R}(\bm{k}, \omega) & 
\Sigma_{{\rm H}, \sigma}^{\rm K}(\bm{k}, \omega) \\[4pt]
0 &
\Sigma_{{\rm H}, \sigma}^{\rm A}(\bm{k}, \omega) 
\end{pmatrix}
\notag\\
&=
\begin{pmatrix}
U n_{{\rm H}, -\sigma} &0  \\
0 &U n_{{\rm H}, -\sigma} 
\end{pmatrix},
\end{align}
where
\begin{equation}
n_{{\rm H}, \sigma} = -\frac{i}{2}\sum_{\bm{k}} \int_{-\infty}^\infty \frac{d\omega}{2\pi} \tilde{G}^{\rm K}_{{\rm neq}, \sigma}(\bm{k}, \omega) +\frac{1}{2}
\label{eq.filling.NMF}
\end{equation}
is the filling fraction of Fermi atoms at each lattice site. Here, $\tilde{G}^{\rm K}_{{\rm neq}, \sigma}(\bm{k}, \omega)$ is the Keldysh component of the nonequilibrium Green's function $\hat{\tilde{G}}_{{\rm neq}, \sigma}(\bm{k}, \omega)$ in the lattice system, which obeys the Dyson equation~\cite{Kawamura2023}
\begin{align}
&
\hat{\tilde{G}}_{{\rm neq}, \sigma}(\bm{k}, \omega) =
\hat{G}_{0, \sigma}(\bm{k}, \omega)
\notag\\[4pt]
&+
\hat{G}_{0, \sigma}(\bm{k}, \omega) 
\big[
\hat{\Sigma}_{{\rm H}, \sigma}(\bm{k}, \omega) + \hat{\Sigma}_{{\rm env}, \sigma}(\bm{k}, \omega)
\big]
\hat{\tilde{G}}_{{\rm neq}, \sigma}(\bm{k}, \omega).
\label{eq.Dyson.lattice}
\end{align}
This equation is also diagrammatically drawn as Fig.~\ref{fig.NENSR.diagram}(a). In the Dyson equation~\eqref{eq.Dyson.lattice}, $\hat{G}_{0, \sigma}(\bm{k}, \omega)$ is the Green's function in the isolated non-interacting Fermi gas loaded on the optical lattice, which is obtained by simply replacing $\ep_{\bm{p}}=p^2/(2m)$ in Eq.~\eqref{eq.G0.Keldysh} with the dispersion $\ep_{\bm{k}}$ in Eq.~\eqref{eq.ep.lattice}. The self-energy correction $\hat{\Sigma}_{{\rm env}, \sigma}(\bm{k}, \omega)$ due to the system-reservoir couplings is given in Eq.~\eqref{eq.self.env}. Solving the Dyson equation~\eqref{eq.Dyson.lattice}, one has
\begin{equation}
\hat{\tilde{G}}_{{\rm neq}, \sigma}(\bm{k}, \omega) =
\begin{pmatrix}
\frac{1}{\omega -\tilde{\ep}_{\bm{k}, \sigma} +2i\gamma} & 
\frac{-4i\gamma [1 -f(\omega -\mu_{\rm L}) -f(\omega -\mu_{\rm R})]}{[\omega -\tilde{\ep}_{\bm{k}, \sigma}]^2 + 4\gamma^2} \\[6pt]
0 & \frac{1}{\omega -\tilde{\ep}_{\bm{k}, \sigma} -2i\gamma}
\end{pmatrix},
\label{eq.Gneq.lattice}
\end{equation}
where the renormalized the kinetic energy $\tilde{\ep}_{\bm{k}, \sigma}$ has the form, 
\begin{equation}
\tilde{\ep}_{\bm{k}, \sigma} = \ep_{\bm{k}} -U n_{{\rm H}, -\sigma}.
\label{eq.tild.ep}
\end{equation}
Since $n_{{\rm H}, -\sigma}$ in Eq.~\eqref{eq.tild.ep} is determined from Eq.~\eqref{eq.filling.NMF}, we need to solve self-consistently Eqs.~\eqref{eq.filling.NMF} and \eqref{eq.Gneq.lattice}. 

\subsection{Nonequilibrium Nozi\`{e}res-Schmitt-Rink strong-coupling theory}

Although the spatial isotropy can be removed by considering the tight-binding model given in Eq.~\eqref{eq.ep.lattice}, it also makes numerical calculation harder than the isotropic case, because we cannot simplify computations by using the spatial isotropy of the main system. Thus, we take into account pairing fluctuations in the present lattice system by extending the strong-coupling theory developed by Nozi\`{e}res and Schmitt-Rink (NSR)~\cite{NSR1985} in the thermal equilibrium state to the nonequilibrium steady state.

The nonequilibrium NSR theory (NENSR) may be viewed as a simplified version of NETMA explained in Sec.~\ref{sec.formalism.NETMA}, because the nonequilibrium Green's function $\hat{\tilde{G}}_{{\rm NENSR}, \sigma}(\bm{k}, \omega)$ in NENSR obeys the truncated Dyson equation,   
\begin{align}
&
\hat{\tilde{G}}_{{\rm NENSR}, \sigma}(\bm{k}, \omega)=
\scalebox{0.98}{$\displaystyle
\begin{pmatrix}
\tilde{G}^{\rm R}_{{\rm NENSR}, \sigma}(\bm{k}, \omega) &  
\tilde{G}^{\rm K}_{{\rm NENSR}, \sigma}(\bm{k}, \omega) \\
0 &
\tilde{G}^{\rm A}_{{\rm NENSR}, \sigma}(\bm{k}, \omega)  
\end{pmatrix}$}
\notag\\
&= \hat{\tilde{G}}_{{\rm neq}, \sigma}(\bm{k}, \omega) 
\notag\\
&\hspace{0.5cm}+
\hat{\tilde{G}}_{{\rm neq}, \sigma}(\bm{k}, \omega)
\hat{\tilde{\Sigma}}_{{\rm NENSR}, \sigma}(\bm{k}, \omega)
\hat{\tilde{G}}_{{\rm neq}, \sigma}(\bm{k}, \omega),
\label{eq.Dyson.NENSR}
\end{align}
which is diagrammatically drawn as Fig.~\ref{fig.NENSR.diagram}(c). [Note that the thick solid line in the last diagram in Fig.~\ref{fig.NETMA.Dyson} is replaced by the double solid line in Fig.~\ref{fig.NENSR.diagram}(c).] The NENSR self-energy $\hat{\tilde{\Sigma}}_{{\rm NENSR}, \sigma}(\bm{k}, \omega)$ is the same as the NETMA self-energy $\hat{\Sigma}_{{\rm NETMA}, \sigma}(\bm{k}, \omega)$ in Eqs.~\eqref{eq.sigR.NETMA} and \eqref{eq.sigK.NETMA}, except that the ladder diagrams consists of $\hat{\tilde{G}}_{{\rm neq}, \sigma}(\bm{k}, \omega)$ in Eq.~\eqref{eq.Gneq.lattice} instead of $\hat{G}_{{\rm neq}, \sigma}(\bm{p}, \omega)$ in Eq.~\eqref{eq.Gneq2}, as diagrammatically shown in Fig.~\ref{fig.NENSR.diagram}(b). Thus, the NENSR self-energy  is obtained by replacing $\hat{G}_{{\rm neq}, \sigma}(\bm{p}, \omega)$ and $\hat{\Gamma}_{\rm neq}(\bm{q}, \omega)$ in the NETMA self-energy with $\hat{\tilde{G}}_{{\rm neq}, \sigma}(\bm{k}, \omega)$ and $\hat{\tilde{\Gamma}}_{\rm neq}(\bm{q}, \omega)$, where
\begin{equation}
\hat{\tilde{\Gamma}}_{\rm neq}(\bm{q}, \nu)= \frac{-U}{1-U\hat{\tilde{\Pi}}_{\rm neq}(\bm{q}, \nu)}
\end{equation}
is the particle-particle scattering matrix in the presence of the optical lattice. Here the pair correlation function  $\hat{\tilde{\Pi}}_{\rm neq}(\bm{q}, \nu)$ in the lattice system is obtained by replacing $\hat{G}_{{\rm neq}, \sigma}(\bm{p}, \omega)$ in Eqs.~\eqref{eq.Pi.R.neq} and \eqref{eq.Pi.K.neq} with $\hat{\tilde{G}}_{{\rm neq}, \sigma}(\bm{k}, \omega)$.

The equation for the filling fraction $n$ per lattice site, which incorporates the effects of both pairing fluctuations and system-reservoir couplings, is obtained from the Keldysh component $\tilde{G}^{\rm K}_{{\rm NENSR}, \sigma}(\bm{k}, \omega)$ of the NENSR Green's function  as
\begin{align}
n
&=
-\frac{i}{2}\sum_{\sigma=\up, \down} 
\sum_{\bm{k}} \int_{-\infty}^\infty \frac{d\omega}{2\pi} \tilde{G}^{\rm K}_{{\rm NENSR}, \sigma}(\bm{k}, \omega) -\frac{1}{2}
\notag\\
&=
\sum_{\sigma=\up, \down}  n_{{\rm H}, \sigma} +\frac{i}{2}\sum_{\sigma=\up, \down}\sum_{\bm{k}} \int_{-\infty}^\infty \frac{d\omega}{2\pi}
\notag\\
&\hspace{2cm}\times
 \big[
\hat{\tilde{G}}_{{\rm neq}, \sigma}(\bm{k}, \omega)
\hat{\tilde{\Sigma}}_{{\rm NENSR}, \sigma}
\hat{\tilde{G}}_{{\rm neq}, \sigma}(\bm{k}, \omega) \big]^{\rm K}
\notag\\[4pt]
&\equiv
\sum_{\sigma=\up, \down} 
n_{{\rm H}, \sigma} +n_{{\rm FL}, \sigma}.
\label{eq. filling.NENSR}
\end{align}
Here, $n_{{\rm H}, \sigma}$ is given in Eq.~\eqref{eq.filling.NMF} and
\begin{equation}
\big[\hat{A}\hat{B}\hat{C} \big]^{\rm K} = A^{\rm R} B^{\rm R} C^{\rm K} +A^{\rm R} B^{\rm K} C^{\rm A} +A^{\rm K} B^{\rm A} C^{\rm A}.
\end{equation}
In Eq.~\eqref{eq. filling.NENSR}, $n_{{\rm FL}, \sigma}$ is the fluctuation correction to the filling fraction.

The nonequilibrium Thouless criterion in the NENSR theory reads
\begin{equation}
\big[\tilde{\Gamma}^{\rm R}_{\rm neq}(\bm{q}=\bm{q}_{\rm pair}, \nu=\mu_{\rm pair})\big]^{-1}=0.
\label{eq.Thouless.NENSR}
\end{equation}
As in the NETMA case, the real part and the imaginary part of Eq.~\eqref{eq.Thouless.NENSR} gives two equations, and the imaginary part gives the solution $\mu_{\rm pair}=2\mu$. Then, substituting this into the real part of Eq.~\eqref{eq.Thouless.NENSR}, one reaches the following $T_{\rm env}^{\rm c}$ equation for the driven-dissipative lattice Fermi gas:
\begin{widetext}
\begin{equation}
\frac{1}{U} = \gamma \sum_{\bm{k}} \int_{-\infty}^\infty \frac{d\omega}{2\pi} \frac{\big[2\omega +\tilde{\ep}_{\bm{k}+\bm{q}_{\rm pair}/2, \up} -\tilde{\ep}_{-\bm{k}+\bm{q}_{\rm pair}/2, \down}\big] \left[\tanh\left(\frac{\omega +\Delta \mu}{2T_{\rm env}}\right) +\tanh\left(\frac{\omega -\Delta \mu}{2T_{\rm env}}\right)\right]}{\big[(\omega +\tilde{\ep}_{\bm{k}+\bm{q}_{\rm pair}/2, \up} -\mu)^2 +4\gamma^2\big]\big[(\omega -\tilde{\ep}_{-\bm{k}+\bm{q}_{\rm pair}/2, \down} +\mu)^2 +4\gamma^2\big]}.
\label{eq.Thouless.NENSR.Re}
\end{equation}
\end{widetext}

In the NENSR theory, one solves the $T_{\rm env}^{\rm c}$ equation~\eqref{eq.Thouless.NENSR.Re}, together with the equation for the filling fraction $n$ in Eq.~\eqref{eq. filling.NENSR}, to self-consistently determine $T_{\rm env}^{\rm c}$ and $\mu(T_{\rm env}^{\rm c})$ for a given parameter set $(n, \gamma, \Delta \mu)$. In the $T_{\rm env}^{\rm c}$ equation~\eqref{eq.Thouless.NENSR.Re}, $\bm{q}_{\rm pair}$ is determined so as to obtain the highest $T_{\rm env}^{\rm c}$. The self-consistent solution with $\bm{q}_{\rm pair}=0$ ($\bm{q}_{\rm pair}\neq 0$)  describes the phase transition into the spatially uniform BCS (nonuniform FFLO) type superfluid.

\subsection{Stabilization of the nonequilibrium FFLO superfluid}

\begin{figure}[t]
\centering
\includegraphics[width=6.5cm]{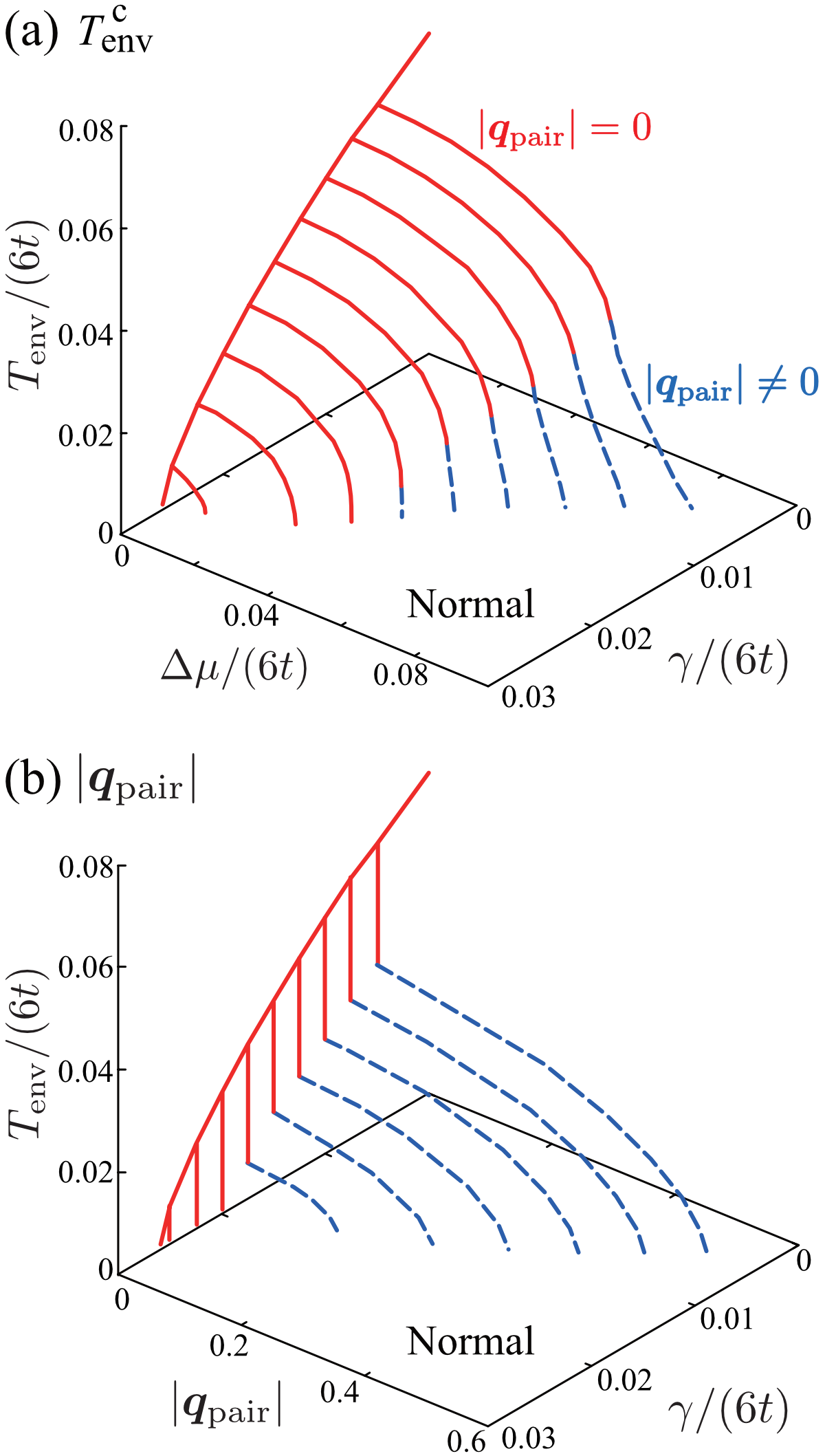}
\caption{(a) Calculated $T_{\rm env}^{\rm c}$ in the driven-dissipative lattice Fermi gas, as functions of the chemical potential bias $\Delta\mu$ and the system-reservoir coupling strength $\gamma$. (b) Magnitude $|\bm{q}_{\rm pair}|$ of the center-of-mass momentum of a Cooper pair at $T_{\rm env}^{\rm c}$. We take $t'=0$, $n=0.3$, and $U/(6t)=0.8$.}
\label{fig.NENSR.Tc} 
\end{figure}

Figure~\ref{fig.NENSR.Tc}(a) shows the NENSR results on $T_{\rm env}^{\rm c}$ in the driven-dissipative lattice Fermi gas. In contrast to the spatially isotropic case without the optical lattice, where the nonequilibrium FFLO superfluid is completely destroyed by pairing fluctuations as shown in Fig.~\ref{fig.NBCS}(a), Figs.~\ref{fig.NENSR.Tc}(a) and (b) clearly show that the nonequilibrium FFLO superfluid survives against their pairing fluctuations in the presence of the optical lattice.

\begin{figure}[t]
\centering
\includegraphics[width=8.5cm]{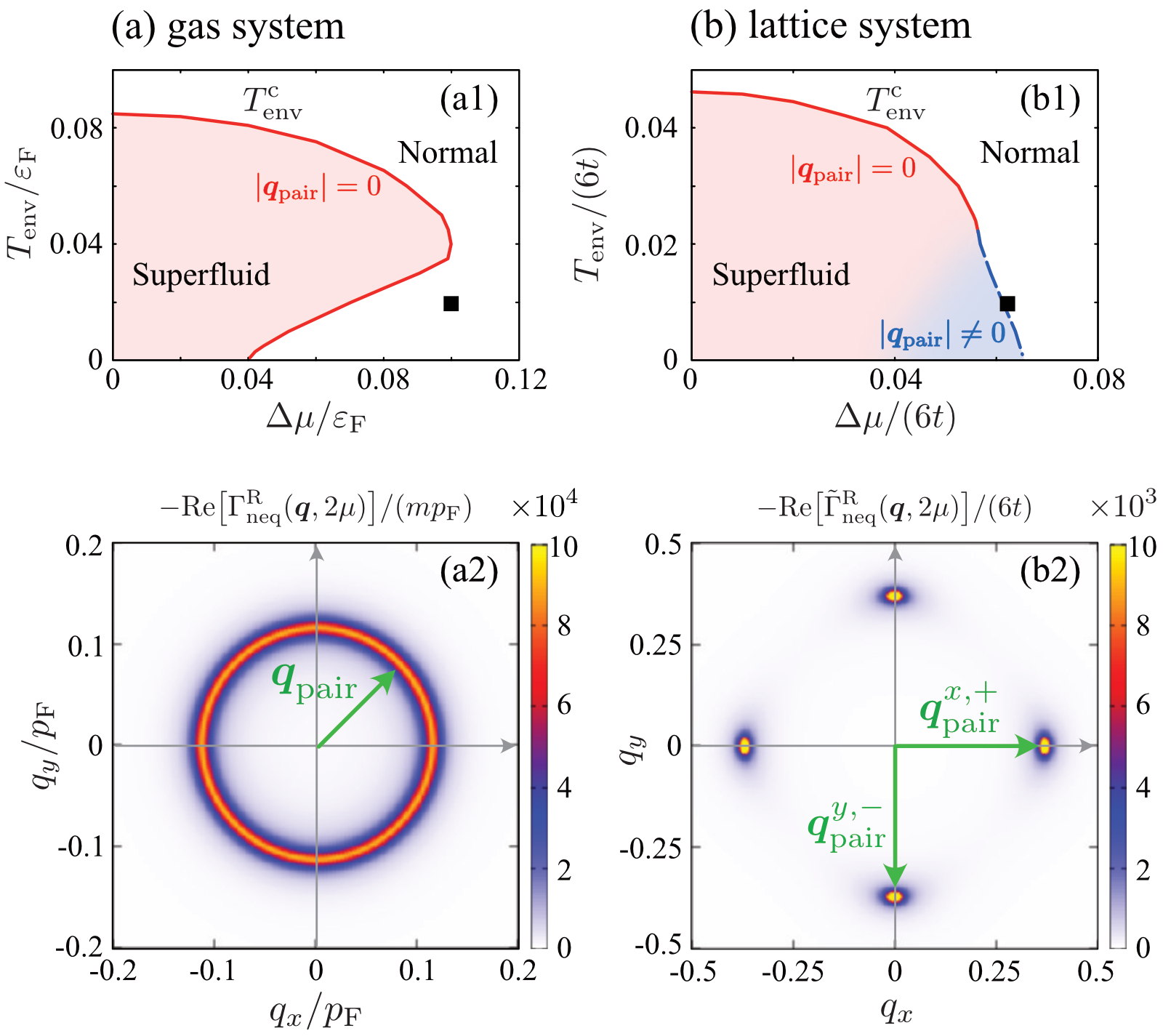}
\caption{(a1) Calculated $T_{\rm env}^{\rm c}$ in the absence of the optical lattice. This is the same plot as Fig.~\ref{fig.NBCS}(a). (a2) Calculated intensity $-{\rm Re}\big[\Gamma^{\rm R}_{\rm neq}(\bm{q}, 2\mu) \big]$ of the particle-particle scattering matrix at the solid square in panel (a1). Panel (b) shows the case in the presence of the optical lattice. In panel (b), we set $U/(6t)=0.8$, $\gamma/(6t)=0.01$, $t'=0$, and $n=0.3$.}
\label{fig.NENSR.gas.lattice}
\end{figure}

Figure~\ref{fig.NENSR.gas.lattice}(a2) shows the intensity $-{\rm Re}\big[\Gamma^{\rm R}_{\rm neq}(\bm{q}, 2\mu) \big]$ of the retarded particle-particle scattering matrix at the solid square in Fig.~\ref{fig.NENSR.gas.lattice}(a1) in the absence of the optical lattice. As explained in Sec.~\ref{sec.NETMA.weak}, this quantity has large intensity around $|\bm{q}|=|\bm{q}_{\rm pair}|\neq 0$, reflecting the enhancement of nonequilibrium FFLO pairing fluctuations. 

Figure~\ref{fig.NENSR.gas.lattice}(b2) shows the result at the solid square in Fig.~\ref{fig.NENSR.gas.lattice}(b1) in the presence of the optical lattice, which is quite different from the gas case shown in Fig.~\ref{fig.NENSR.gas.lattice}(a2). Since the spatial isotropy is explicitly broken by the optical lattice, the ring structure seen in Fig.~\ref{fig.NENSR.gas.lattice}(a2) is not obtained. Instead, Fig.~\ref{fig.NENSR.gas.lattice}(b2) shows that $-{\rm Re}\big[\tilde{\Gamma}^{\rm R}_{\rm neq}(\bm{q}, 2\mu) \big]$ has peaks at $\bm{q}=\bm{q}^{j, \eta}_{\rm pair}$ ($j=x,y,z$ and $\eta=\pm$), reflecting the {\it discreate} rotational symmetry of the optical lattice potential. 

The above difference between the gas case and the lattice case results in a significant difference in the fluctuation correction term $n_{{\rm FL}, \sigma}$ in Eq.~\eqref{eq. filling.NENSR}: In the presence of the optical lattice, noting that the $\Gamma^{\rm R}_{\rm neq}(\bm{q}, 2\mu)$ is enhanced around $(\bm{q}, \nu)= (\bm{q}_{\rm piar}^{j, \eta}, 2\mu)$ near the nonequilibrium FFLO transition, we can approximate the NENSR self-energy to
\begin{align}
&
\hat{\tilde{\Sigma}}_{{\rm NENSR}, \sigma}(\bm{k}, \omega)
\simeq 
-\Delta_{\rm PG}^2 \sum_{\eta=\pm} \sum_{j=x,y,z}
\notag\\
&
\scalebox{0.93}{$\displaystyle
\times
\begin{pmatrix}
\tilde{G}^{\rm A}_{{\rm neq}, -\sigma}(\bm{q}^{j, \eta}_{\rm pair} -\bm{k}, 2\mu -\omega) & 
\tilde{G}^{\rm K}_{{\rm neq}, -\sigma}(\bm{q}^{j, \eta}_{\rm pair} -\bm{k}, 2\mu -\omega)  \\[6pt]
0 &
\tilde{G}^{\rm R}_{{\rm neq}, -\sigma}(\bm{q}^{j, \eta}_{\rm pair} -\bm{k}, 2\mu -\omega) 
\end{pmatrix}$}
\notag\\[4pt]
&=
-\Delta_{\rm PG}^2 \sum_{\eta=\pm} \sum_{j=x,y,z}\hat{\tilde{G}}_{{\rm neq}, -\sigma}^*(\bm{q}^{j, \eta}_{\rm pair} -\bm{k}, 2\mu -\omega),
\label{eq.self.NENSR.app}
\end{align}
where
\begin{equation}
\Delta_{\rm PG}^2 = \frac{i}{2} \sum_{\bm{q}} \int_{-\infty}^\infty \frac{d\nu}{2\pi} \tilde{\Gamma}^{\rm K}_{\rm neq}(\bm{q}, \nu)
\label{eq.PG.NENSR}
\end{equation}
is the pseudogap parameter. Substituting Eq.~\eqref{eq.self.NENSR.app} into $n_{{\rm FL}, \sigma}$ in Eq.~\eqref{eq. filling.NENSR}, one has
\noindent
\begin{align}
&
n_{{\rm FL}, \sigma}= 
-\frac{i\Delta_{\rm PG}^2}{2} 
\sum_{\eta=\pm} 
\sum_{j =x, y, z}
\sum_{\bm{k}} 
\notag\\
&\hspace{0.5cm}\times
\int_{-\infty}^\infty \frac{d\omega}{2\pi} \big[\hat{\tilde{G}}_{{\rm neq}, \sigma}(\bm{k}, \omega)\hat{\tilde{G}}_{{\rm neq}, -\sigma}^*(\bm{q}_{\rm pair}^{j, \eta} -\bm{k}, 2\mu -\omega)
\notag\\
&\hspace{1.9cm}\times
\hat{\tilde{G}}_{{\rm neq}, \sigma}(\bm{k}, \omega) \big]^{\rm K}.
\label{eq.NFL.NENSR.app}
\end{align}
To evaluate the pseudogap parameter $\Delta_{\rm PG}$ in Eq.~\eqref{eq.PG.NENSR}, we also employ the following approximation:
\begin{equation}
\tilde{\Gamma}^{\rm R}_{\rm neq}(\bm{q}, \nu) \simeq \sum_{\eta=\pm} \sum_{j =x,y,z} \frac{-U}{C\big[\bm{q}-\bm{q}_{\rm pair}^{j, \eta}\big]^2 -i\lambda \big[\nu -2\mu\big]},
\label{eq.Tmat.app}
\end{equation}
where
\begin{align}
&
C = \left.\frac{U}{2} \nabla_{\bm{q}}^2 \tilde{\Pi}^{\rm R}_{\rm neq}(\bm{q}, 2\mu) \right|_{\bm{q}=\bm{q}_{\rm pair}^{j, \eta}}, 
\\[4pt]
&
\lambda = \frac{\pi U}{8 T_{\rm env}} N(\mu) {\rm sech}^2 \left(\frac{\Delta \mu}{2T_{\rm env}} \right),
\end{align}
with $N(\mu)$ being the density of states in the main system at $\omega=\mu$.  In obtaining Eq.~\eqref{eq.Tmat.app}, for simplicity, we have taken the limit $\gamma\to 0^+$ in $\tilde{\Pi}^{\rm R}_{\rm neq}(\bm{q}, \nu)$ as,
\begin{align}
&
\lim_{\gamma \to 0^+} \tilde{\Pi}^{\rm R}_{\rm neq}(\bm{q}, \nu) 
\notag\\
&=
\scalebox{0.95}{$\displaystyle
\sum_{\bm{p}} \frac{1 -f(\tilde{\ep}_{\bm{p}+\bm{q}/2, \up} -\mu -\Delta\mu) -f(\tilde{\ep}_{-\bm{p}+\bm{q}/2, \down} -\mu +\Delta\mu)}{\nu +i\delta -\tilde{\ep}_{\bm{p}+\bm{q}/2, \up} -\tilde{\ep}_{-\bm{p}+\bm{q}/2, \down}}$},
\label{eq.PiR.neq.limit}
\end{align}
and have expanded Eq.~\eqref{eq.PiR.neq.limit} around $(\bm{q}, \nu) = (\bm{q}_{\rm pair}^{j, \eta}, 2\mu)$. Substituting Eq.~\eqref{eq.Tmat.app} into Eq.~\eqref{eq.PG.NENSR}, one has
\begin{align}
\Delta_{\rm PG}^2 
&=
\frac{i}{2} \sum_{\bm{q}} \int_{-\infty}^\infty \frac{d\nu}{2\pi} |\tilde{\Gamma}^{\rm R}_{\rm neq}(\bm{q}, \nu) |^2 \tilde{\Pi}^{\rm K}_{\rm neq}(\bm{q}, \nu)
\notag\\
&\simeq 
\sum_{\eta=\pm} \sum_{j=x, y, z} \frac{i U^2 \tilde{\Pi}^{\rm K}_{\rm neq}(\bm{q}_{\rm pair}^{j, \eta} , 2\mu)}{2}
\notag\\
&\hspace{0.5cm}\times
\sum_{\bm{q}} \int_{-\infty}^\infty \frac{d\nu}{2\pi}  \frac{1}{C^2\big[\bm{q} -\bm{q}_{\rm pair}^{j, \eta}\big]^4 +\lambda^2 \big[\nu -2\mu\big]^2}
\notag\\[4pt]
&=
\sum_{\eta=\pm} \sum_{j =x, y, z} 
\frac{i U^2 \tilde{\Pi}^{\rm K}_{\rm neq}(\bm{q}_{\rm pair}^{j, \eta}, 2\mu)}{4 \lambda C}\sum_{\bm{q}} \frac{1}{\big[\bm{q} -\bm{q}_{\rm pair}^{j, \eta} \big]^2}.
\label{eq.PG.app.lattice}
\end{align}
In the same manner, we can evaluate the pseudogap parameter in the spatially isotropic gas system, which reads~\cite{Kawamura2023}
\begin{equation}
\Delta^2_{\rm PG} \simeq 
\frac{i U^2 \Pi^{\rm K}_{\rm neq}(\bm{q}_{\rm pair}, 2\mu)}{4 \lambda C}
\int_0^{q_{\rm c}} \frac{dq}{2\pi^2}
 \frac{q^2}{\big[|\bm{q}| -|\bm{q}_{\rm pair}|\big]^2},
\label{eq.PG.app.gas}
\end{equation}
with $q_{\rm c}$ being a cutoff momentum. 

The crucial difference between Eq.~\eqref{eq.PG.app.lattice} and Eq.~\eqref{eq.PG.app.gas} is that the factor $1/[|\bm{q}| - |\bm{q}_{\rm pair}|]^2$ in the gas case is replaced by $1/[\bm{q} -\bm{q}^{j, \eta}_{\rm pair}]^2$ in the lattice Fermi gas. Then, the $q$ integral in the spatially isotropic case in Eq.~\eqref{eq.PG.app.gas} always {\it diverges} as far as $\bm{q}_{\rm pair} \neq 0$, irrespective of the system dimension. As seen from Eq.~\eqref{eq.NFL.NENSR.app}, the fluctuation correction term $n_{{\rm FL}, \sigma}$ in Eq.~\eqref{eq.NFL.NENSR.app} diverges when the pseudogap parameter $\Delta^2_{\rm PG}$ diverges. Because of this singularity, the nonequilibrium FFLO superfluid is prohibited in the spatially isotropic gas system. On the other hand, replacing $\bm{q}-\bm{q}_{\rm pair}^{j, \eta}$ by $\bm{q}$ in Eq.~\eqref{eq.PG.app.lattice}, one finds that the pseudogap parameter $\Delta^2_{\rm PG}$ in the lattice system converges even when $|\bm{q}_{\rm pair}^{j, \eta}|\neq 0$. Thus, in the presence of the optical lattice, the NENSR coupled equations \eqref{eq. filling.NENSR} with \eqref{eq.Thouless.NENSR} can be satisfied simultaneously at the nonequilibrium FFLO phase transition, where $|\bm{q}_{\rm pair}|\neq 0$.

\begin{figure}[t]
\centering
\includegraphics[width=6.5cm]{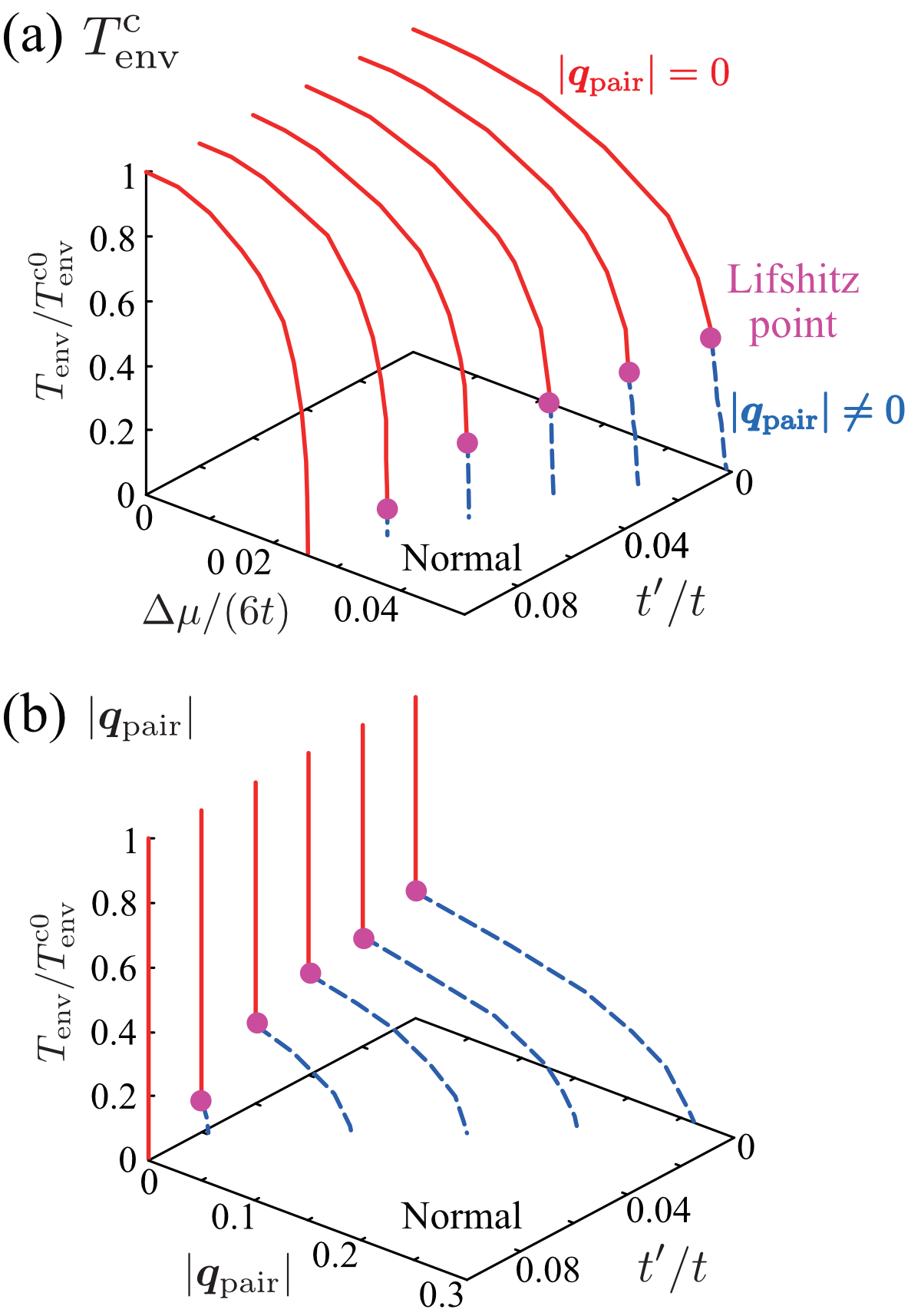}
\caption{(a) Calculated $T_{\rm env}^{\rm c}$ and effects of the NNN hopping amplitude $t'$. (b) $|\bm{q}_{\rm pair}|$ at $T_{\rm env}^{\rm c}$. We take $n=0.3$, $U/(6t)=0.8$, and $\gamma/(6t)=0.015$. The temperature $T_{\rm env}$ is normalized by the value at $\Delta\mu=0$ ($\equiv T_{\rm env}^{\rm c0}$). The solid circle shows the Lifshitz point, at which the normal, the nonequilibrium BCS, and the nonequilibrium FFLO phases meet one another.}
\label{fig.NENSR.NNN}
\end{figure}

The essence of the stabilization of the nonequilibrium FFLO-type long-range order is, strictly speaking, not the optical lattice itself, but rather the resulting anisotropy of the Fermi surface edges in momentum space.

 To demonstrate this, we show in Fig.~\ref{fig.NENSR.NNN} how the stabilization of the nonequilibrium FFLO state is affected by the anisotropy of the Fermi surface. (We recall that, as shown in Fig.~\ref{fig.FS}, the Fermi surface becomes more spherical for larger $t'$.) As expected, the nonequilibrium FFLO phase transition is gradually suppressed as the Fermi surface shape becomes close to the sphere by increasing the value of $t'$. In particular, the temperature at the boundary between the nonequilibrium BCS and the FFLO phase (solid circle in Fig.~\ref{fig.NENSR.NNN}), which is referred to as the Lifshitz point in the literature \cite{ChaikinBook, Hornreich1975}, decreases with increasing $t'$. This is because stronger nonequilibrium FFLO pairing fluctuations are introduced by more spherical Fermi surface edges.

In the current stage of cold Fermi gas physics, a superfluid $^6{\rm Li}$ Fermi gas in an optical lattice has been realized only when the lattice potential is very shallow~\cite{Chin2006}. Although this experimental situation is somehow different from the simple Hubbard model in Eq.~\eqref{eq.Hsys.lattice},  Fig.~\ref{fig.NENSR.NNN} indicates that the crucial key to stabilize the nonequilibrium FFLO superfluid is not the detailed lattice potential but the resulting anisotropic Fermi surface edge. Thus, if such a shallow optical lattice can still deform the Fermi surface enough to suppress the nonequilibrium FFLO pairing fluctuations, the nonequilibrium FFLO superfluid would be realized there.

\begin{figure*}[t]
\centering
\includegraphics[width=15cm]{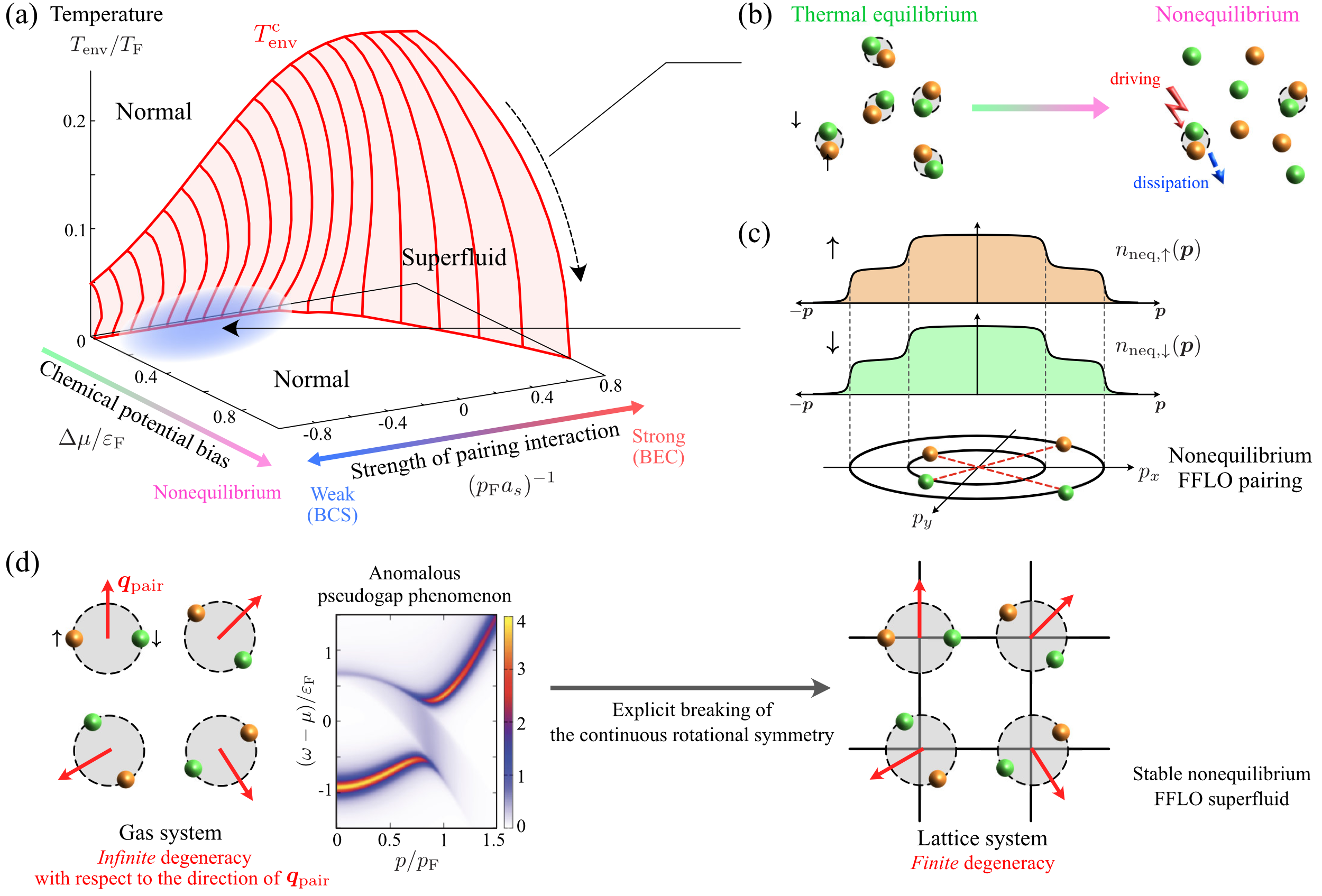}
\caption{Summary of this article. (a) Phase diagram of the driven-dissipative Fermi gas in the nonequilibrium BCS-BEC crossover regime (see Sec.~\ref{sec.NETMA.neq.BCS.BEC}). (b) In the BEC regime, as the chemical potential bias $\Delta\mu$ increases, the diatomic molecules are destroyed and the superfluid phase is monotonically suppressed (see Sec.~\ref{sec.NETMA.BEC}). (c) In the BCS regime, the nonequilibrium momentum distribution $n_{{\rm neq},\sigma}(\bm{p})$ induces FFLO-type Cooper pairs with nonzero center-of-mass momentum $\bm{q}_{\rm pair}$ (see Sec.~\ref{sec.NFFLO.NMF}). (d) In the gas system with continuous rotational symmetry, the nonequilibrium FFLO superfluid is unstable against paring fluctuations due to the infinite degeneracy with respect to the direction of $\bm{q}_{\rm pair}$ (see Sec.~\ref{sec.NETMA.weak}). This instability causes the re-entrant behavior of $T_{\rm env}^{\rm c}$ and the anomalous pseudogap phenomenon. In the presence of the optical lattice, the infinite degeneracy is lifted, so that the nonequilibrium FFLO superfluid can survive against paring fluctuations (see Sec.~\ref{sec.stable.NFFLO}).}
\label{fig.summary}
\end{figure*}

\section{Summary}

Figure~\ref{fig.summary} summarizes the nonequilibrium BCS-BEC crossover physics in the driven-dissipative Fermi gas. We have calculated the superfluid transition temperature $T_{\rm env}^{\rm c}$ in the nonequilibrium BCS-BEC crossover regime, as shown in Fig.~\ref{fig.summary}(a). This is done by extending the $T$-matrix approximation, developed in thermal equilibrium BCS-BEC crossover physics, to the driven-dissipative nonequilibrium Fermi gas, by employing the Keldysh Green's function technique.

The behavior of $T_{\rm env}^{\rm c}$ is quite different between the weak-coupling BCS and the strong-coupling BEC regime: In the BEC regime, $T_{\rm env}^{\rm c}$ is monotonically suppressed with increasing of the chemical potential bias $\Delta\mu$. This is because the strong pumping and decay of atoms destroy the diatomic molecules, as schematically shown in Fig.~\ref{fig.summary}(b). On the other hand, $T_{\rm env}^{\rm c}$ shows the re-entrant behavior in the BCS regime. This phenomenon is caused by strong FFLO-type pairing fluctuations, being enhanced by the nonequilibrium momentum distribution with the two-step structure, as shown in Fig.~\ref{fig.summary}(c). The instability of the nonequilibrium FFLO superfluid against pairing fluctuations results in the re-entrant behavior of $T_{\rm env}^{\rm c}$, as well as the anomalous pseudogap phenomenon.

A possible route to realize the long-range order of this nonequilibrium FFLO state is the removal of the infinite degeneracy with respect to the direction of the center-of-mass momentum $\bm{q}_{\rm pair}$ of the nonequilibrium FFLO Cooper pairs in the isotropic gas system. Indeed, when the main system is loaded on the cubic optical lattice, the lifting of this infinite degeneracy suppresses pairing fluctuations, which stabilizes the long-range order of the nonequilibrium FFLO superfluid in the driven-dissipative lattice Fermi gas.

The recent progress in cold Fermi gas experiments has enabled us to study BCS-BEC crossover physics in a variety of nonequilibrium situations. With this experimental progress, the theoretical understanding of nonequilibrium BCS-BEC crossover physics would become increasingly important in this research field.

\begin{acknowledgments}
T.K. was supported by MEXT and JSPS KAKENHI Grant-in-Aid for JSPS fellows Grant No. JP24KJ0055. Y.O. was supported by a Grant-in-aid for Scientific Research from MEXT and JSPS in Japan (No. JP22K03486).
\end{acknowledgments}

\bibliography{Review}%

\end{document}